\documentclass[aip, jcp, reprint, final]{revtex4-1}

\usepackage{graphicx}
\usepackage{amsmath}
    \allowdisplaybreaks[1]
\usepackage{hyperref}
\usepackage[T1]{fontenc}
\usepackage[utf8]{inputenc}
\usepackage{upgreek}
\usepackage{siunitx}
\usepackage{multirow}
\usepackage{xspace}
\usepackage{physics}
\usepackage[version=4]{mhchem}
\usepackage{mathptmx}

\newcommand{\ASigma}{{$\mathrm{A}\,{}^2\Sigma^+$}\xspace}

\newcommand{\BPi}{{$\mathrm{B}\,{}^2\Pi$}\xspace}
\newcommand{\CPi}{{$\mathrm{C}\,{}^2\Pi$}\xspace}
\newcommand{\XPi}{{$\mathrm{X}\,{}^2\Pi$}\xspace}
\newcommand{\DSigma}{{$\mathrm{D}\,{}^2\Sigma^+$}\xspace}
\newcommand{\LPi}{{$\mathrm{L}\,{}^2\Pi$}\xspace}
\newcommand{\KPi}{{$\mathrm{K}\,{}^2\Pi$}\xspace}
\newcommand{\LpPhi}{{$\mathrm{L}'\,{}^2\Phi$}\xspace}
\newcommand{\HpPi}{$\mathrm{H}'\,{}^2\Pi$\xspace}

\newcommand{\Ldoubling}{{$\varLambda$\,-\,doubling}\xspace}
\newcommand{\abinitio}{\textit{ab~initio}\xspace}
\newcommand{\NO}{\ce{^{14}N^{16}O}\xspace}

\newcommand{\red}[1]{{\color{red} #1}}
\newcommand{\blue}[1]{{\color{blue} #1}}

\usepackage[commentmarkup=uwave]{changes}
    \setaddedmarkup{\blue{#1}}
    \setdeletedmarkup{\red{\sout{#1}}}

\newcommand{\duo}{\textsc{Duo}\xspace}

\newcommand{\eg}{\textit{e.g.}\xspace}
\newcommand{\etal}{\textit{et~al.}\xspace}
\newcommand{\etc}{\textit{etc.}\xspace}
\newcommand{\ie}{\textit{i.e.}\xspace}

\draft 

\begin{document}


\title{A spectroscopic model for the low-lying electronic states of NO}



\author{Qianwei Qu, Bridgette Cooper, Sergei N. Yurchenko and Jonathan Tennyson}
\email[]{j.tennyson@ucl.ac.uk}
\affiliation{Department of Physics and Astronomy, University College London, London WC1E 6BT, United Kingdom}


\date{20 January 2021}

\begin{abstract}
    The rovibronic structure of \ASigma, \BPi\ and \CPi\
    states of nitric oxide (NO)
    is studied with the aim of producing comprehensive line lists for its near ultraviolet spectrum.
    \replaced{Empirical}{Empricial} energy levels for the three electronic states are determined using the a combination of the empirical MARVEL procedure
    and \abinitio\ calculations, and the available experimental data are critically evaluated. {\it Ab inito} methods which deal simultaneously with the Rydberg-like  \ASigma\ and \CPi,
    and the valence \BPi\ state are tested. 
    Methods of modeling the sharp avoided crossing between the \BPi\ and \CPi\ states are tested.
    A rovibronic Hamiltonian matrix is constructed using variational nuclear motion program \duo
    whose eigenvalues are fitted to the MARVEL energy levels.
    The matrix also includes coupling terms obtained from the refinement of the \abinitio\
    potential energy and spin-orbit coupling curves.
    Calculated and observed energy levels
    agree well with each other, validating the applicability of our method and providing a useful
    model for this open shell system.
\end{abstract}

\pacs{\red{The article has been accepted by \emph{The Journal of Chemical Physics}.}}

\maketitle 

\section{Introduction}
\label{sec:intro}
    
    Nitric oxide (NO)
    is one of the principle oxides of nitrogen.
    It plays a significant role in
    the nitrogen cycle of our atmosphere \cite{10CaGlFa.NO,97ViAbRo.NO}
    but also causes problems of air pollution and acid rain \cite{94ChKaYi.NO,96LiDrBu.NO,07SiAgxx.NO}.
    Therefore,
    scientists are devoting increasing
    attention to reducing
    NO in combustion processes
    \cite{00HuNaKo.NO,98LiLuRu.NO}.
    NO is a biological messenger for both animals and plants \cite{92BrSnxx.NO, 94LoDiSn.NO, 07ArFlxx.NO}
    but it may be harmful or even deadly as well \cite{97MaHexx.NO, 02EsJoxx.NO}.
    Apart from on Earth,
    NO was also observed 
    in the interstellar environments and 
    atmospheres of other planets \cite{91ZiMcMi.NO, 08CoSaGe.NO,08GeCoSa.NO,09GeGoSo.NO}.
    
    The importance of
    NO has aroused the interest of academia and industry 
    since it was prepared 
    by van Helmont in the 17th century \cite{36Partin.NO}
    and then studied by Priestley in 1772 \cite{72Priest.NO}.
    In numerous theoretical and experimental works,
    there are large number of spectroscopic investigations,
    as spectra provide a powerful weapon to reveal the physical and chemical
    properties of the molecule.
    For instance, 
    as a stable open shell molecule,
    the electronically excited Rydberg states
    of NO have been extensively studies, 
    see the paper of Deller and Hogan\cite{20DeHoxx.NO}
    and references therein.
    The spectrum of NO was
    also of great value in many applications, 
    such as temperature measurements
    by laser induced fluorescence\cite{04BeScxx.NO,13VaHrJa.NO}.

    The {ExoMol} project\cite{jt528} computes
    molecular line lists studies of exoplanet and (other) hot atmospheres.
    The ExoMol database was formally released in  2016  \cite{jt631}. The most recent
    2020 version  \cite{jt810}
    covers the line lists of 80 molecules 
    and 190 isotopologues, totaling 700 billion transitions.
    It includes 
    an accurate infrared (IR) line list of NO, called \texttt{NOname},
    which contains the rovibrational transitions within the ground electronic state \cite{jt686}.
    The rovibronic transitions of NO in the ultraviolet (UV) region are not included in \texttt{NOname}.
    These bands are strong,  atmospherically
    important 
    and have been observed  in many studies
    \cite{58LaMixx.NO, 97DaDoKe.NO, 06YoThMu.NO}.
    There is no NO UV line list  in
    well-known databases such as 
    {HITRAN} \cite{jt691} and {GEISA} \cite{jt636}
    either.
    
    Luque and Crosley have
    investigated spectra of diatomic molecules
    over a long period \cite{95LuCrxx.NO,99LuCrxx.NO,00LuCrxx.NO}.
    Based on their works,
    they developed a spectral simulation program, {LIFBASE} \cite{99LIFBASE},
    providing a database of
    OH, OD, CH \etc, and NO as well.
    {LIFBASE} contains
    the positions and relative probabilities 
    of UV transitions in four spectral
    systems of NO, \ie,
    $\upgamma$ (\ASigma to \XPi),
    $\upbeta$ (\BPi to \XPi),
    $\updelta$ (\CPi to \XPi) and
    $\upvarepsilon$ (\DSigma to \XPi) systems.
    The upper vibrational energy levels 
    for \BPi and \CPi of NO in {LIFBASE}
    are limited
    \added{to}
    below $v=7$
    and $v=1$, respectively.
    However,
    the observed $\upbeta$
    and $\updelta$ transitions
    corresponding to higher upper vibrational energy levels
    are even stronger \cite{98YoEsPa.NO,06YoThMu.NO}.
    There is a need  to develop a  comprehensive UV line list for NO to
    cover these band systems. To do this one first needs to 
    construct a spectroscopic model which requires overcoming
    a number of theoretical difficulties. The purpose of this
    paper is to present our model and explain how we resolve these difficulties.
    
    A major issue in generating a UV line list for NO results from
    the difficulty of modelling the interaction
    between \BPi\ and \CPi\ states,
    which is caused by the particular electronic structure of NO.
    To understand this fifteen electrons system  one must
    analyse the electron configuration of these states
    from the perspective of molecular orbitals.
    On one hand,
    excitation of inner paired electrons to higher valence orbitals
    leads to valence states such as \BPi.
    On the other hand,
    the outermost unpaired electron may be excited to
    Rydberg orbitals, 
    yielding a series of Rydberg states such \ASigma\ or \CPi.
    These Rydberg \replaced{states}{state} lie close in energy to the 
    valence ones. 
    Furthermore, as \ce{NO+}
    has a shorter equilibrium bondlength than NO \cite{79AlScZa.NOplus}, 
    Rydberg states tend to be lower in energy at short bondlengths, $r$, while
    valence states are lower at larger $r$.
    Thus, in NO,
    Rydberg-valence interactions are
    densely distributed in the neighbourhood of the
    equilibrium bond length of its ground state,
    where large Franck-Condon factors exist.
    The \BPi\,-\,\CPi\ interaction is
    the lowest one 
    and has attracted the most attention.
    As described by Lagerqivst and Miescher \cite{58LaMixx.NO},
    the two states show a strong and extended mutual perturbation.
    They proposed a
    `deperturbation' method to
    explain the vibrational and rotational perturbation
    of \BPi\,-\,\CPi interaction.
    Further analysis was made by
    Gallusser and Dressler \cite{82GaDrxx.NO},
    who set up a vibronic interaction matrix
    of five $^2\Pi$ states and
    fitted the eigenvalues of the matrix
    to experimental
    data in the determination of RKR potential curves 
    and off-diagonal electronic energies.
    As a consequence,
    \replaced{they predicted vibrational states of the \BPi electronic state up to 
    $v = 37$.}{
    they assigned
    vibrational quantum number of \BPi state up to $v = 37$.}
    
    In this paper,
    we propose a method based on
    directly diagonalizing a rovibronic matrix 
    to resolve the energy structures of \BPi\,-\,\CPi coupled states.
    This matrix is based on the use of full variational solution of the rovibronic nuclear motion Hamiltonian rather than
    perturbation theory. This method is general and can be used to predict spectra, for example at elevated temperatures.
    
\begin{figure}
        \centering
        \includegraphics{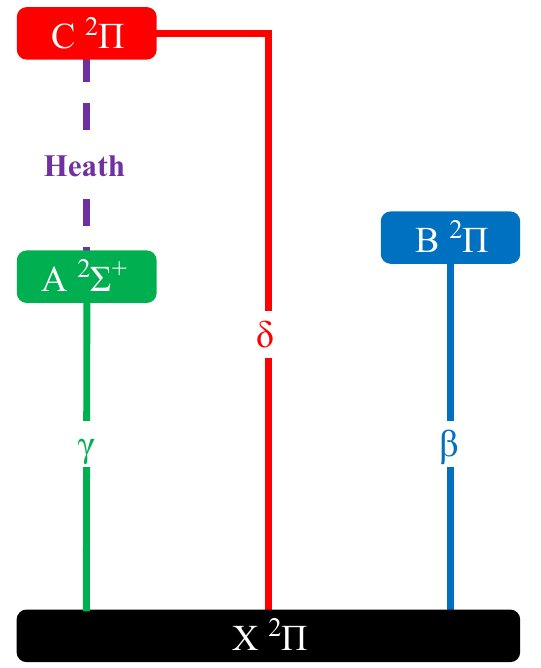}
        \caption{The band systems of 
        NO involved in this work and their names.
        The $\upgamma$,
        $\upbeta$ and $\updelta$
        systems mainly cover the UV transitions of NO.
        Jenkins \etal recorded many visible lines
        from \BPi state to higher vibrational levels of \XPi,
        \eg, those of $\upbeta(3, 16)$ band \cite{27JeBaMu.NO}.
        The high-accuracy IR transitions of
        $\mathrm{Heath}(0,0)$ band were measured by
        Amiot and Verges \cite{82Amiot.NO}.
        For a comprehensive band system diagram,
        see the work of  
        Cartwright \etal \cite{00CaBrCa.NO}
        }
        \label{fig:bandSystem}
    \end{figure}

    In addition to the vibronic matrix elements
    (\eg, spin doublets) considered in the previous studies,
    more fine structure terms, 
    such as \Ldoubling and spin-rotational coupling,
    are used to construct the rovibronic matrix.
    The eigenvalues of the matrix
    are fitted to rovibronic energies
    obtained using a {MARVEL} (measured active rotation-vibration energy levels) procedure \cite{jt412,jt750} 
    analysis of the observed NO IR/visible/UV transitions 
    to ensure a quantitatively accurate result.
    Figure\,\ref{fig:bandSystem} summarizes
    the band systems involved in our MARVEL analysis.
    The objective functions were constrained with 
    the \abinitio curves produced using Molpro \cite{MOLPRO2015}
    to avoid overfitting problems.
    The above procedures
    are also applied to the \ASigma state of NO
    to get a self-consistent description of the 
    doublet electronic states up to and including \CPi.

    This work forms the foundation of
    our future study on the generation of UV line list of NO.
    The modeling of \BPi\,-\,\CPi 
    paves the way
    for the investigations of molecules
    with similar avoided crossing structures.

\section{Theoretical study of the low-lying electronic states of NO}
\label{sec:abinitio}

\begin{figure}[!htb]
            \includegraphics{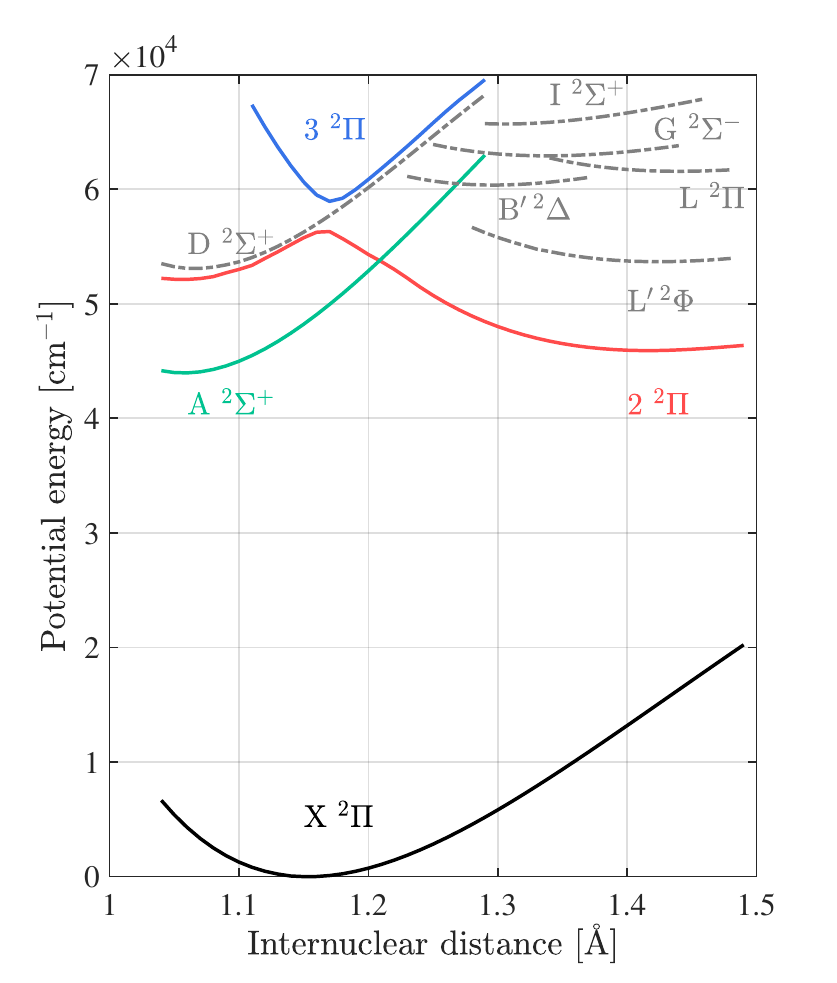}
            \caption{NO PECs calculated by Shi and East \cite{06ShEaxx.NO}.
                The states of interest are plotted by solid curves.
                Here, $2\,^2\Pi$ is the \BPi to \CPi PEC
                while $3\,^2\Pi$ is the \CPi to \BPi PEC.}
            \label{fig:ShEaxxPEC}
        \end{figure}

    Complete active space self-consistent field (CASSCF)
    and multireference configuration interaction (MRCI)
    calculations were performed in the quantum chemistry package Molpro 2015 \cite{MOLPRO}
    to get the potential energy and spin-orbit curves
    of the \XPi, \ASigma, \BPi and \CPi states. A major issue in the
    calculation is achieving a balance between representations of the Rydberg, A and C, states
    and the valence, X and B, states. Figure\,\ref{fig:ShEaxxPEC} presents an overview of the low-lying PECs
    and illustrates the importance of the \CPi -- \BPi Rydberg -- Valence avoided crossing.

    The history of high quality CI calculation for the excited states of NO 
    can be tracked back to 1982,
    when Grein and Kapur reported
    their work on the states with 
    the minimum electronic energies lower than \SI{6.58}{\eV} \cite{82GrKaxx.NO}.
    Several years later,
    a comprehensive theoretical study on NO 
    were presented and discussed 
    by de Vivie and Peyerimhoff \cite{88DePexx.NO}.
    The results of this paper was 
    further improved by Shi and East in 2006 \cite{06ShEaxx.NO}.
    More accurate curves 
    were obtained with extended basis set and active space
    in the recent \replaced{works}{work} of Cheng \etal \cite{17ChZhCh.NO, 17ChZhChS.NO}.
    Although the previous works 
    \cite{82GrKaxx.NO, 82Cooper.NO, 88LaBaPa.NO, 88DePexx.NO, 91LaPaBa.NO, 03PoFixx.NO, 06ShEaxx.NO, 17ChZhCh.NO}
    provide us strong inspiration,
    the task is still challenging due to
    the interactions between Rydberg and valence states of NO.
    
    \subsection{Active space and basis set}
    
        For heteronuclear diatomic molecules, 
        Molpro executes calculations in 
        four irreducible representations
        $a_1$, $b_1$, $b_2$ and $a_2$ of the $C_\mathrm{2v}$ point group.
        Here, we use
        $[(n_1, n_2, n_3, n_4) - (n'_1, n'_2, n'_3, n'_4)]$
        to represent occupied orbitals excluding closed orbitals,
        \ie the calculation active space.
        A typical active space for the lower electronic states calculation of 
        NO is $[(8, 3, 3, 0) - (2, 0, 0, 0)]$,
        as suggested by Shi and East \cite{06ShEaxx.NO}.
        Although only a few of the PECs are of direct interest here,
        we had to include extra states to achieve correct calculation.
        We also adjusted the active space to get smooth curves.
        
        A Dunning aug-cc-pV(n)Z
        basis set \cite{89Dunning.ai}
        was used in both CASSCF and MRCI calculation.
        This basis set
        has an additional shell of diffuse functions compare to the cc-pV(n)Z basis set,
        which benefits the calculation of Rydberg states.
        Too many diffuse functions,
        \eg, those of the d-aug-cc-pV(n)Z basis set, 
        may have negative effects on the calculation
        because of the overemphasis of the Rydberg states relative to the valence states.
            
    \subsection{CASSCF calculation}
\begin{figure*}[!htb]
            \includegraphics{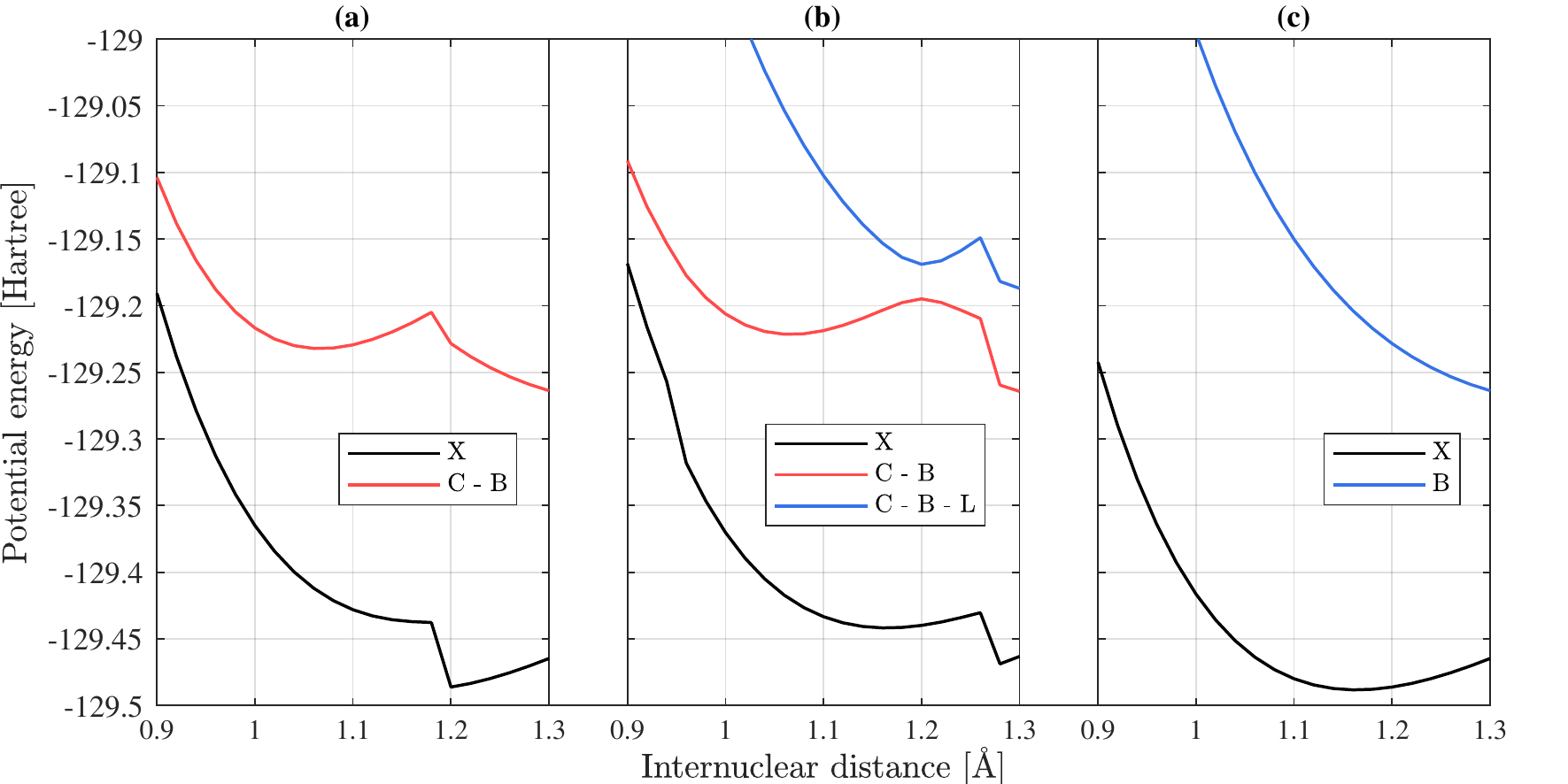}
            \caption{The PECs in 
                the active space of $[(8, 3, 3, 0) - (2, 0, 0, 0)]$
                with the basis set of aug-cc-pVTZ.
                (a) Two $^2\Pi$ states averaged CASSCF calculation
                starting from \SI{0.9}{\angstrom}.
                (b) Three $^2\Pi$ states averaged CASSCF calculation
                starting from \SI{0.9}{\angstrom}.
                (c) Two $^2\Pi$ states averaged CASSCF calculation
                starting from \SI{1.3}{\angstrom}.}
            \label{fig:casPi}
        \end{figure*}

        Our calculations started with 
        a  $[(8, 3, 3, 0) - (2, 0, 0, 0)]$ active space
        in which
        the interactions between the Rydberg
        and valence states are inescapable. However,
        representing the avoided crossing points caused by \CPi
        and the valence $^2\Pi$ states proved to be a 
        huge obstacle to obtaining satisfactory results.
        Panel (a) of Fig.\,\ref{fig:casPi} shows the terrible
        behavior of B\,-\,C interaction near \SI{1.18}{\angstrom}.
        The potential energy curve (PEC)
        of \CPi suddenly jumps to that of \BPi,
        producing discontinuity in the PEC of \XPi too.
        To get the exited states,
        we used the state average algorithm
        but the average energy of the two $^2\Pi$
        states changed when traversing the crossing point of \CPi and \BPi.
        
        A valid way to smooth the curves
        is to increase the number of averaged states.
        For example,
        the discontinuities near \SI{1.18}{\angstrom}
        disappears when introducing a third $^2\Pi$ state in CASSCF calculation,
        as shown in Panel (b) of Fig.\,\ref{fig:casPi}.
        Nevertheless,
        similar phenomenon arises when the third state comes across \LPi.
        Alternatively,
        smooth curves can be obtained in limited active space.
        For example,
        we can get a continuous curves of \CPi
        in the active space $[(6, 3, 3, 0) - (4, 1, 1, 0)]$ 
        from \SI{0.9}{\angstrom} to \SI{1.28}{\angstrom}.
        
        We always started a new CASSCF iteration from 
        the orbitals of a nearby geometry to stabilize and accelerate the calculation.
        The PECs in Panels (a) and (b) of Fig.\,\ref{fig:casPi},
        are obtained by increasing the internuclear distance from \SI{0.9}{\angstrom}
        to \SI{1.3}{\angstrom}.
        Interestingly, 
        with a initial geometry at \SI{1.3}{\angstrom},
        reversing the calculation direction gives a
        completely different result in the same active space,
        \ie, two smooth valence PECs of \XPi and \BPi states in
        Panel (c) of Fig.\,\ref{fig:casPi}.
        Due to the 
        limitation of nonlinear programming,
        CASSCF iterations may fall into local minima.
        To get the target states,
        the numerical optimization must be properly initialized.
        For the NO molecule,
        the iterations which begin with valence orbitals usually end with valence orbitals
        but it is uncertain for those begin with Rydberg orbitals.
        The results
        imply that there are at least two kinds of local minimums 
        in the \abinitio calculation of NO with Molpro:
        pure valence orbitals
        (corresponding to Panel (c) of Fig.\,\ref{fig:casPi}) 
        and Rydberg-valence hybrid orbitals
        (corresponding to Panels (a) and (b) of Fig.\,\ref{fig:casPi}).
        To verify the conjecture:
        initializing a calculation of two $^2\Pi$ states average
        with the CASSCF orbitals of the \XPi state in the single state calculation,
        one can get almost the same curves
        as those in Panel (c) of Fig.\,\ref{fig:casPi},
        starting from \SI{0.9}{\angstrom}.

        In Section \ref{sec:refine},
        we use diabatic potentials
        in modeling interaction between \BPi and \CPi states.
        We describe the curves as `adiabatic'
        if they contain the B\,-\,C avoided crossing feature,
        \eg, those in Panel (b) of Fig.\,\ref{fig:casPi}.
        If not,
        we call the curves `diabatic',
        \eg, those in Panel (c) of Fig.\,\ref{fig:casPi}.

    \subsection{MRCI calculation}  
\begin{figure*}[htb]
            \centering
            \includegraphics{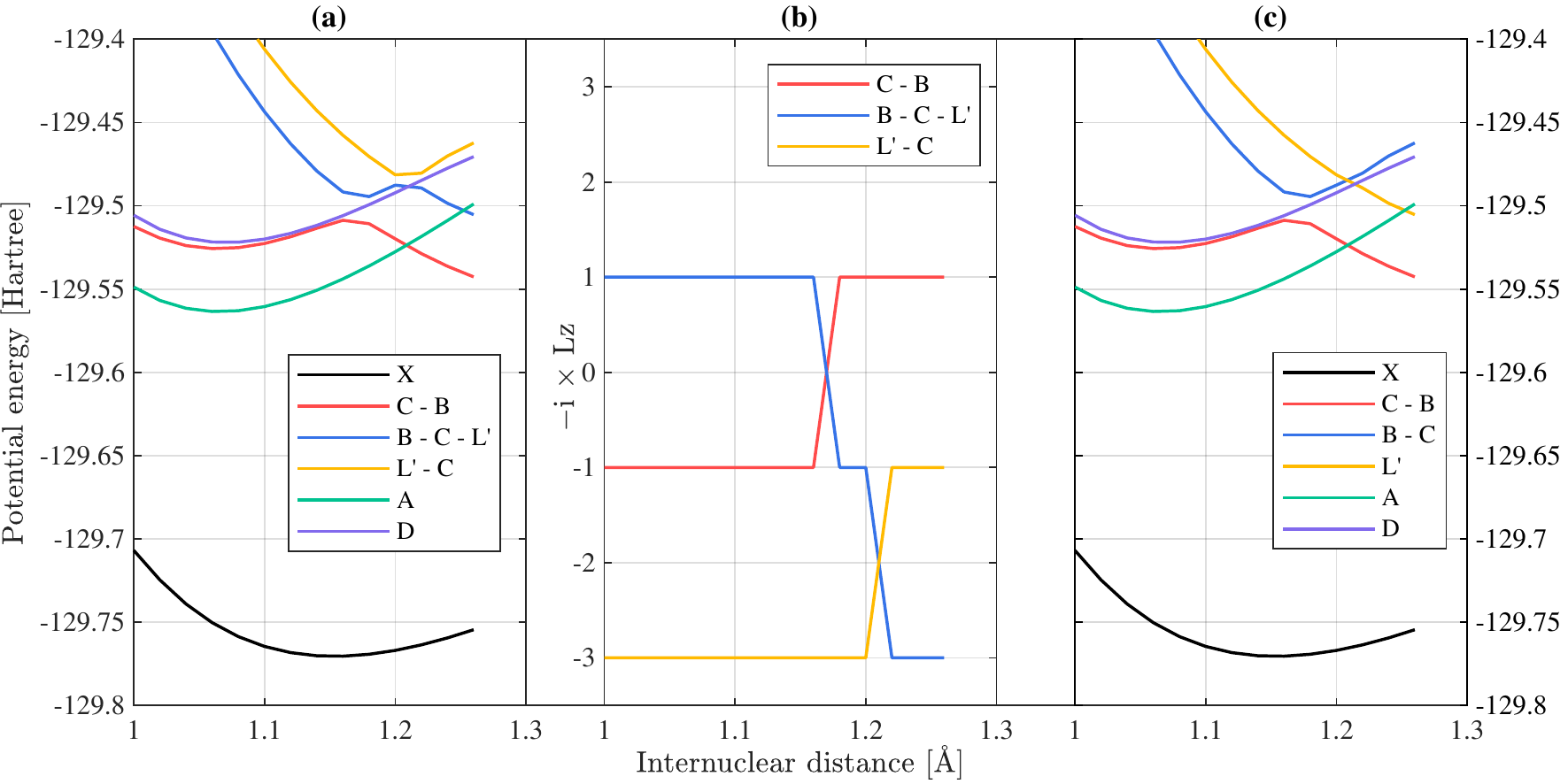}
            \caption{The PECs of the \XPi, \ASigma, 
                \BPi, \CPi, \DSigma and \LpPhi states,
                obtained by CASSCF \& {{MRCI\,+\,Q}} calculation
                starting from \SI{1.06}{\angstrom}
                to both sides
                in the active space $[(8, 3, 3, 0) - (2, 0, 0, 0)]$
                with the aug-cc-pV5Z basis set.
                The third $^2\Pi$ curve
                and the \LpPhi curve in Panel (a)
                were manually switched in Panel (c)
                on the right of \SI{1.2}{\angstrom},
                according to the value of
                \texttt{Lz},
                shown in Panel (b). 
                The phase of \texttt{Lz},
                in the Cartesian representation,
                is random.
                To distinguish different electronic states,
                the yellow curve in Panel (b)
                is smoothed for  
                internuclear distances less than 
                \SI{1.1}{\angstrom}. The
                \XPi
                state is not shown in the Panel as 
                the \texttt{Lz} values 
                obtained are all $-\mathrm{i}$.}
            \label{fig:ciAllPECs}
        \end{figure*}

        Although consuming many more computational resources,
        the MRCI calculation in Molpro is straightforward.
        Molpro automatically takes the CASSCF orbitals as the references
        and performs an internally contracted configuration interaction calculation
        based on single or double excitation.
        The spin-orbit coupling terms
        were also produced.
        To compensate the error brought by truncated configuration interaction expansion,
        the energies were modified by Davision correction,
        \ie, {{MRCI\,+\,Q}} calculation.
        Panel (a) of Fig.\,\ref{fig:ciAllPECs}
        demonstrates the results of CASSCF \& {{MRCI\,+\,Q}} calculation 
        of the \XPi, \ASigma, \BPi, \CPi, \DSigma and \LpPhi states,
        in $[(8, 3, 3, 0)-(2, 0, 0, 0)]$
        active space with aug-cc-pV5Z basis set.
        
        In the CASSCF routine,
        the projection of angular momentum of 
        a diatomic molecule on its internuclear axis,
        $\varLambda$, 
        can be assigned to specify the expected states.
        However,
        the MRCI routine does not have the option
        and always finds the lowest energy states of the same spin.
        As a result,
        the PECs of \CPi and \LpPhi exchange with each other
        at their crossing point
        although the avoided crossing principle
        is not applicable for the two states,
        as shown by the blue curve in Panel (a) of Fig.\,\ref{fig:ciAllPECs}.
        It is feasible to 
        calculate and output the $\varLambda$ quantum numbers
        (technically, \texttt{Lz}, 
        which is defined as a non-diagonal matrix element
        between two degenerate components, 
        \eg $\mel{\Pi_x}{\hat{L}_z}{\Pi_y}$) 
        in MRCI calculations, 
        which helps to distinguish the \CPi, \LpPhi and \LPi states.
        The blue and yellow curves 
        on the right of their crossing point were manually switched,
        as shown in Panel (c) of Fig.\,\ref{fig:ciAllPECs},
        according to their $\varLambda$ quantum numbers shown in Panel (b).
        \added{The $T_e$ values of \ASigma, \BPi and \CPi
        states are compared with those calculated
        by Shi and East in Table\,\ref{tab:compareTe}.}
        
        \begin{table}[!htb]
        \centering
        \caption{\added{Comparison of $T_e$ values of
        the MRCI + Q calculation}}
        \label{tab:compareTe}
        \begin{ruledtabular}
        \begin{tabular}{lllll}
        State & \multicolumn{2}{c}{CASSCF \& MRCI + Q} & \multicolumn{2}{c}{Empirical} \\
              & \multicolumn{1}{l}{Shi and East \cite{06ShEaxx.NO} } & \multicolumn{1}{l}{This work} & \multicolumn{1}{l}{Huber and Herzberg \cite{79HeHuxx.book}} & \multicolumn{1}{l}{This work\footnotemark[2]} \\ 
        \hline
        \ASigma     & \num{43558} &  \num{45410.2}      & \num{43965.7} & \num{43902.99} \\
        \BPi     & \num{44803} &  \num{46260.3}\footnotemark[1]     & \num{45913.6} & \num{45867.05} \\
        \CPi     & \num{51808} &  \num{53709.5}     & \num{52126} & \num{52081.97} \\
        \end{tabular}%
        \end{ruledtabular}
        \footnotetext[1]{Two-state average CASSCF \& MRCI + Q calculation.}
        \footnotetext[2]{See Section\,\ref{sec:refine}.}
        \end{table}

        The PECs
        in Fig.\,\ref{fig:ciAllPECs} range
        from \SIrange{1.0}{1.26}{\angstrom}.
        The curves were deliberately truncated at the right endpoint
        because of the \CPi\,-\,\LPi interaction
        as shown in Panel (b) of Fig.\,\ref{fig:casPi}.
        On the left endpoint,
        The MRCI program exited with 
        an `\texttt{INSUFFICIENT OVERLAP}' error.
        The error is triggered by interactions with another $^2\Pi$ state, \HpPi, which lies below
         \BPi near the \SI{1.06}{\angstrom} and
        which cannot be
        described by the reference space.
        A solution to the problem is
        to perform MRCI calculations using a larger active space such as
        $[(8, 4, 4, 0)-(2, 0, 0, 0)]$.
    
        \vspace{1em}
        It is not  quantitatively accurate
        to generate line lists with the \abinitio curves; however, the curves and couplings
        provide a suitable starting point for work.
        These curves and couplings need to be refined 
        using experimental data,
        which is the content of the subsequent two sections.
    
\section{MARVEL analysis of the rovibronic energy levels of \NO}
\label{sec:marvel}
    The rovibronic energy levels
    of \ASigma, \BPi and \CPi states were 
    reconstructed by MARVEL analysis of
    the experimental transitions of the
    $\upgamma$, $\upbeta$, $\updelta$,
    and $\mathrm{Heath}$ systems
    and those inside the ground state.
    
\begin{figure}[!htb]
        \includegraphics{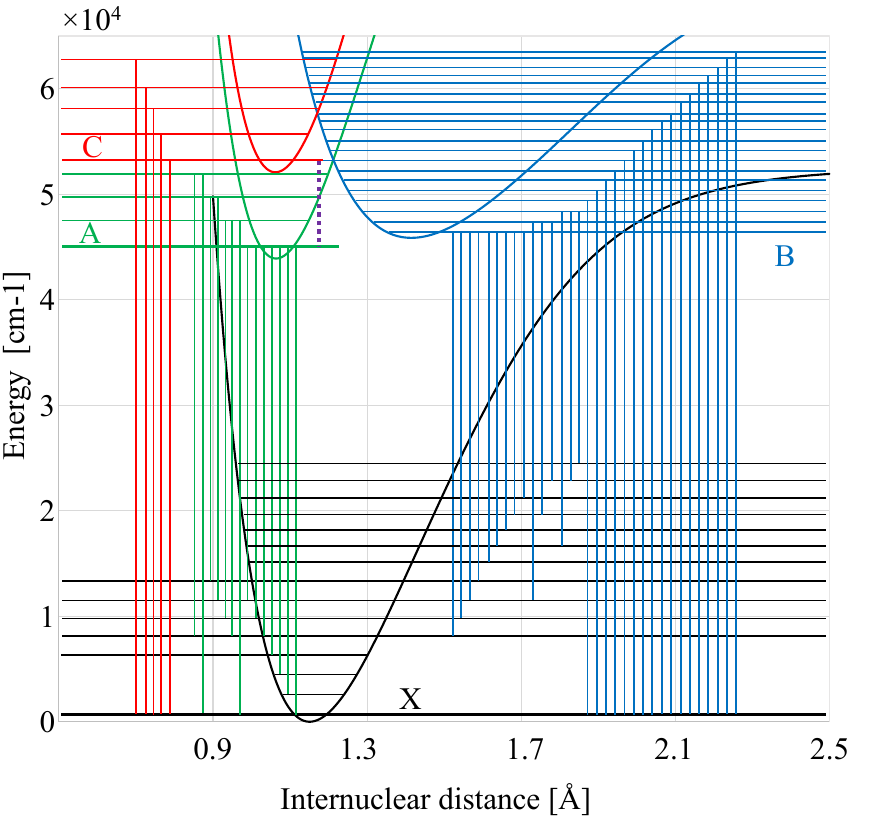}
        \caption{Vibronic structure of the MARVEL analysis.
            The vertical solid lines in green, blue and red illustrate the 
            bands of $\upgamma$, $\upbeta$ and $\updelta$ systems, respectively.
            The vertical dash line in purple represents
            the $\mathrm{Heath}(0,0)$ band.
            }
        \label{fig:marvel_network}
    \end{figure}

    In the previous work by Wong \etal \cite{jt686},
    \num{11136} IR transitions were collected,
    yielding a spectroscopic network of \num{4106} energy levels.
    To retrieve the energy levels of \ASigma,
    \BPi and \CPi states, 
    we extracted a further \num{9861} transitions
    (including \num{3393} $\upgamma$,
    \num{5103} $\upbeta$,
    \num{1004} $\updelta$
    and \num{361} Heath transitions)
    from the data sources
    listed in Table\,\ref{tab:marvel_source}.
    The vibronic structure of the 
    spectroscopic network is illustrated in Fig.\,\ref{fig:marvel_network}.
    
    Although there are studies which report measured transition frequencies for the four band  systems of interest,
    only the most reliable data sets were included in
    our MARVEL analysis.
    For example,
    Lagerqvist and Miescher published the line position data of
    20 bands of the $\upbeta$ and $\updelta$ systems
    ($\upbeta(5,0)$ to $\upbeta(19,0)$ and $\updelta(0, 0)$ to $\updelta(4, 0)$, respectively)
    in 1958 (58LaMi \cite{58LaMixx.NO}),
    but half of them were replaced by more accurate line lists
    measured by Yoshino \etal around 2000 
    (94MuYoEs \cite{94MuYoEs.NO}, 
    98YoEsPa \cite{98YoEsPa.NO},
    00ImYoEs \cite{00ImYoEs.NO},
    02ChLoLe \cite{02ChLoLe.NO},
    02RuYoTh \cite{02RuYoTh.NO},
    06YoThMu \cite{06YoThMu.NO}).
    
    \begin{table}
      \caption{ Data sources used in the final MARVEL analysis}
      \label{tab:marvel_source}
      \begin{ruledtabular}
      \begin{tabular}{llrrlrr}
        \multirow{2}[0]{*}{Source} & \multirow{2}[0]{*}{Band} & \multirow{2}[0]{*}{ $J_\textrm{min}''$} & \multirow{2}[0]{*}{$J_\textrm{max}''$} &{Uncertainty} & \multicolumn{2}{c}{Trans.\footnotemark[1]} \\
          &       &       &       &  [$\mathrm{cm}^{-1}$]    & (A)     & (V) \\
        \hline
        97DaDoKe \cite{97DaDoKe.NO} & $\upgamma(0,0)$   & 0.5   & 41.5 & 0.04 - 0.15 & 304& 277\\
        97DaDoKe  & $\upgamma(0,1)$   & 0.5   & 40.5 & 0.04 - 0.15 & 277& 245\\
        97DaDoKe & $\upgamma(0,2)$     & 1.5   & 39.5 & 0.04 - 0.15 & 339& 317\\
        97DaDoKe & $\upgamma(0,3)$    & 1.5   & 38.5 & 0.04 - 0.1 & 289& 279\\
        97DaDoKe & $\upgamma(0,4)$    & 1.5   & 42.5 & 0.04 - 0.1 & 294& 283\\
        97DaDoKe & $\upgamma(0,5)$    & 1.5   & 37.5 & 0.04 - 0.1 & 266& 249\\
        97DaDoKe & $\upgamma(0,6)$    & 1.5   & 31.5 & 0.04 - 0.15& 158& 142\\
        97DaDoKe & $\upgamma(1,0)$    & 0.5   & 30.5 & 0.04 - 0.15& 302& 275\\
        97DaDoKe & $\upgamma(1,4)$   & 0.5   & 41.5 & 0.04 - 0.15  & 295& 277\\
        97DaDoKe & $\upgamma(1,5)$    & 1.5   & 39.5 & 0.04 - 0.15 & 142& 135\\
        97DaDoKe & $\upgamma(2,6)$     & 1.5   & 40.5 & 0.04 - 0.15 & 277& 246\\
        97DaDoKe & $\upgamma(2,7)$    & 2.5   & 41.5 & 0.04 - 0.15 & 160& 155\\
        02ChLoLe \cite{02ChLoLe.NO} & $\upgamma(3,0)$   & 0.5   & 24.5 & 0.03 - 0.05 & 227& 205\\
        97DaDoKe & $\upgamma(3,4)$    & 4.5   & 32.5 & 0.04 - 0.2 & 63& 56\\
        27JeBaMu \cite{27JeBaMu.NO} & $\upbeta(0,4)$   & 0.5   & 24.5 & 0.2 & 122 & 52\\
        27JeBaMu  & $\upbeta(0,5)$   & 0.5   & 24.5 & 0.2 & 152 & 143\\
        27JeBaMu  & $\upbeta(0,6)$   & 0.5   & 24.5 & 0.2 & 126 & 124\\
        27JeBaMu  & $\upbeta(0,7)$   & 0.5   & 29.5 & 0.2 & 202 & 200\\
        27JeBaMu  & $\upbeta(0,8)$   & 0.5   & 31.5 & 0.2 & 206 & 204\\
        27JeBaMu  & $\upbeta(0,9)$   & 0.5   & 31.5 & 0.2 & 192 & 188\\
        27JeBaMu  & $\upbeta(0,10)$   & 0.5   & 31.5 & 0.2 & 208 & 202\\
        27JeBaMu  & $\upbeta(0,11)$   & 0.5   & 31.5 & 0.2 & 184 & 180\\
        27JeBaMu  & $\upbeta(0,12)$   & 0.5   & 22.5 & 0.2 & 138 & 138\\
        27JeBaMu  & $\upbeta(1,6)$   & 0.5   & 19.5 & 0.2 & 123 & 119\\
        27JeBaMu  & $\upbeta(1,11)$   & 0.5   & 24.5 & 0.2 & 148 & 142\\
        27JeBaMu  & $\upbeta(1,13)$   & 0.5   & 23.5 & 0.2 & 154 & 150\\
        27JeBaMu  & $\upbeta(2,9)$   & 0.5   & 22.5 & 0.2 & 138 & 130\\
        27JeBaMu  & $\upbeta(2,13)$   & 0.5   & 21.5 & 0.2 & 128 & 128\\
        27JeBaMu  & $\upbeta(2,14)$   & 0.5   & 21.5 & 0.2 & 144 & 139\\
        27JeBaMu  & $\upbeta(2,15)$   & 0.5   & 24.5 & 0.2 & 102 & 99\\
        92FaCo \cite{92FaCoxx.NO} & $\upbeta(3,0)$   & 0.5   & 31.5 & 0.05 - 0.1 & 432 & 426\\
        96DrWo \cite{96DrWoxx.NO} & $\upbeta(4,0)$   & 0.5   & 8.5 & 0.003 - 0.004 & 66 & 66\\
        96DrWo & $\upbeta(5,0)$   & 0.5   & 7.5 & 0.003 - 0.005 &52&52\\
        58LaMi \cite{58LaMixx.NO} & $\upbeta(5,0)$   & 8.5   & 14.5 & 0.2 &36 &36\\
        02ChLoLe & $\upbeta(6,0)$  & 0.5   & 17.5 & 0.03 - 0.1 &138&135\\
        94MuYoEs \cite{94MuYoEs.NO} & $\upbeta(7,0)$   & 0.5   & 7.5 & 0.03 - 0.1 &76 &60\\
        58LaMi & $\upbeta(7,0)$   & 6.5   & 16.5 & 0.2 - 0.25 &70 &64\\
        58LaMi & $\upbeta(8,0)$    & 0.5   & 16.5 & 0.2 &124&120\\
        98YoEsPa \cite{98YoEsPa.NO} & $\upbeta(9,0)$ & 0.5   & 23.5 & 0.02 - 0.03 &188&178\\
        06YoThMu \cite{06YoThMu.NO} & $\upbeta(10,0)$   & 0.5   & 12.5 & 0.03 - 0.15 &218 &193\\
        02RuYoTh \cite{02RuYoTh.NO} & $\upbeta(11,0)$    & 0.5   & 17.5 & 0.03 - 0.08 &134&125\\
        06YoThMu  & $\upbeta(12,0)$     & 0.5   & 20.5 & 0.03 - 0.15 &188 &173\\
        58LaMi & $\upbeta(13,0)$    & 11.5  & 18.5 & 0.2 & 97& 97\\
        06YoThMu & $\upbeta(14,0)$  & 0.5   & 20.5 & 0.03 - 0.08 & 196& 153\\
        58LaMi & $\upbeta(15,0)$     & 0.5   & 17.5 & 0.2 - 0.5 & 239 & 215\\
        58LaMi & $\upbeta(16,0)$     & 0.5   & 14.5 & 0.2 - 0.3 & 138& 133\\
        58LaMi & $\upbeta(17,0)$   & 0.5   & 11.5 & 0.2 - 0.5 &42&42\\
        58LaMi & $\upbeta(18,0)$    & 0.5   & 12.5 & 0.2 - 0.5 & 120 & 108\\
        58LaMi & $\upbeta(19,0)$    & 0.5   & 12.5 & 0.2 - 0.5 & 82 & 80\\
        94MuYoEs & $\updelta(0,0)$    & 0.5   & 20.5 & 0.03 - 0.1 &225&217\\
        00ImYoEs \cite{00ImYoEs.NO} & $\updelta(1,0)$   & 0.5  & 18.5 & 0.03 - 0.1 &261&205\\
        06YoThMu & $\updelta(2,0)$    & 0.5   & 21.5 & 0.03 - 0.15 &250 &210\\
        06YoThMu & $\updelta(3,0)$    & 0.5   & 18.5 & 0.03 - 0.08 &138&109\\
        58LaMi & $\updelta(4,0)$   & 0.5   & 11.5 & 0.2 - 0.6 &130&120\\
        82AmVe \cite{82AmVexx.NO} & $\mathrm{Heath}(0, 0)$   & 0.5   & 11.5 & 0.01  &361&360\\
        \end{tabular}
        \end{ruledtabular}
        \footnotetext[1]{Number of measured (A)  and validated (V) Transitions}
    \end{table}   
    
    The spectroscopic network in MARVEL \cite{MARVEL}
    is established in accordance with the upper and lower
    quantum numbers of the transitions.
    We used five quantum numbers,
    as shown in Table \ref{tab:marvel_quantum_number},
    to uniquely label the rovibronic energy levels.
    The quantum numbers of some transitions were improperly assigned.
    New assignments plus some other comments on the sources are given below:
    \begin{itemize}
        \item \deleted{  Most measurements did not resolve \Ldoubling or spin-rotational fine structures.} 
            In some cases  (\eg 
            for the \ASigma state,
            the $Q_{21}$ branch
            is indeed a copy of $R_{11}$ branch
            as listed in
            97DaDoKe \cite{97DaDoKe.NO}) duplicate
            transition are provided in source data.
            \added{In
            27JeBaMu \cite{27JeBaMu.NO},
            58LaMi \cite{58LaMixx.NO},
            \etc,
            \Ldoubling fine structures
            of many transitions
            are not resolved;} therefore
            we simply created two transitions differing
            in $e/f$ parity with the same frequency in the MARVEL dataset.
            
        \item The uncertainties of 
            the transitions of 27JeBaMu \cite{27JeBaMu.NO}
            and 58LaMi \cite{58LaMixx.NO} were given by combination difference
            test, referring to the energies of \XPi state \cite{jt686}.
            
        \item The uncertainties of most validated transitions
            are close to the lower bounds listed in Table\,\ref{tab:marvel_source}
            (see the supplementary material).
            
        \item The transitions of $\upgamma(3, 0)$, $\upbeta(6, 0)$ and
            $\upbeta(11, 0)$ bands extracted from 02ChLoLe
            \cite{02ChLoLe.NO}, 02ChLoLe \cite{02ChLoLe.NO} and 02RuYoTh \cite{02RuYoTh.NO} were increased by \SI{+0.083}{\per\cm},
            \SI{+0.083}{\per\cm} and \SI{+0.067}{\per\cm}, respectively,
            as suggested in 05ThRuYo \cite{05ThRuYo.NO}. 
            The uncertainties of these transitions
            should be \SI{0.1}{\per\cm}
            because the absolute
            frequencies were not calibrated \cite{05ThRuYo.NO}.
            However,
            we used relative accuracy, \ie, \SI{0.03}{\per\cm},
            as the lower bound of uncertainty
            to \replaced{constrain}{constraint} the MARVEL analysis.
            The uncertainties should be adjusted to \SI{0.1}{\per\cm}
            if data of higher accuracy are included in the future.
        \item In the $\upbeta(10, 0)$ band of 06YoThMu \cite{06YoThMu.NO},
            $R_{11}(3.5)$ and $P_{11}(3.5)$ were exchanged;
            the $R_{21}$ and $P_{21}$ branches were exchanged.
        \item In the $\updelta(0, 0)$ band of 94MuYoEs \cite{94MuYoEs.NO}, 
            $P_{12}(15.5)_e$ and $P_{12}(16.5)_f$ should be
            $P_{22}(15.5)_e$ and $P_{22}(16.5)_e$, respectively.
        \item In the $\updelta(1, 0)$ band of 00ImYoEs \cite{00ImYoEs.NO},
            the frequencies of $R_{12}(15.5)_e$ and $R_{12}(15.5)_f$ should be exchanged;
            the frequencies of $P_{11}(5.5)$ and $P_{11}(16.5)_f$ should be \SI{54668.636}{\per\cm}.
        \item In the $\updelta(2, 0)$ band of 06YoThMu \cite{06YoThMu.NO},
            the frequencies of $Q_{22}(5.5)_e$ and $Q_{22}(6.5)_e$ should be
            \num{56967.72} and \SI{56966.61}{\per\cm}, respectively.
        \item \added{The transitions,
        $R_{22}(0.5)_{ff}$,
        $Q_{22}(0.5)_{fe}$,
        $R_{12}(0.5)_{ee}$ 
        of 97DaDoKe \cite{97DaDoKe.NO}
        and
        $R_{22}(0.5)$ of
        02ChLoLe \cite{02ChLoLe.NO},
        are related to
        unknown lower states ($J=0.5$ and $\varOmega = 1.5$).
        Those transitions were
        not validated.
        }
        \end{itemize}
        
        The most serious issue we encountered 
        concerned the 2020 measurements of
        Ventura and Fellows (20VeFe\cite{20VeFexx.NO}) 
        who published a new line list for the $\upgamma$ system
        containing \num{6436} transitions. 
        The transitions of 20VeFe disagree with those 
        measured by Danielak \etal (97DaDoKe) \cite{97DaDoKe.NO}.
        MARVEL and combination difference analysis indicates that
        their data set is self-consistent within the
        claimed accuracy, \ie \SIrange{0.005}{0.06}{\per\cm}.
        However,
        it is inconsistent with the ground state MARVEL energies of Wong \etal \cite{jt686}. 
        Combination difference test shows that the standard deviations of
        most energy levels calculated by the data set
        are greater than \SI{0.1}{\per\cm}.
            
\begin{figure}
            \centering
            \includegraphics{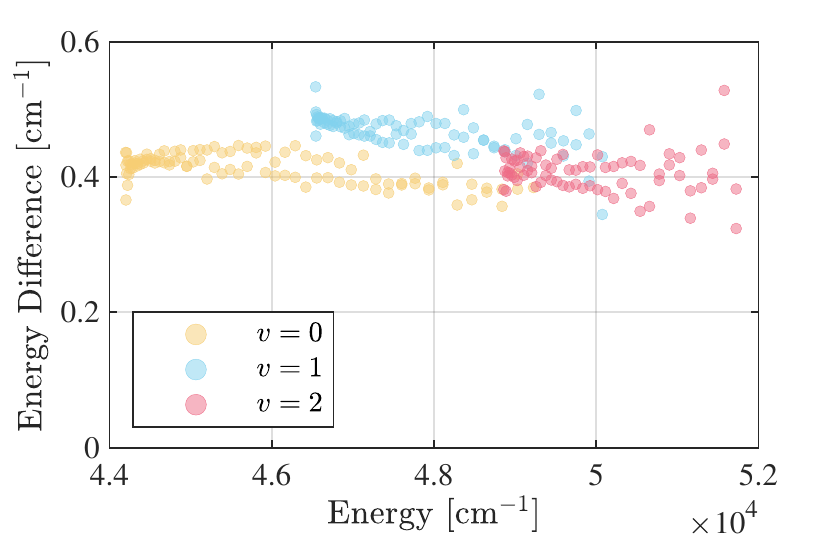}
            \caption{The differences
            between the \ASigma state MARVEL energies
            corresponding to 20VeFe and
            those obtained from the sources of 
            Table\,\ref{tab:marvel_source}.
            The average energy shift is
            \SI{0.43}{\per\cm}.}
            \label{fig:shift20Ve}
        \end{figure}

        In contrast,  
        the line list of 97DaDoKe \cite{97DaDoKe.NO}
        is consistent with others.  
        The measurements of 20VeFe differ
        from those of 97DaDoKeby up to \SI{0.7}{\per\cm}, 
        as acknowledged by 20VeFe.
        The transitions $\upgamma(3,4)$ band measured by 97DaDoKe
        are consistent with the transitions in $\upgamma(3,0)$
        band measured by Cheung \etal (02ChLoLe) \cite{02ChLoLe.NO}.
        Furthermore, use of Heath band potential provides 
        a closed loop or cycle by following
        $\upgamma(0,0)$-$\mathrm{Heath}(0,0)$-$\updelta(0,0)$. 
        The measurements of 97DaDoKe
        gave consistency in this cycle, 
        within the stated uncertainties of the various measurements,
        but 20VeFe did not. 
        Analyzing the ground state data
        and 20VeFe individually,
        we observed an average \SI{0.43}{\per\cm}
        shift for the lower three vibrational levels of the \ASigma state;
        these energy differences
        are plotted in Fig.\,\ref{fig:shift20Ve}.
        We were therefore forced to conclude
        that the measurements of 20VeFe are not consistent with the
        other measurements and 
        these data were excluded from our MARVEL analysis.
    
    \begin{table}[!htb]
        \centering
        \caption{Quantum numbers used in the MARVEL analysis}
        \label{tab:marvel_quantum_number}
        \begin{ruledtabular}
        \begin{tabular}{p{1.5cm}p{6cm}}
             Quan. No. & Meaning \\ 
            \hline
            State & Electronic state label, \eg, X stands for \XPi \\
            $J$ &  Total angular momentum \\
            parity & + or - \\
            $v$ & Vibration quantum number\\
            $\varOmega$ & Projection of the total angular momentum on the internuclear axis\\
        \end{tabular}
        \end{ruledtabular}
    \end{table}

\begin{figure}[!htb]
        \centering
        \includegraphics{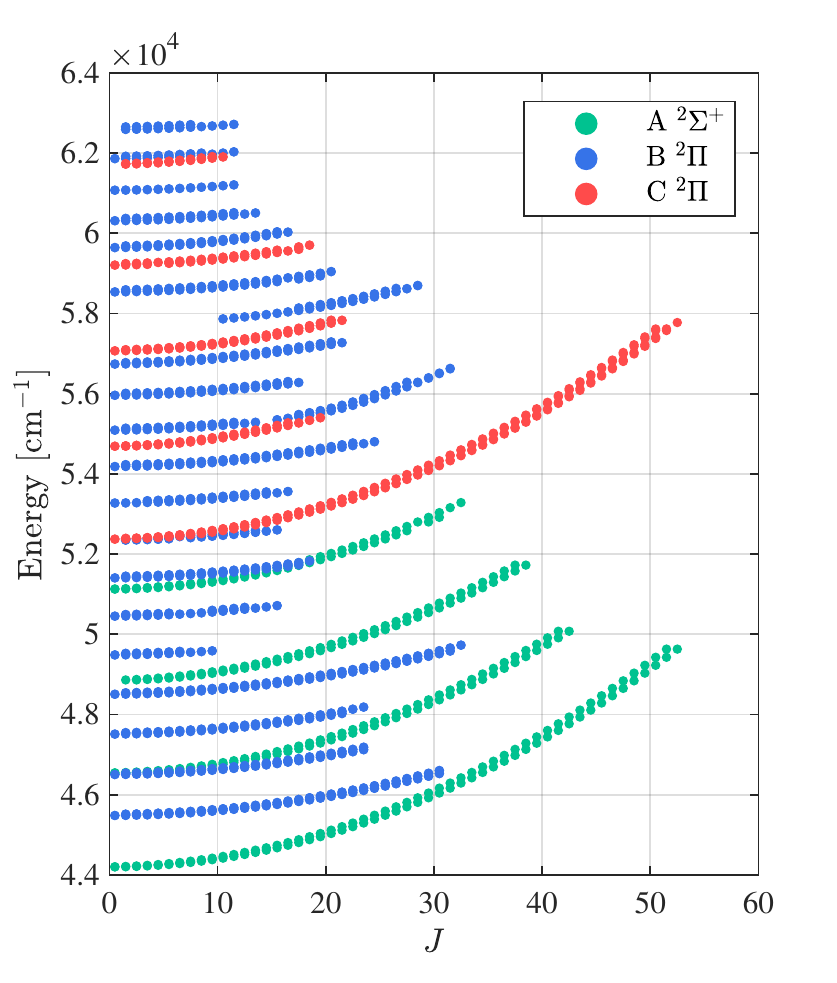}
        \caption{\ASigma, \BPi and \CPi state energy levels generated by MARVEL analysis.}
        \label{fig:marvel_energies}
    \end{figure}

    The \num{20293} validated transitions 
    (including \num{3141} $\upgamma$,
    \num{4795} $\upbeta$,
    \num{861} $\updelta$
    and \num{360} Heath transitions)
    yielded
    \num{327}, \num{1400} and \num{466}
    energy levels of the \ASigma, \BPi and \CPi states, respectively. These levels  are plotted as a function of
    total angular momentum $J$ in Fig.\,\ref{fig:marvel_energies}. 
    The MARVEL transitions (input) file and energies (output) file
    are given as part of the supplementary data. 
    
\begin{figure}[!htb]
        \centering
        \includegraphics{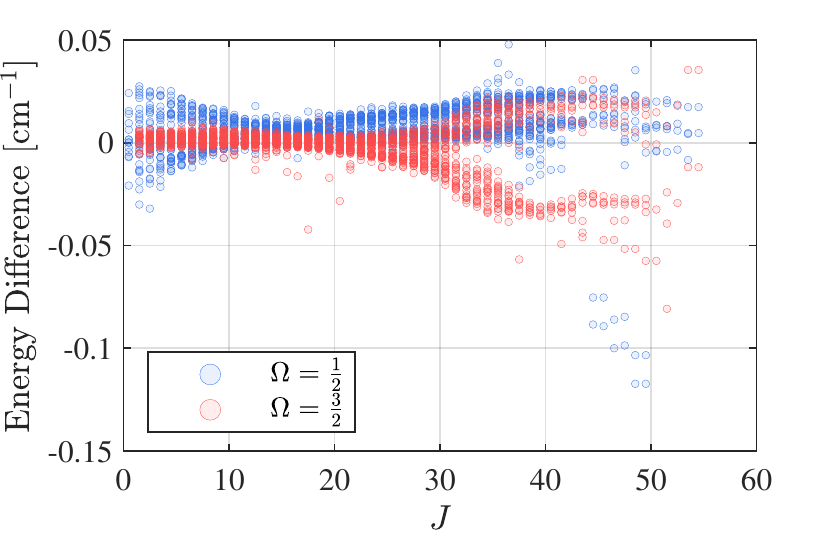}
        \caption{
        \added{Energy difference
        of the \XPi state between MARVEL result in
        this paper
        and the RITZ result in the work
        of Sulakshina and Borkov \cite{18SuBoxx.NO}.}
        }
        \label{fig:diff18SuBo}
    \end{figure}

    \added{
    Sulakshina and Borkov compared
    the ground state energies calculated 
    by their RITZ code \cite{18SuBoxx.NO}
    with our previous MARVEL result \cite{jt686}.
    The MARVEL analysis here
    updates the energy values
    of the \XPi state by including new rovibronic transitions;    as shown in Fig.\ref{fig:diff18SuBo},
    the energy gaps between the 
    results of the MARVEL and RITZ analysis are 
    narrowed as a result of this. This is 
    especially true for high $J$ levels
    belonging to the $\varOmega = \frac{3}{2}$ series 
    (see Fig.\,8(b) of Sulakshina and Borkov's \cite{18SuBoxx.NO}).
    The majority of levels agree within the uncertainty of their determination.}

\section{Refinement of curves for \NO}
\label{sec:refine}
\subsection{Calculation setup}
    The PECs of \ASigma, \BPi and \CPi,
    as well as other coupling curves,
    were refined based on the empirical energy levels
    yielded by the MARVEL analysis in Section\,\ref{sec:marvel}; the PEC for the \XPi\ state was left
    unchanged from that of Wong \etal\cite{jt686}
    The refinement was executed in \duo\
    which is a general variational nuclear-motion program for calculating spectra of diatomic molecules \cite{jt609}.
    
    \duo solves the diatomic molecular {Schr\"{o}dinger}
    equation in two steps.
    Firstly
    the rotation-free radial equation 
    of each electronic
    state is solved to get the vibrational energy levels, $E_v$,
    and wavefunctions, $\psi_v(r)$:
    \begin{equation}
        -\frac{\hbar^{2}}{2 \mu} \frac{\mathrm{d}^{2}}{\mathrm{d} r^{2}}\psi_{v}(r)
            +V_{\mathrm{state}}(r)\psi_{v}(r) 
        =E_{v} \psi_v(r) \, ,
        \label{eq:radialEq}
    \end{equation}
    where $\mu$ is the reduced mass of the molecule
    and $V_{\mathrm{state}}(r)$ is the potential energy curve.
    This step creates vibrational basis functions, $\ket{\mathrm{state},v}$.
    Secondly,
    the fully-coupled, rovibronic Hamiltonian is diagonalized 
    under the Hund's case (a) rovibronic basis set defined by:
    \begin{equation}
        \ket{\mathrm{state}, J, \varOmega, \varLambda, S, \varSigma, v}
        =\ket{\mathrm{state}, \varLambda, S, \varSigma}
        \ket{J, \varOmega, M}\ket{\mathrm{state},v}\, ,
    \end{equation}
    where $\ket{\mathrm{state}, \varLambda, S, \varSigma}$ and
    $\ket{J, \varOmega, M}$
    represent the electronic and rotational
    basis functions, respectively.\cite{jt632}
    The quantum number $M$ is
    the projection of the total angular momentum
    along the laboratory {$Z$-axis}. 
    
    Users are asked to set up some super-parameters
    to get the correct solution.
    The calculation setup for the refinement of \NO is summarized below.
    More details can be found in the
    \duo input file which is given as supplementary material and includes the PEC parameters.
    \begin{itemize}
        \item Equation\,(\ref{eq:radialEq}) was solved by the Sinc-DVR method \cite{92CoMixx.method}.
        \item The calculation range was from
        \replaced{0.6 to 4.0}{0.9 to 2.5} \si{\angstrom}.
        \item The number of grids points was \replaced{701}{401}, uniformly spaced.
        \item The numbers of vibrational basis sets for
            \XPi, \ASigma, \BPi and \CPi were 10, 10, 30 and 10, respectively.
        \item The maximum total angular momentum considered here was $52\frac{1}{2}$.
        \item The upper bound of the total energy was 
        \replaced{\num{65000}}{\num{73000}} \si{\per\cm}.
    \end{itemize}

\subsection{Refinement results of the \ASigma state}  
\begin{figure*}
        \centering
        \includegraphics{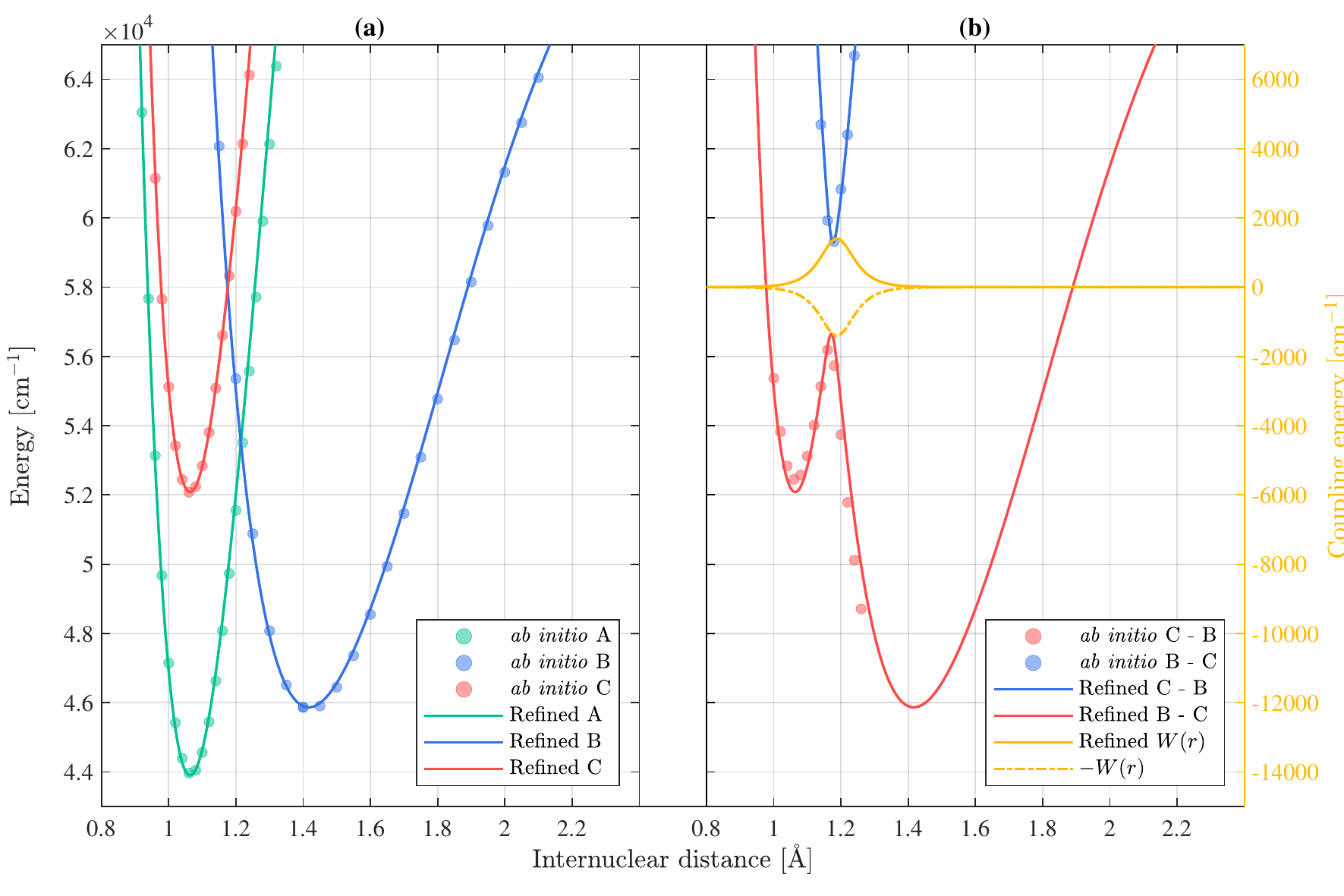}
        \caption{
        The \abinitio and refined PECs of the \ASigma, 
        \BPi and \CPi states as well as the refined B\,-\,C
        interaction term $W(r)$.
        (a) diabatic and
        (b) adiabatic representations.
        The refined potential curves in Panel (b)
        is calculated using
        Eqs.\,(\ref{eq:adiabticPEC1}) and (\ref{eq:adiabticPEC2}).
        The \abinitio curves
        are shifted using empirical
        $T_e$ values.}
        \label{fig:comparePEC}
    \end{figure*}

    The PEC of \ASigma state represented by a 
    fourth-order Extended Morse Oscillator (EMO) function \cite{EMO}.
    The EMO is defined as
    a function of internuclear distance,
    $r$:
    \begin{equation}
        V(r)=T_{\mathrm{e}}+\left(D_{\mathrm{e}}-T_{\mathrm{e}}\right)
        \left[1-\exp \left(-\beta_{\mathrm{EMO}}(r)\left(r-r_{\mathrm{e}}\right)\right)\right]^{2},
        \label{eq:emo}
    \end{equation}
    where the distance-dependent coefficient $\beta_\mathrm{EMO}$ is expressed as
    \begin{equation}
        \beta_{\mathrm{EMO}}(r)=\sum_{i=0}^N b_{i} y_{p}^{\mathrm{eq}}(r)^{i}.
    \end{equation}
    The reduced variable $y_{p}^{\mathrm{eq}}(r)$ has the formula:
    \begin{equation}
        y_{p}^{\mathrm{eq}}(r)=\frac{r^{p}-(r_\mathrm{e})^{p}}{r^{p}+(r_\mathrm{e})^{p}} \,,
        \label{eq:surkus}
    \end{equation}
    where $p$ controls the shape of $y_{p}^{\mathrm{eq}}(r)$.
    The programmed EMO function in \duo is not exactly 
    the same as defined by Eq.\,\ref{eq:emo}.
    A reference point $R_\mathrm{ref}$ 
    (usually the equilibrium internuclear distance)
    divides the curve into left and right parts.
    The numbers of terms $N$, as well as $p$,
    for the left and right parts
    can be assigned different values,
    \ie, $N_\mathrm{L}$, $N_\mathrm{R}$,
    $p_\mathrm{L}$ and $p_\mathrm{R}$.
    The unknown dissociation energy of the state is regarded as
    a dummy parameter in the refinement.
    The initial guess of $D_e$ was given by a pure Morse function
    and the value was fine-tuned in each iteration.
    The optimal parameters of the EMO function is 
    listed in Table\,\ref{tab:emoParameterABC}.
    The \abinitio and refined PECs of the \ASigma state
    are compared in Panel (a) of Fig.\,\ref{fig:comparePEC}.
    
    \begin{table*}
        \caption{The optimized EMO parameters of the PECs of \ASigma,
        \BPi, \CPi states and the spin-orbit (SO) coupling within the \BPi state. The pararmeters are given
        electronically in the \duo\ input which is given in the supplementary information.}
        \label{tab:emoParameterABC}
        \begin{ruledtabular}
        \begin{tabular}{lcccc}
        Parameter & {\ASigma} & {\BPi} & {\CPi} & {$\mel{\mathrm{B}\,^2\Pi}{\hat{H}_\mathrm{SO}}{\mathrm{B}\,^2\Pi}$}\\
        \hline
        $T_\mathrm{e}$ [\si{\per\cm}] & \tablenum{4.39029927730943E+04} & \tablenum{4.58670450676095E+04} & \tablenum{5.20819735839884E+04}  & \tablenum{8.49742400431892E+00} \\
        $r_\mathrm{e}, r_\mathrm{ref} $ [\si{\angstrom}] &\tablenum{1.06366600836862} &\tablenum{1.41663977245069} & \tablenum{1.06370470837254} & \tablenum{1.1} \\
        $D_\mathrm{e}$ [\si{\per\cm}]&\tablenum{1.29205139030394E+05} & \tablenum{7.1627E+04} & \tablenum{1.27177318888436E+05} &\tablenum{5.77407792042591E+01}\\
        $p$ &\tablenum{4} & \tablenum{4} & \tablenum{4} & \tablenum{4}\\
        $N_\mathrm{l}$ &\tablenum{2} & \tablenum{4} & \tablenum{2} & \tablenum{4}\\
        $N_\mathrm{r}$ &\tablenum{4} & \tablenum{8} & \tablenum{4} & \tablenum{4}\\
        $b_0$  [\si{\per\angstrom}]& \tablenum{2.70491398179678E+00} & \tablenum{2.15014413975452E+00} & \tablenum{2.86398560325524E+00} &\tablenum{2.01598755938854E+00}\\
        $b_1$ [\si{\per\angstrom}]& \tablenum{2.07390344060448E-02} & \tablenum{9.83590099793413E-02} & \tablenum{5.52527533543132E-01} &\tablenum{0}\\
        $b_2$ [\si{\per\angstrom}]& \tablenum{8.14901009782977E-02} & \tablenum{4.11898373004704E-01}& \tablenum{3.45263996521417E+00} & \tablenum{-3.76897684548242E+00}\\
        $b_3$ [\si{\per\angstrom}]& \tablenum{-9.53970289289683E-01} & \tablenum{0}                 & \tablenum{-3.78870019311205E+01} &\tablenum{0}\\
        $b_4$ [\si{\per\angstrom}]& \tablenum{2.05067738507637E+00} & \tablenum{-4.43639675058521E-01}& \tablenum{8.28631743949508E+01}&\tablenum{6.67251152788665E+00}\\
        $b_5$ [\si{\per\angstrom}]&                                 & \tablenum{1.21571927232752E+01} &     \\
        $b_6$ [\si{\per\angstrom}]&                                 & \tablenum{-1.78479535757777E+01} &        \\
        $b_7$ [\si{\per\angstrom}]&                                 & \tablenum{3.12778454452573E+00} &     \\
        $b_8$ [\si{\per\angstrom}]&                                  & \tablenum{5.40307772958800E+00} &   \\
        \end{tabular}
        \end{ruledtabular}
    \end{table*}
    
    In addition, 
    our model of the \ASigma state
    contains a spin-rotational term.
    In \duo, the nonzero diagonal and 
    off-diagonal matrix elements of spin-rotational operator
    ${\hat{H}_{\mathrm{SR}}}$\cite{jt632}
    are given by
    \begin{align}
      \mel**{\varLambda, S, \varSigma}{\hat{H}_{\mathrm{SR}}}{ \varLambda, S, \varSigma} 
     & \, = \frac{\hbar^{2}}{2 \mu r^{2}} \gamma^{\mathrm{SR}}(r)\left[\varSigma^{2}-S(S+1)\right] \, , \\
     \mel**{\varLambda, S, \varSigma}{\hat{H}_{\mathrm{SR}}}{\varLambda, S, \varSigma \pm 1}  
     &\,=\frac{\hbar^{2}}{4 \mu r^{2}} \gamma^{\mathrm{SR}}(r)
        \sqrt{S(S+1)-\varSigma(\varSigma \pm 1)} \notag \\
    &  \quad \times \sqrt{J(J+1)-\varOmega(\varOmega \pm 1)} \, .
    \end{align}
    The dimensionless spin-rotational coefficient $\gamma^\mathrm{SR}(r)$ of \ASigma state
    was modeled as a constant whose value
    is 
    \begin{equation}
        \gamma^\mathrm{SR}_\mathrm{A}(r) =\num{-2.08043004478781E-03}\,. 
    \end{equation}

\subsection{Refinement results of the \BPi\,-\,\CPi coupled states}

    \subsubsection{Deperturbation of the \BPi\,-\,\CPi interaction}
        
        For this work we only consider coupling between two electronic states.
        The interaction between two electronic states
        belong to the same irreducible representation of the molecular point group
        directly depends on the avoided crossing of their diabatic PECs.
        Thus,
        it is possible to model the coupled states
        by introducing two adiabatic potentials \cite{18KaBeAv}.
        This could be accomplished by diagonalizing the 
        matrix:
        \begin{equation}
            \begin{pmatrix}
                V_1(r) & W(r) \\
                W(r) & V_2(r) 
            \end{pmatrix} \, ,
        \end{equation}
        where $V_1(r)$ and $V_2(r)$
        are two diabatic potentials and $W(r)$ is the coupling curves.
        The adiabatic PECs, \ie the eigenvalues of the matrix,
        are
        \begin{align}
            V_{\text {low }}(r) &=\frac{V_{1}(r)+V_{2}(r)}{2}-\frac{\sqrt{\left[V_{1}(r)-V_{2}(r)\right]^{2}+4 W^{2}(r)}}{2} \, , 
            \label{eq:adiabticPEC1}\\
            V_{\text {upp }}(r) &=\frac{V_{1}(r)+V_{2}(r)}{2}+\frac{\sqrt{\left[V_{1}(r)-V_{2}(r)\right]^{2}+4 W^{2}(r)}}{2} \, .
            \label{eq:adiabticPEC2}
        \end{align}
        EMO potential functions are used to model $V_1(r)$ and $V_2(r)$ in
        Eqs.\,(\ref{eq:adiabticPEC1}) and (\ref{eq:adiabticPEC2}) 
        while  $W(r)$ is given by:
        \begin{equation}
            W(r)=\frac{\sum_{i \geq 0} w_{i}\left(r-r_{0}\right)^{i}}{\cosh \left(b\left(r-r_{0}\right)\right)} \, .
        \end{equation}
        The function rapidly decreases to $W_0$
        when $r$ moves away from $r_0$.
        
        The coupled PECs of
        $\mathrm{X}\,^1\Sigma_\mathrm{g}^+$ and $\mathrm{B}'\,^1\Sigma_\mathrm{g}^+$
        states of \ce{C_2}
        were represented by adiabatic potential
        in our previous work \cite{jt736},
        producing accurate line list.
        Nevertheless,
        this method is not optimal for NO where the avoided crossing between the B and C states is very sharp.
        Thus, for example, the  adiabatic B -- X and C -- X transition dipole moment curves (TDMCs)
        change dramatically around the crossing point making them hard to use in any reliable calculation of transition intensities and a slight shift of the crossing point, $r_0$, during refinement may
        significantly change
        the intensities of nearby lines.
        We therefore adopt the following procedure for the
        generating line lists involving these  coupled electronic states:
        \begin{enumerate}
            \item Solve the radial equations  
                set up with diabatic PECs
                of different electronic states
                to get vibrational basis.
            \item Construct rovibronic Hamiltonian matrix
                with all necessary elements,
                including the electronic interaction terms.
            \item Diagonalize the matrix
                under rovibronic basis set 
                to get the rovibronic energy levels
                and the corresponding wavefunctions.
            \item Refine the diabatic PECs,
                electronic interaction terms
                and other coupling curves 
                by fitting the energies to observed energy levels.
            \item Calculate the Einstein A coefficient
                with the diabatic TDMCs
                and let the wavefunctions determine the
                weights of TDMCs for each 
                rovibronic state at different geometries.
        \end{enumerate}
        The method not only rescues us from
        the dilemma of constructing adiabatic TDMCs
        but also improves the 
        flexibility
        of our program.
        For instance,
        it is convenient
        to model the \BPi\,-\,\CPi\,-\,\LPi 
        coupled states of NO
        by adding new definitions of the
        potential of \LPi and 
        coupled term between \CPi and \LPi
        in the input file of \duo,
        without changing its code.

    \subsubsection{Refined curves}  
        
        The diabatic PECs of \BPi and \CPi
        states were modeled using EMO functions
        whose optimal parameters are listed in Table\,\ref{tab:emoParameterABC}.
        The \abinitio and refined PECs of
        \BPi and \CPi states are compared in Panel (a) Fig.\,\ref{fig:comparePEC}.
        Its optimal parameters 
        of the function are listed in Table\,\ref{tab:lorentzParameter}.
        Although not used in this work,
        the adiabatic curves were calculated as defined
        by Eqs. (\ref{eq:adiabticPEC1}) and (\ref{eq:adiabticPEC2}).
        They are compared with the \abinitio
        adiabatic PECs in Panel (b) of Fig.\,\ref{fig:comparePEC}.
        The dissociation energy 
        of \CPi state is also a dummy parameter.
        The refined PECs of \ASigma and \CPi states 
        are physically meaningless outside 
        the our calculation range
        (\ie, when energy is greater than \SI{73000}{\per\cm}).

        \begin{table}
            \caption{Optimized Lorentz parameters for the {B\,-\,C} interaction curve.}
            \label{tab:lorentzParameter}
            \begin{ruledtabular}
            \begin{tabular}{lc}
            {Parameter}  &  {Value}\\
            \hline
            $b$ [\si{\per\angstrom}] &\tablenum{2.21707630646740E+01}\\
            $r_0$ [\si{\angstrom}] & \tablenum{1.18808573722698} \\
            $w_0$  [\si{\per\cm}]   & \tablenum{1.40173178754200E+03} \\
            \end{tabular}
            \end{ruledtabular}
        \end{table}

\begin{figure}[htb]
            \centering
            \includegraphics{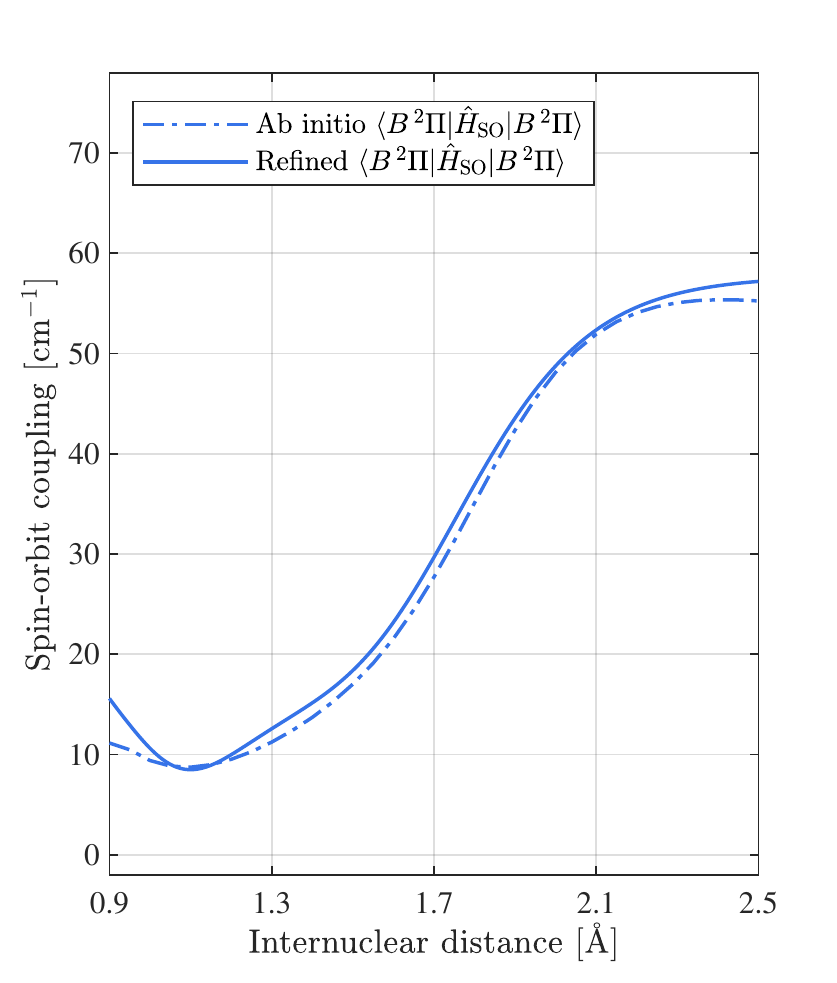}
            \caption{The spin-orbit coupling curves of \BPi states.}
            \label{fig:compareSOB}
        \end{figure}

        The spin-orbit coupling curve (SOC) of \BPi state 
        was also fitted to an EMO function
        whose optimal parameters are listed in the last column of Table\,\ref{tab:emoParameterABC}.
        Figure\,\ref{fig:compareSOB} compares the \abinitio
        and refined SOCs.
        The diagonal spin-orbital term of \CPi state
        and the off-diagonal term between \BPi and \CPi
        were determined empirically by
        fitting to constants.
        The spin-rotational coefficient of \CPi state
        was also model on a constant.
        The values of these terms are listed in Table\,\ref{tab:SOSRconstants}.
        
        \begin{table}
            \caption{Optimized values of the spin-orbit (SO) and spin-rotation (SR) coupling terms.}
            \label{tab:SOSRconstants}
            \begin{ruledtabular}
            \begin{tabular}{lc}
            {Term}  &  {Value}\\
            \hline
            $\mel{\mathrm{C}\,^2\Pi}{\hat{H}_\mathrm{SO}}{\mathrm{C}\,^2\Pi}$ [\si{\cm^{-1}}]& \tablenum{1.81766772228787E+00} \\
            $\mel{\mathrm{C}\,^2\Pi}{\hat{H}_\mathrm{SO}}{\mathrm{B}\,^2\Pi} $ [\si{\per\cm}] &\tablenum{2.28206375746938E+00} \\
            $\quad\gamma_\mathrm{C}^\mathrm{SR}$ &\tablenum{2.70593061789197E-03}\\
            \end{tabular}
            \end{ruledtabular}
        \end{table}
        
        The \Ldoubling fine structures of $\upbeta$ and
        $\updelta$ system bands were observed in most of the work
        listed in Table\,\ref{tab:marvel_source}.
        \duo calculates the \Ldoubling matrix elements,
        \ie, $\mel{\varLambda' \varSigma' J' \varOmega'}
                {\hat{H}_\mathrm{LD}}
                {\varLambda'' \varSigma'' J'' \varOmega''}
                $,
        according to the terms given by Brown and Merer, \cite{79BrMexx.methods}:
        \begin{align}
            &\mel**{\mp 1, \varSigma \pm 2, J, \varOmega}
                {\hat{H}_\mathrm{LD}}
                {\pm 1, \varSigma, J, \varOmega}
                =  \frac{1}{2}\left(o_{v}+p_{v}+q_{v}\right) \times \notag \\
             &  \quad   \sqrt{[S(S+1)-\varSigma(\varSigma \pm 1)]
                    [S(S+1)-(\varSigma \pm 1)(\varSigma \pm 2)]} \, ,
                \label{eq:Ldoubling1}\\
            &\mel**{\mp 1, \varSigma \pm 1, J, \varOmega \mp 1}
                {\hat{H}_\mathrm{LD}}{\pm 1, \varSigma, J, \varOmega}
                =  -\frac{1}{2}\left(p_{v}+2 q_{v}\right) \times \notag \\
            & \quad  \sqrt{[S(S+1)-\varSigma(\varSigma \pm 1)]
                    [J(J+1)-\varOmega(\varOmega \mp 1)]} \, ,
                \label{eq:Ldoubling2}\\
           & \mel**{\mp 1, \varSigma, J, \varOmega \mp 2}{\hat{H}_\mathrm{LD}}
                {\pm 1, \varSigma, J ,\varOmega}
                = \frac{1}{2} q_{v} \times \notag \\
            & \quad   \sqrt{[J(J+1)-\varOmega(\varOmega \mp 1)]
                    [J(J+1)-(\varOmega \mp 1)(\varOmega \mp 2)]} \, .
                \label{eq:Ldoubling3}
        \end{align}
        For \BPi and \CPi, 
        $\varSigma = \pm 1/2$.
        Therefore,
        the matrix elements described in Eq.\,(\ref{eq:Ldoubling1}) are zero
        and only the coefficient curves of Eqs. (\ref{eq:Ldoubling2}) and (\ref{eq:Ldoubling3}) were fitted to
        polynomials, \ie,
        \begin{equation}
            P(r) = a_0 + \sum_{i\geq 0} a_i(r-r_0)^i.
        \end{equation}
        The optimized parameters of the \Ldoubling terms are listed in Table\,\ref{tab:ldoublingParameters}.
        
        \begin{table*}
            \caption{Optimized polynomial parameters
            of the \Ldoubling curves of the \BPi and \CPi states}
            \label{tab:ldoublingParameters}
            \begin{ruledtabular}
            \begin{tabular}{lcccc}
            \multirow{2}[0]{*}{Parameter} & \multicolumn{2}{c}{\BPi} & \multicolumn{2}{c}{\CPi} \\
                  & \mbox{$p_v+2q_v$}  & \mbox{$q_v$}     & \mbox{$p_v+2q_v$}  & \mbox{$q_v$} \\
             \hline
               $r_\mathrm{0} $ [\si{\angstrom}] & \tablenum{1.41650470352337E+00 } &\tablenum{1.41650470352337E+00} &\tablenum{0} &\tablenum{1.06443605941232E+00}\\
               $a_0$ [\si{\per\cm}]&\tablenum{1.06551670346650E-02} &\tablenum{6.45332691633386E-05} & \tablenum{-3.66039401364346E-02}  &\tablenum{-1.61243738150496E-02}\\
               $a_1$ [\si{\per\cm\per\angstrom}]&\tablenum{-2.92114281362927E-01} &\tablenum{-1.18974108983174E-02} & \tablenum{0} &\tablenum{3.00321609791786E-02} \\
               $a_2$ [\si{\cm^{-1}\angstrom^{-2}}]&\tablenum{5.09517016483691E-01} &\tablenum{3.04077180915239E-02} & \tablenum{0} &\tablenum{0}\\
            \end{tabular}
            \end{ruledtabular}
        \end{table*}

    \subsubsection{Fitting residues of the rovibronic energy levels}  
    
\begin{figure*}
            \centering
            \includegraphics{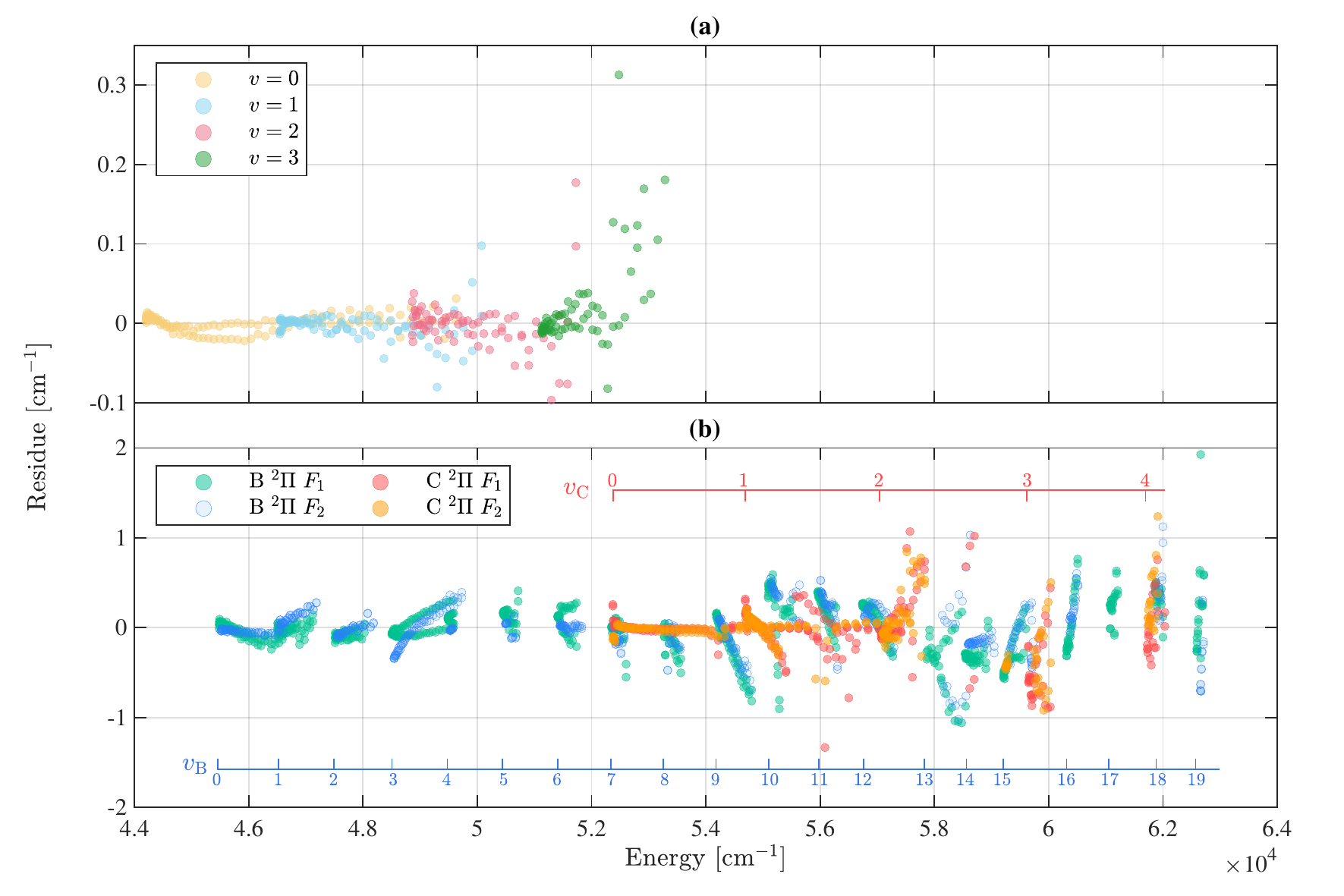}
            \caption{Fitting residues of (a) \ASigma state and
            (b) \BPi\,-\,\CPi coupled states.}
            \label{fig:residueABC}
        \end{figure*}

        The fitting residues of the \ASigma state
        are shown in Panel (a) of Fig.\,\ref{fig:residueABC}.
        The high-$J$ energies of $v=3$ vibrational levels
        are mainly determined by blended lines of 97DaDoKe.NO \cite{97DaDoKe.NO}.
        The fitting residues of the \BPi and \CPi states
        are shown
        in Panel (b) of Fig.\,\ref{fig:residueABC},
        where the cold colors represent the \BPi state
        and the warm ones represent the \CPi state.
        The $F_1$ (\ie, $\varOmega = \frac{1}{2}$) and $F_2$ (\ie, $\varOmega = \frac{3}{2}$) 
        levels are also distinguishable. 
        The residue distributions indicate $J$-dependent
        systematic error of our model,
        which may result from some off-diagonal couplings,
        \eg, the coupling between \CPi and \DSigma states \cite{82AmVexx.NO}.
        
        The residues of all rovibronic energy levels
        are plotted against their corresponding uncertainties.
        The root-mean-square and
        average value of uncertainties and residues are compare\added{d}
        in Table\,\ref{tab:RMS}.
        
        \begin{figure}
            \centering
            \includegraphics{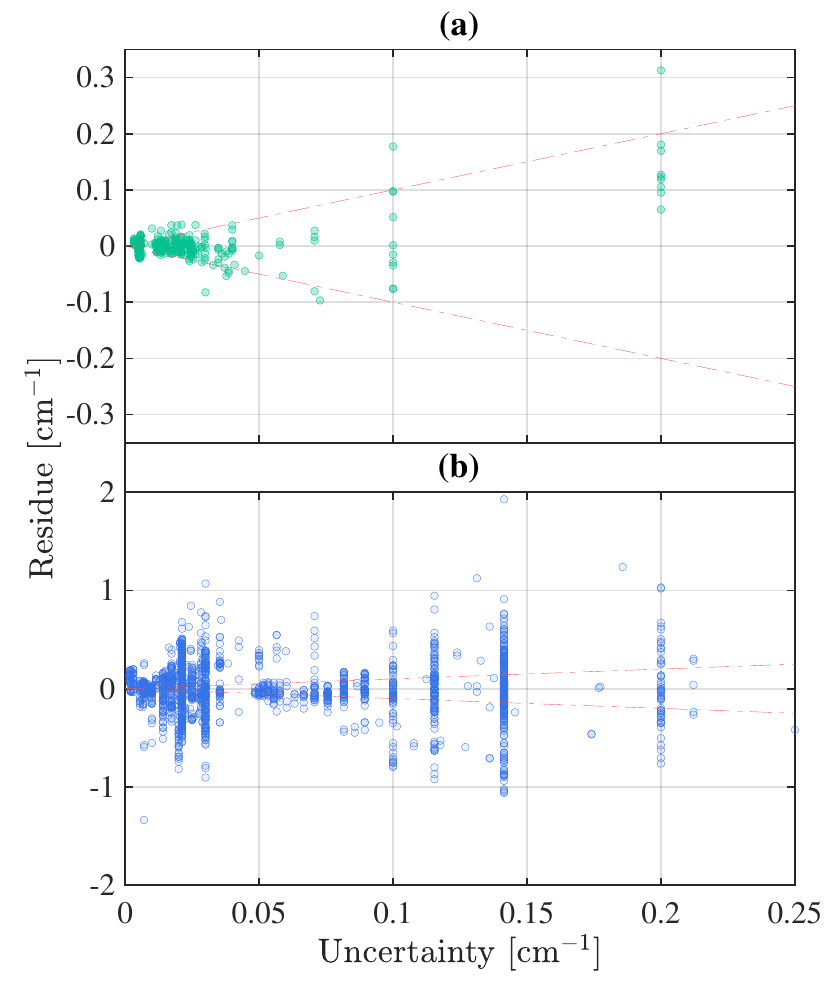}
            \caption{Residues against uncertainties
            of (a) \ASigma state and (b) B\,-\,C coupled states with vibrational 
            states given in the bars.}
            \label{fig:compareError}
        \end{figure}
        
        \begin{table}
            \caption{Overall 
                comparison of
                uncertainty and 
                residue.}
            \label{tab:RMS}
            \begin{ruledtabular}
            \begin{tabular}{lll}
            All in \si{\per\cm}& \ASigma &\BPi\,-\,\CPi\\
            \hline
            RMS uncertainty & 0.04284 & 0.07927\\
            RMS residue  & 0.03390 & 0.27217\\
            Average uncertainty & 0.02453&0.05753\\
            Average absolute residue & 0.01599&0.18603\\
            \end{tabular}
            \end{ruledtabular}
        \end{table}
 
        \added{
        The accuracy of our model
        is definitely higher than those
        of Lagerqvist and Miescher \cite{58LaMixx.NO},
        or Gallusser and
        Dressler \cite{82GaDrxx.NO},
         On one hand,
        the most recent measurements 
        (\eg, the works of Yoshino \etal \cite{06YoThMu.NO})
        and spectroscopic analysis techniques  (MARVEL \cite{MARVEL})
        helped us 
        reconstruct reliable spectroscopic 
        network and energy levels.
        On the other hand,
        our model
        was directly fitted to the
        observed
        rovibronic levels.
        The vibronic residues given by
        Gallusser and Dressler \cite{82GaDrxx.NO} 
        are  greater than our 
        rovibronic residues.
        Unlike Gallusser and Dressler,
        we did not include higher
        electronic states, such as \LPi and \KPi,
        in our model,
        which reduces its range of applicability
        where the state energy is greater than \SI{63000}{\per\cm}.
        However,
        thanks to diabatic coupling strategy of \duo,
        the model can easily  be updated in a future study.
        }
        
\begin{figure*}
            \centering
            \includegraphics{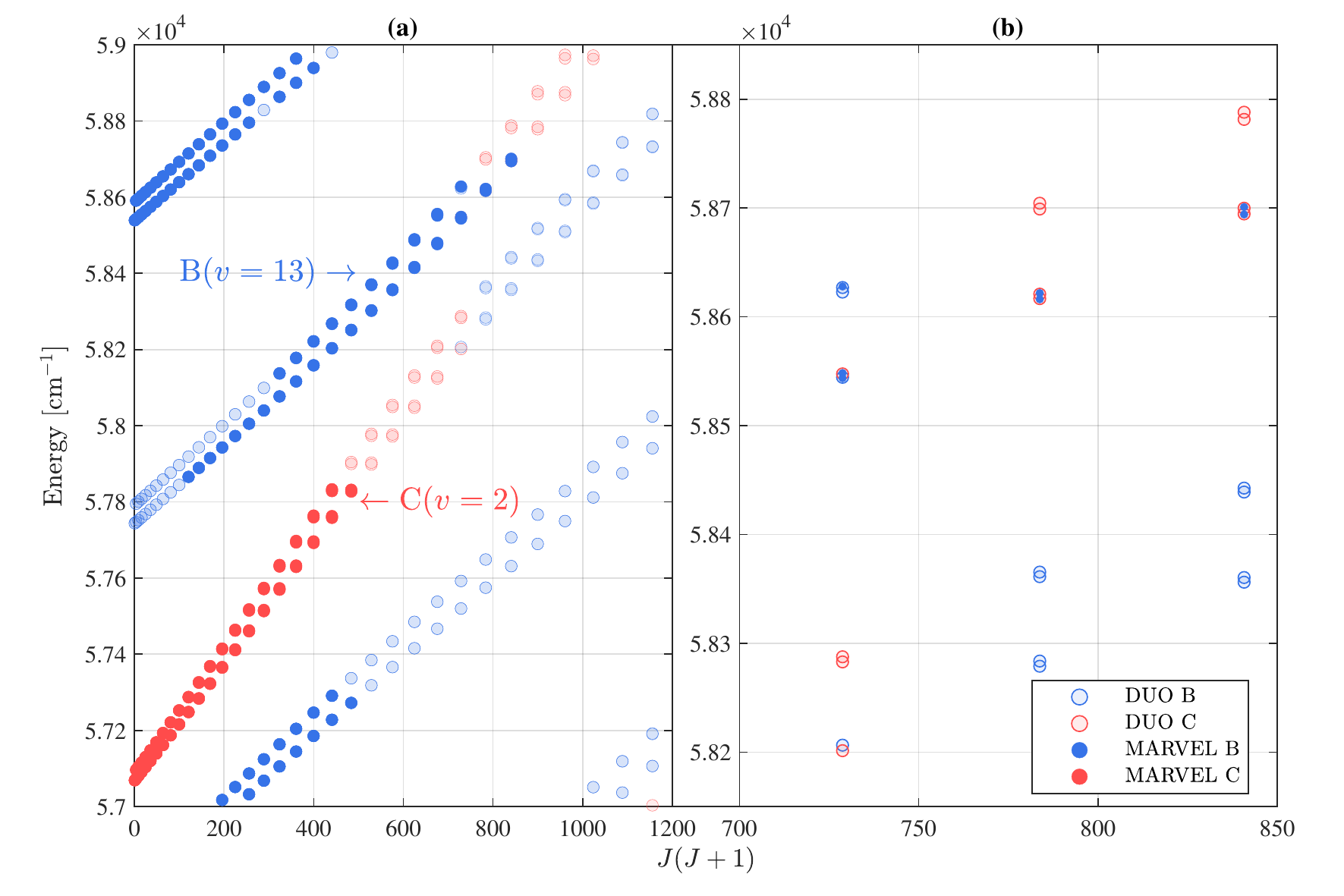}
            \caption{Calculated and observed energy levels of the $\mathrm{B}(v=13)$\,-\,$\mathrm{C}(v=2)$ coupled series. 
            The right hand panel
            is a blow up of the avoided crossing between the states which
            gives a clearer view of the
            \Ldoubling splitting and
            the difference between
            the quantum numbers given by MARVEL and \duo.
            }
            
            \label{fig:compareEnergy}
        \end{figure*}

        We note that some of the assignments to B or C electronic states differ between \duo\ and
        our MARVEL analysis. \duo\ uses
        three good quantum number, namely
        the total angular momentum $J$, the total parity and
        the counting number of the levels with the same values of $J$ and parity.
        The other quantum numbers such as state, $v$, $\varOmega$,
        are estimated using the contribution of the basis functions to
        a given wavefunction. 
        It is to be
        anticipated that in regions of heavily mixed
        wavefunctions this may lead to differences compared
        to other assignment methods.
       The MARVEL and \duo energy levels of B ($v=13$)\,-\,C ($v=2$)
        coupled series are plotted in Fig.\,\ref{fig:compareEnergy}.
        Table\,\ref{tab:duoEnergy} lists
        some energy levels in the output \texttt{.en} file of \duo.
        Both of them demonstrate the differences between
        the quantum numbers of MARVEL and \duo results.

        \begin{table*}
            \caption{Sample lines extracted from the output \texttt{.en} file of \duo.}
            \label{tab:duoEnergy}
            \tt
            \begin{ruledtabular}   
            \begin{tabular}{cclcllrrcrrrrcrrrr}
            \multicolumn{1}{c}{\textrm{\duo}} & \multicolumn{1}{c}{\textrm{Assigned}} &       &       & \multicolumn{1}{c}{\textrm{MARVEL}} & \multicolumn{1}{c}{\textrm{\duo}} &       &       & \multicolumn{5}{c}{\textrm{\duo}}               & \multicolumn{5}{c}{\textrm{MARVEL}} \\
            \multicolumn{1}{c}{$N$} & \multicolumn{1}{c}{$N$\footnotemark[1]} & \multicolumn{1}{c}{$J$} & \multicolumn{1}{c}{\textrm{Parity}} & \multicolumn{1}{c}{\textrm{Energy}} & \multicolumn{1}{c}{\textrm{Energy}} & \multicolumn{1}{c}{\textrm{Residue}} & \multicolumn{1}{c}{\textrm{Weight}} & \multicolumn{1}{c}{\textrm{state}\footnotemark[2]} & \multicolumn{1}{c}{$v$} & \multicolumn{1}{c}{$\varLambda$} & \multicolumn{1}{c}{$\varSigma$} & \multicolumn{1}{c}{$\varOmega$} & \multicolumn{1}{c}{\textrm{state}\footnotemark[2]} & \multicolumn{1}{c}{$v$} & \multicolumn{1}{c}{$\varLambda$} & \multicolumn{1}{c}{$\varSigma$} & \multicolumn{1}{c}{$\varOmega$} \\
            \hline
            39    & 39    & 1.5   & +     & 52349.0418 & 52349.0274 & 0.0144 & 9.50E-05 & 3     & 7     & 1     & -0.5  & 0.5   & 3     & 7     & 1     & -0.5  & 0.5 \\
            40    & 40    & 1.5   & +     & 52373.2372 & 52373.3626 & -0.1255 & 1.40E-03 & 4     & 0     & 1     & 0.5   & 1.5   & 3     & 7     & 1     & 0.5   & 1.5 \\
            41    & 41    & 1.5   & +     & 52380.1912 & 52380.1101 & 0.0810 & 1.30E-03 & 4     & 0     & 1     & -0.5  & 0.5   & 4     & 0     & 1     & -0.5  & 0.5 \\
            42    & 42    & 1.5   & +     & 52392.3007 & 52392.3172 & -0.0165 & 1.30E-03 & 3     & 7     & 1     & 0.5   & 1.5   & 4     & 0     & 1     & 0.5   & 1.5 \\
            64    & 64    & 2.5   & -     & 59217.4976 & 59217.9730 & -0.4754 & 9.50E-05 & 3     & 15    & -1    & 0.5   & -0.5  & 4     & 3     & -1    & 0.5   & -0.5 \\
            65    & 65    & 2.5   & -     & 59250.3720 & 59250.8248 & -0.4528 & 9.50E-05 & 4     & 3     & -1    & -0.5  & -1.5  & 4     & 3     & -1    & -0.5  & -1.5 \\
            66    & 66    & 2.5   & -     & 59654.3005 & 59654.8551 & -0.5546 & 3.70E-06 & 4     & 3     & -1    & 0.5   & -0.5  & 3     & 15    & -1    & 0.5   & -0.5 \\
            67    & 67    & 2.5   & -     & 59692.2845 & 59692.6292 & -0.3447 & 4.80E-06 & 3     & 15    & -1    & -0.5  & -1.5  & 3     & 15    & -1    & -0.5  & -1.5 \\
            \end{tabular}
            \end{ruledtabular}
            \footnotetext[1]{The counting numbers ($N$) were 
                manually assigned to match the
                corresponding MARVEL energy level. }
            \footnotetext[2]{In these columns, 
                `\texttt{3}' and `\texttt{4}' indicate the
                \BPi and \CPi states, respectively.}
        \end{table*}

\section{Conclusion}
\label{sec:conclusion}

    In this paper, potential energy curves and coupling for
    the low-lying
    electronic state of NO are 
    calculated
    using quantum chemistry package Molpro.
    The strong interaction between Rydberg and 
    valence states makes
    the \abinitio calculation challenging.
    We obtain both adiabatic and diabatic
    PECs and SOCs for the \ASigma, \BPi and \CPi states.
    The curves were refined
    by fitting the rovibronic energy levels
    calculated by variational nuclear motion program \duo to 
    those reconstructed by MARVEL analysis.
    The RMS error of the \ASigma state fitting
    and \BPi\,-\,\CPi coupled states fitting
    are \SI{0.03390}{\per\cm}
    and \SI{0.27217}{\per\cm},
    respectively, which 
    energies were determined by our use of a MARVEL procedure and the best available measurements.
    The success of \BPi\,-\,\CPi coupled states fitting
    validates our
    deperturbation method for treating
    the coupled electronic state.
    This work, when combined with the earlier \XPi study of Wang \etal,\cite{jt686} provides
    a comprehensive spectroscopic model four the lowest for electronic states of NO and thus
    a good start point
    for the generation of a NO UV line list. 
    This line list will be presented elsewhere.

\begin{acknowledgments}
    We are indebted to Dr. Rafał Hakalla
    (University of Rzeszów)
    for valuable discussions.
    Qianwei Qu acknowledges the financial support from University College London
    and China Scholarship Council.
    This work was supported by the STFC Projects No. ST/M001334/1 and ST/R000476/1, and ERC Advanced Investigator Project 883830.
    The authors acknowledge the use of the UCL Myriad, Grace and Kathleen High Performance Computing Facilities and associated support services in the completion of this work.
\end{acknowledgments}

\section*{Data Availability}
The data that supports the findings of this study are available within the article and its
supplementary material.

\section*{Supplementary material}
Three text files are provided as supplementary material to the article:\\
MARVEL\_Transitions.txt: input transitions file used with MARVEL.\\
MARVEL\_Energies.txt: energy levels file generated by MARVEL using the file MARVEL\_Transitions.txt.\\
NO\_XABC.model.txt: a Duo input file which full specifies our spectroscopic model including the associated potential energy and coupling curves.


\bibliography{journals_iso, NO, abinitio, methods, programs, jtj, Books}

\begin{thebibliography}{73}%
\makeatletter
\providecommand \@ifxundefined [1]{%
 \@ifx{#1\undefined}
}%
\providecommand \@ifnum [1]{%
 \ifnum #1\expandafter \@firstoftwo
 \else \expandafter \@secondoftwo
 \fi
}%
\providecommand \@ifx [1]{%
 \ifx #1\expandafter \@firstoftwo
 \else \expandafter \@secondoftwo
 \fi
}%
\providecommand \natexlab [1]{#1}%
\providecommand \enquote  [1]{``#1''}%
\providecommand \bibnamefont  [1]{#1}%
\providecommand \bibfnamefont [1]{#1}%
\providecommand \citenamefont [1]{#1}%
\providecommand \href@noop [0]{\@secondoftwo}%
\providecommand \href [0]{\begingroup \@sanitize@url \@href}%
\providecommand \@href[1]{\@@startlink{#1}\@@href}%
\providecommand \@@href[1]{\endgroup#1\@@endlink}%
\providecommand \@sanitize@url [0]{\catcode `\\12\catcode `\$12\catcode
  `\&12\catcode `\#12\catcode `\^12\catcode `\_12\catcode `\%12\relax}%
\providecommand \@@startlink[1]{}%
\providecommand \@@endlink[0]{}%
\providecommand \url  [0]{\begingroup\@sanitize@url \@url }%
\providecommand \@url [1]{\endgroup\@href {#1}{\urlprefix }}%
\providecommand \urlprefix  [0]{URL }%
\providecommand \Eprint [0]{\href }%
\providecommand \doibase [0]{http://dx.doi.org/}%
\providecommand \selectlanguage [0]{\@gobble}%
\providecommand \bibinfo  [0]{\@secondoftwo}%
\providecommand \bibfield  [0]{\@secondoftwo}%
\providecommand \translation [1]{[#1]}%
\providecommand \BibitemOpen [0]{}%
\providecommand \bibitemStop [0]{}%
\providecommand \bibitemNoStop [0]{.\EOS\space}%
\providecommand \EOS [0]{\spacefactor3000\relax}%
\providecommand \BibitemShut  [1]{\csname bibitem#1\endcsname}%
\let\auto@bib@innerbib\@empty
\bibitem [{\citenamefont {Canfield}, \citenamefont {Glazer},\ and\
  \citenamefont {Falkowski}(2010)}]{10CaGlFa.NO}%
  \BibitemOpen
  \bibfield  {author} {\bibinfo {author} {\bibfnamefont {D.~E.}\ \bibnamefont
  {Canfield}}, \bibinfo {author} {\bibfnamefont {A.~N.}\ \bibnamefont
  {Glazer}}, \ and\ \bibinfo {author} {\bibfnamefont {P.~G.}\ \bibnamefont
  {Falkowski}},\ }\href {\doibase 10.1126/science.1186120} {\bibfield
  {journal} {\bibinfo  {journal} {Science}\ }\textbf {\bibinfo {volume}
  {330}},\ \bibinfo {pages} {192} (\bibinfo {year} {2010})}\BibitemShut
  {NoStop}%
\bibitem [{\citenamefont {Vitousek}\ \emph {et~al.}(1997)\citenamefont
  {Vitousek}, \citenamefont {Aber}, \citenamefont {Howarth}, \citenamefont
  {Likens}, \citenamefont {Matson}, \citenamefont {Schindler}, \citenamefont
  {Schlesinger},\ and\ \citenamefont {Tilman}}]{97ViAbRo.NO}%
  \BibitemOpen
  \bibfield  {author} {\bibinfo {author} {\bibfnamefont {P.~M.}\ \bibnamefont
  {Vitousek}}, \bibinfo {author} {\bibfnamefont {J.~D.}\ \bibnamefont {Aber}},
  \bibinfo {author} {\bibfnamefont {R.~W.}\ \bibnamefont {Howarth}}, \bibinfo
  {author} {\bibfnamefont {G.~E.}\ \bibnamefont {Likens}}, \bibinfo {author}
  {\bibfnamefont {P.~A.}\ \bibnamefont {Matson}}, \bibinfo {author}
  {\bibfnamefont {D.~W.}\ \bibnamefont {Schindler}}, \bibinfo {author}
  {\bibfnamefont {W.~H.}\ \bibnamefont {Schlesinger}}, \ and\ \bibinfo {author}
  {\bibfnamefont {D.~G.}\ \bibnamefont {Tilman}},\ }\href {\doibase
  10.1890/1051-0761(1997)007[0737:HAOTGN]2.0.CO;2} {\bibfield  {journal}
  {\bibinfo  {journal} {Ecol. Appl.}\ }\textbf {\bibinfo {volume} {7}},\
  \bibinfo {pages} {737} (\bibinfo {year} {1997})}\BibitemShut {NoStop}%
\bibitem [{\citenamefont {Chameides}\ \emph {et~al.}(1994)\citenamefont
  {Chameides}, \citenamefont {Kasibhatla}, \citenamefont {Yienger},\ and\
  \citenamefont {Levy}}]{94ChKaYi.NO}%
  \BibitemOpen
  \bibfield  {author} {\bibinfo {author} {\bibfnamefont {W.~L.}\ \bibnamefont
  {Chameides}}, \bibinfo {author} {\bibfnamefont {P.~S.}\ \bibnamefont
  {Kasibhatla}}, \bibinfo {author} {\bibfnamefont {J.}~\bibnamefont {Yienger}},
  \ and\ \bibinfo {author} {\bibfnamefont {H.}~\bibnamefont {Levy}},\ }\href
  {\doibase 10.1126/science.264.5155.74} {\bibfield  {journal} {\bibinfo
  {journal} {Science}\ }\textbf {\bibinfo {volume} {264}},\ \bibinfo {pages}
  {74} (\bibinfo {year} {1994})}\BibitemShut {NoStop}%
\bibitem [{\citenamefont {Likens}, \citenamefont {Driscoll},\ and\
  \citenamefont {Buso}(1996)}]{96LiDrBu.NO}%
  \BibitemOpen
  \bibfield  {author} {\bibinfo {author} {\bibfnamefont {G.~E.}\ \bibnamefont
  {Likens}}, \bibinfo {author} {\bibfnamefont {C.~T.}\ \bibnamefont
  {Driscoll}}, \ and\ \bibinfo {author} {\bibfnamefont {D.~C.}\ \bibnamefont
  {Buso}},\ }\href {\doibase 10.1126/science.272.5259.244} {\bibfield
  {journal} {\bibinfo  {journal} {Science}\ }\textbf {\bibinfo {volume}
  {272}},\ \bibinfo {pages} {244} (\bibinfo {year} {1996})}\BibitemShut
  {NoStop}%
\bibitem [{\citenamefont {Singh}\ and\ \citenamefont
  {Agrawal}(2007)}]{07SiAgxx.NO}%
  \BibitemOpen
  \bibfield  {author} {\bibinfo {author} {\bibfnamefont {A.}~\bibnamefont
  {Singh}}\ and\ \bibinfo {author} {\bibfnamefont {M.}~\bibnamefont
  {Agrawal}},\ }\href@noop {} {\bibfield  {journal} {\bibinfo  {journal} {J.
  Environ. Biol.}\ }\textbf {\bibinfo {volume} {29}},\ \bibinfo {pages} {15}
  (\bibinfo {year} {2007})}\BibitemShut {NoStop}%
\bibitem [{\citenamefont {Hu}\ \emph {et~al.}(2000)\citenamefont {Hu},
  \citenamefont {Naito}, \citenamefont {Kobayashi},\ and\ \citenamefont
  {Hasatani}}]{00HuNaKo.NO}%
  \BibitemOpen
  \bibfield  {author} {\bibinfo {author} {\bibfnamefont {Y.}~\bibnamefont
  {Hu}}, \bibinfo {author} {\bibfnamefont {S.}~\bibnamefont {Naito}}, \bibinfo
  {author} {\bibfnamefont {N.}~\bibnamefont {Kobayashi}}, \ and\ \bibinfo
  {author} {\bibfnamefont {M.}~\bibnamefont {Hasatani}},\ }\href {\doibase
  10.1016/S0016-2361(00)00047-8} {\bibfield  {journal} {\bibinfo  {journal}
  {Fuel}\ }\textbf {\bibinfo {volume} {79}},\ \bibinfo {pages} {1925} (\bibinfo
  {year} {2000})}\BibitemShut {NoStop}%
\bibitem [{\citenamefont {Li}, \citenamefont {Lu},\ and\ \citenamefont
  {Rudolph}(1998)}]{98LiLuRu.NO}%
  \BibitemOpen
  \bibfield  {author} {\bibinfo {author} {\bibfnamefont {Y.}~\bibnamefont
  {Li}}, \bibinfo {author} {\bibfnamefont {G.}~\bibnamefont {Lu}}, \ and\
  \bibinfo {author} {\bibfnamefont {V.}~\bibnamefont {Rudolph}},\ }\href
  {\doibase 10.1016/S0009-2509(97)87569-0} {\bibfield  {journal} {\bibinfo
  {journal} {Chem. Eng. Sci.}\ }\textbf {\bibinfo {volume} {53}},\ \bibinfo
  {pages} {1} (\bibinfo {year} {1998})}\BibitemShut {NoStop}%
\bibitem [{\citenamefont {Bredt}\ and\ \citenamefont
  {Snyder}(1992)}]{92BrSnxx.NO}%
  \BibitemOpen
  \bibfield  {author} {\bibinfo {author} {\bibfnamefont {D.~S.}\ \bibnamefont
  {Bredt}}\ and\ \bibinfo {author} {\bibfnamefont {S.~H.}\ \bibnamefont
  {Snyder}},\ }\href {\doibase 10.1016/0896-6273(92)90104-L} {\bibfield
  {journal} {\bibinfo  {journal} {Neuron}\ }\textbf {\bibinfo {volume} {8}},\
  \bibinfo {pages} {3} (\bibinfo {year} {1992})}\BibitemShut {NoStop}%
\bibitem [{\citenamefont {Bredt}\ and\ \citenamefont
  {Snyder}(1994)}]{94LoDiSn.NO}%
  \BibitemOpen
  \bibfield  {author} {\bibinfo {author} {\bibfnamefont {D.~S.}\ \bibnamefont
  {Bredt}}\ and\ \bibinfo {author} {\bibfnamefont {S.~H.}\ \bibnamefont
  {Snyder}},\ }\href {\doibase 10.1146/annurev.bi.63.070194.001135} {\bibfield
  {journal} {\bibinfo  {journal} {Annu. Rev. Biochem.}\ }\textbf {\bibinfo
  {volume} {63}},\ \bibinfo {pages} {175} (\bibinfo {year} {1994})}\BibitemShut
  {NoStop}%
\bibitem [{\citenamefont {Arasimowicz}\ and\ \citenamefont
  {Floryszak-Wieczorek}(2007)}]{07ArFlxx.NO}%
  \BibitemOpen
  \bibfield  {author} {\bibinfo {author} {\bibfnamefont {M.}~\bibnamefont
  {Arasimowicz}}\ and\ \bibinfo {author} {\bibfnamefont {J.}~\bibnamefont
  {Floryszak-Wieczorek}},\ }\href {\doibase 10.1016/j.plantsci.2007.02.005}
  {\bibfield  {journal} {\bibinfo  {journal} {Plant Sci.}\ }\textbf {\bibinfo
  {volume} {172}},\ \bibinfo {pages} {876} (\bibinfo {year}
  {2007})}\BibitemShut {NoStop}%
\bibitem [{\citenamefont {Mayer}\ and\ \citenamefont
  {Hemmens}(1997)}]{97MaHexx.NO}%
  \BibitemOpen
  \bibfield  {author} {\bibinfo {author} {\bibfnamefont {B.}~\bibnamefont
  {Mayer}}\ and\ \bibinfo {author} {\bibfnamefont {B.}~\bibnamefont
  {Hemmens}},\ }\href {\doibase 10.1016/S0968-0004(97)01147-X} {\bibfield
  {journal} {\bibinfo  {journal} {Trends Biochem. Sci.}\ }\textbf {\bibinfo
  {volume} {22}},\ \bibinfo {pages} {477} (\bibinfo {year} {1997})}\BibitemShut
  {NoStop}%
\bibitem [{\citenamefont {Est{\'{e}}vez}\ and\ \citenamefont
  {Jord{\'{a}}n}(2002)}]{02EsJoxx.NO}%
  \BibitemOpen
  \bibfield  {author} {\bibinfo {author} {\bibfnamefont {A.~G.}\ \bibnamefont
  {Est{\'{e}}vez}}\ and\ \bibinfo {author} {\bibfnamefont {J.}~\bibnamefont
  {Jord{\'{a}}n}},\ }\href {\doibase 10.1111/j.1749-6632.2002.tb04069.x}
  {\bibfield  {journal} {\bibinfo  {journal} {Ann. N.Y. Acad. Sci.}\ }\textbf
  {\bibinfo {volume} {962}},\ \bibinfo {pages} {207} (\bibinfo {year}
  {2002})}\BibitemShut {NoStop}%
\bibitem [{\citenamefont {Ziurys}\ \emph {et~al.}(1991)\citenamefont {Ziurys},
  \citenamefont {McGonagle}, \citenamefont {Minh},\ and\ \citenamefont
  {Irvine}}]{91ZiMcMi.NO}%
  \BibitemOpen
  \bibfield  {author} {\bibinfo {author} {\bibfnamefont {L.~M.}\ \bibnamefont
  {Ziurys}}, \bibinfo {author} {\bibfnamefont {D.}~\bibnamefont {McGonagle}},
  \bibinfo {author} {\bibfnamefont {Y.}~\bibnamefont {Minh}}, \ and\ \bibinfo
  {author} {\bibfnamefont {W.~M.}\ \bibnamefont {Irvine}},\ }\href {\doibase
  10.1086/170072} {\bibfield  {journal} {\bibinfo  {journal} {Astrophys. J.}\
  }\textbf {\bibinfo {volume} {373}},\ \bibinfo {pages} {535} (\bibinfo {year}
  {1991})}\BibitemShut {NoStop}%
\bibitem [{\citenamefont {Cox}\ \emph {et~al.}(2008)\citenamefont {Cox},
  \citenamefont {Saglam}, \citenamefont {Gerard}, \citenamefont {Bertaux},
  \citenamefont {Gonzalez-Galindo}, \citenamefont {Leblanc},\ and\
  \citenamefont {Reberac}}]{08CoSaGe.NO}%
  \BibitemOpen
  \bibfield  {author} {\bibinfo {author} {\bibfnamefont {C.}~\bibnamefont
  {Cox}}, \bibinfo {author} {\bibfnamefont {A.}~\bibnamefont {Saglam}},
  \bibinfo {author} {\bibfnamefont {J.-C.}\ \bibnamefont {Gerard}}, \bibinfo
  {author} {\bibfnamefont {J.-L.}\ \bibnamefont {Bertaux}}, \bibinfo {author}
  {\bibfnamefont {F.}~\bibnamefont {Gonzalez-Galindo}}, \bibinfo {author}
  {\bibfnamefont {F.}~\bibnamefont {Leblanc}}, \ and\ \bibinfo {author}
  {\bibfnamefont {A.}~\bibnamefont {Reberac}},\ }\href {\doibase
  10.1029/2007JE003037} {\bibfield  {journal} {\bibinfo  {journal} {J. Geophys.
  Res.-Planets}\ }\textbf {\bibinfo {volume} {113}},\ \bibinfo {pages} {E08012}
  (\bibinfo {year} {2008})}\BibitemShut {NoStop}%
\bibitem [{\citenamefont {G{\'{e}}rard}\ \emph {et~al.}(2008)\citenamefont
  {G{\'{e}}rard}, \citenamefont {Cox}, \citenamefont {Saglam}, \citenamefont
  {Bertaux}, \citenamefont {Villard},\ and\ \citenamefont
  {Nehm{\'{e}}}}]{08GeCoSa.NO}%
  \BibitemOpen
  \bibfield  {author} {\bibinfo {author} {\bibfnamefont {J.-C.}\ \bibnamefont
  {G{\'{e}}rard}}, \bibinfo {author} {\bibfnamefont {C.}~\bibnamefont {Cox}},
  \bibinfo {author} {\bibfnamefont {A.}~\bibnamefont {Saglam}}, \bibinfo
  {author} {\bibfnamefont {J.-L.}\ \bibnamefont {Bertaux}}, \bibinfo {author}
  {\bibfnamefont {E.}~\bibnamefont {Villard}}, \ and\ \bibinfo {author}
  {\bibfnamefont {C.}~\bibnamefont {Nehm{\'{e}}}},\ }\href {\doibase
  10.1029/2008JE003078} {\bibfield  {journal} {\bibinfo  {journal} {J. Geophys.
  Res.}\ }\textbf {\bibinfo {volume} {113}},\ \bibinfo {pages} {E00B03}
  (\bibinfo {year} {2008})}\BibitemShut {NoStop}%
\bibitem [{\citenamefont {G{\'{e}}rard}\ \emph {et~al.}(2009)\citenamefont
  {G{\'{e}}rard}, \citenamefont {Cox}, \citenamefont {Soret}, \citenamefont
  {Saglam}, \citenamefont {Piccioni}, \citenamefont {Bertaux},\ and\
  \citenamefont {Drossart}}]{09GeGoSo.NO}%
  \BibitemOpen
  \bibfield  {author} {\bibinfo {author} {\bibfnamefont {J.-C.}\ \bibnamefont
  {G{\'{e}}rard}}, \bibinfo {author} {\bibfnamefont {C.}~\bibnamefont {Cox}},
  \bibinfo {author} {\bibfnamefont {L.}~\bibnamefont {Soret}}, \bibinfo
  {author} {\bibfnamefont {A.}~\bibnamefont {Saglam}}, \bibinfo {author}
  {\bibfnamefont {G.}~\bibnamefont {Piccioni}}, \bibinfo {author}
  {\bibfnamefont {J.-L.}\ \bibnamefont {Bertaux}}, \ and\ \bibinfo {author}
  {\bibfnamefont {P.}~\bibnamefont {Drossart}},\ }\href {\doibase
  10.1029/2009JE003371} {\bibfield  {journal} {\bibinfo  {journal} {J. Geophys.
  Res.}\ }\textbf {\bibinfo {volume} {114}},\ \bibinfo {pages} {E00B44}
  (\bibinfo {year} {2009})}\BibitemShut {NoStop}%
\bibitem [{\citenamefont {Partington}(1936)}]{36Partin.NO}%
  \BibitemOpen
  \bibfield  {author} {\bibinfo {author} {\bibfnamefont {J.}~\bibnamefont
  {Partington}},\ }\href {\doibase 10.1080/00033793600200291} {\bibfield
  {journal} {\bibinfo  {journal} {Ann. Sci.}\ }\textbf {\bibinfo {volume}
  {1}},\ \bibinfo {pages} {359} (\bibinfo {year} {1936})}\BibitemShut {NoStop}%
\bibitem [{\citenamefont {Priestley}(1772)}]{72Priest.NO}%
  \BibitemOpen
  \bibfield  {author} {\bibinfo {author} {\bibfnamefont {J.}~\bibnamefont
  {Priestley}},\ }\href {\doibase 10.1098/rstl.1772.0021} {\bibfield  {journal}
  {\bibinfo  {journal} {Philos. Trans. R. Soc. Lond.}\ }\textbf {\bibinfo
  {volume} {62}},\ \bibinfo {pages} {147} (\bibinfo {year} {1772})}\BibitemShut
  {NoStop}%
\bibitem [{\citenamefont {Deller}\ and\ \citenamefont
  {Hogan}(2020)}]{20DeHoxx.NO}%
  \BibitemOpen
  \bibfield  {author} {\bibinfo {author} {\bibfnamefont {A.}~\bibnamefont
  {Deller}}\ and\ \bibinfo {author} {\bibfnamefont {S.~D.}\ \bibnamefont
  {Hogan}},\ }\href {\doibase {10.1063/5.0003092}} {\bibfield  {journal}
  {\bibinfo  {journal} {J. Chem. Phys.}\ }\textbf {\bibinfo {volume} {{152}}},\
  \bibinfo {pages} {144305} (\bibinfo {year} {{2020}})}\BibitemShut {NoStop}%
\bibitem [{\citenamefont {Bessler}\ and\ \citenamefont
  {Schulz}(2004)}]{04BeScxx.NO}%
  \BibitemOpen
  \bibfield  {author} {\bibinfo {author} {\bibfnamefont {W.~G.}\ \bibnamefont
  {Bessler}}\ and\ \bibinfo {author} {\bibfnamefont {C.}~\bibnamefont
  {Schulz}},\ }\href {\doibase 10.1007/s00340-004-1421-x} {\bibfield  {journal}
  {\bibinfo  {journal} {Appl. Phys. B-Lasers Opt.}\ }\textbf {\bibinfo {volume}
  {78}},\ \bibinfo {pages} {519} (\bibinfo {year} {2004})}\BibitemShut
  {NoStop}%
\bibitem [{\citenamefont {{Van Gessel}}\ \emph {et~al.}(2013)\citenamefont
  {{Van Gessel}}, \citenamefont {Hrycak}, \citenamefont {Jasi{\'{n}}ski},
  \citenamefont {Mizeraczyk}, \citenamefont {{Van Der Mullen}},\ and\
  \citenamefont {Bruggeman}}]{13VaHrJa.NO}%
  \BibitemOpen
  \bibfield  {author} {\bibinfo {author} {\bibfnamefont {A.~F.}\ \bibnamefont
  {{Van Gessel}}}, \bibinfo {author} {\bibfnamefont {B.}~\bibnamefont
  {Hrycak}}, \bibinfo {author} {\bibfnamefont {M.}~\bibnamefont
  {Jasi{\'{n}}ski}}, \bibinfo {author} {\bibfnamefont {J.}~\bibnamefont
  {Mizeraczyk}}, \bibinfo {author} {\bibfnamefont {J.~J.}\ \bibnamefont {{Van
  Der Mullen}}}, \ and\ \bibinfo {author} {\bibfnamefont {P.~J.}\ \bibnamefont
  {Bruggeman}},\ }\href {\doibase 10.1088/0022-3727/46/9/095201} {\bibfield
  {journal} {\bibinfo  {journal} {J. Phys. D-Appl. Phys.}\ }\textbf {\bibinfo
  {volume} {46}},\ \bibinfo {pages} {095201} (\bibinfo {year}
  {2013})}\BibitemShut {NoStop}%
\bibitem [{\citenamefont {Tennyson}\ and\ \citenamefont
  {Yurchenko}(2012)}]{jt528}%
  \BibitemOpen
  \bibfield  {author} {\bibinfo {author} {\bibfnamefont {J.}~\bibnamefont
  {Tennyson}}\ and\ \bibinfo {author} {\bibfnamefont {S.~N.}\ \bibnamefont
  {Yurchenko}},\ }\href {\doibase 10.1111/j.1365-2966.2012.21440.x} {\bibfield
  {journal} {\bibinfo  {journal} {Mon. Not. Roy. Astron. Soc.}\ }\textbf
  {\bibinfo {volume} {425}},\ \bibinfo {pages} {21} (\bibinfo {year}
  {2012})}\BibitemShut {NoStop}%
\bibitem [{\citenamefont {Tennyson}\ \emph
  {et~al.}(2016{\natexlab{a}})\citenamefont {Tennyson}, \citenamefont
  {Yurchenko}, \citenamefont {Al-Refaie}, \citenamefont {Barton}, \citenamefont
  {Chubb}, \citenamefont {Coles}, \citenamefont {Diamantopoulou}, \citenamefont
  {Gorman}, \citenamefont {Hill}, \citenamefont {Lam}, \citenamefont {Lodi},
  \citenamefont {McKemmish}, \citenamefont {Na}, \citenamefont {Owens},
  \citenamefont {Polyansky}, \citenamefont {Rivlin}, \citenamefont
  {Sousa-Silva}, \citenamefont {Underwood}, \citenamefont {Yachmenev},\ and\
  \citenamefont {Zak}}]{jt631}%
  \BibitemOpen
  \bibfield  {author} {\bibinfo {author} {\bibfnamefont {J.}~\bibnamefont
  {Tennyson}}, \bibinfo {author} {\bibfnamefont {S.~N.}\ \bibnamefont
  {Yurchenko}}, \bibinfo {author} {\bibfnamefont {A.~F.}\ \bibnamefont
  {Al-Refaie}}, \bibinfo {author} {\bibfnamefont {E.~J.}\ \bibnamefont
  {Barton}}, \bibinfo {author} {\bibfnamefont {K.~L.}\ \bibnamefont {Chubb}},
  \bibinfo {author} {\bibfnamefont {P.~A.}\ \bibnamefont {Coles}}, \bibinfo
  {author} {\bibfnamefont {S.}~\bibnamefont {Diamantopoulou}}, \bibinfo
  {author} {\bibfnamefont {M.~N.}\ \bibnamefont {Gorman}}, \bibinfo {author}
  {\bibfnamefont {C.}~\bibnamefont {Hill}}, \bibinfo {author} {\bibfnamefont
  {A.~Z.}\ \bibnamefont {Lam}}, \bibinfo {author} {\bibfnamefont
  {L.}~\bibnamefont {Lodi}}, \bibinfo {author} {\bibfnamefont {L.~K.}\
  \bibnamefont {McKemmish}}, \bibinfo {author} {\bibfnamefont {Y.}~\bibnamefont
  {Na}}, \bibinfo {author} {\bibfnamefont {A.}~\bibnamefont {Owens}}, \bibinfo
  {author} {\bibfnamefont {O.~L.}\ \bibnamefont {Polyansky}}, \bibinfo {author}
  {\bibfnamefont {T.}~\bibnamefont {Rivlin}}, \bibinfo {author} {\bibfnamefont
  {C.}~\bibnamefont {Sousa-Silva}}, \bibinfo {author} {\bibfnamefont {D.~S.}\
  \bibnamefont {Underwood}}, \bibinfo {author} {\bibfnamefont {A.}~\bibnamefont
  {Yachmenev}}, \ and\ \bibinfo {author} {\bibfnamefont {E.}~\bibnamefont
  {Zak}},\ }\href {\doibase 10.1016/j.jms.2016.05.002} {\bibfield  {journal}
  {\bibinfo  {journal} {J. Mol. Spectrosc.}\ }\textbf {\bibinfo {volume}
  {327}},\ \bibinfo {pages} {73} (\bibinfo {year}
  {2016}{\natexlab{a}})}\BibitemShut {NoStop}%
\bibitem [{\citenamefont {Tennyson}\ \emph {et~al.}(2020)\citenamefont
  {Tennyson}, \citenamefont {Yurchenko}, \citenamefont {Al-Refaie},
  \citenamefont {Clark}, \citenamefont {Chubb}, \citenamefont {Conway},
  \citenamefont {Dewan}, \citenamefont {Gorman}, \citenamefont {Hill},
  \citenamefont {Lynas-Gray}, \citenamefont {Mellor}, \citenamefont
  {McKemmish}, \citenamefont {Owens}, \citenamefont {Polyansky}, \citenamefont
  {Semenov}, \citenamefont {Somogyi}, \citenamefont {Tinetti}, \citenamefont
  {Upadhyay}, \citenamefont {Waldmann}, \citenamefont {Wang}, \citenamefont
  {Wright},\ and\ \citenamefont {Yurchenko}}]{jt810}%
  \BibitemOpen
  \bibfield  {author} {\bibinfo {author} {\bibfnamefont {J.}~\bibnamefont
  {Tennyson}}, \bibinfo {author} {\bibfnamefont {S.~N.}\ \bibnamefont
  {Yurchenko}}, \bibinfo {author} {\bibfnamefont {A.~F.}\ \bibnamefont
  {Al-Refaie}}, \bibinfo {author} {\bibfnamefont {V.~H.~J.}\ \bibnamefont
  {Clark}}, \bibinfo {author} {\bibfnamefont {K.~L.}\ \bibnamefont {Chubb}},
  \bibinfo {author} {\bibfnamefont {E.~K.}\ \bibnamefont {Conway}}, \bibinfo
  {author} {\bibfnamefont {A.}~\bibnamefont {Dewan}}, \bibinfo {author}
  {\bibfnamefont {M.~N.}\ \bibnamefont {Gorman}}, \bibinfo {author}
  {\bibfnamefont {C.}~\bibnamefont {Hill}}, \bibinfo {author} {\bibfnamefont
  {A.~E.}\ \bibnamefont {Lynas-Gray}}, \bibinfo {author} {\bibfnamefont
  {T.}~\bibnamefont {Mellor}}, \bibinfo {author} {\bibfnamefont {L.~K.}\
  \bibnamefont {McKemmish}}, \bibinfo {author} {\bibfnamefont {A.}~\bibnamefont
  {Owens}}, \bibinfo {author} {\bibfnamefont {O.~L.}\ \bibnamefont
  {Polyansky}}, \bibinfo {author} {\bibfnamefont {M.}~\bibnamefont {Semenov}},
  \bibinfo {author} {\bibfnamefont {W.}~\bibnamefont {Somogyi}}, \bibinfo
  {author} {\bibfnamefont {G.}~\bibnamefont {Tinetti}}, \bibinfo {author}
  {\bibfnamefont {A.}~\bibnamefont {Upadhyay}}, \bibinfo {author}
  {\bibfnamefont {I.}~\bibnamefont {Waldmann}}, \bibinfo {author}
  {\bibfnamefont {Y.}~\bibnamefont {Wang}}, \bibinfo {author} {\bibfnamefont
  {S.}~\bibnamefont {Wright}}, \ and\ \bibinfo {author} {\bibfnamefont {O.~P.}\
  \bibnamefont {Yurchenko}},\ }\href {\doibase 10.1016/j.jqsrt.2020.107228}
  {\bibfield  {journal} {\bibinfo  {journal} {J. Quant. Spectrosc. Radiat.
  Transf.}\ }\textbf {\bibinfo {volume} {255}},\ \bibinfo {pages} {107228}
  (\bibinfo {year} {2020})}\BibitemShut {NoStop}%
\bibitem [{\citenamefont {Wong}\ \emph {et~al.}(2017)\citenamefont {Wong},
  \citenamefont {Yurchenko}, \citenamefont {Bernath}, \citenamefont {Mueller},
  \citenamefont {McConkey},\ and\ \citenamefont {Tennyson}}]{jt686}%
  \BibitemOpen
  \bibfield  {author} {\bibinfo {author} {\bibfnamefont {A.}~\bibnamefont
  {Wong}}, \bibinfo {author} {\bibfnamefont {S.~N.}\ \bibnamefont {Yurchenko}},
  \bibinfo {author} {\bibfnamefont {P.}~\bibnamefont {Bernath}}, \bibinfo
  {author} {\bibfnamefont {H.~S.~P.}\ \bibnamefont {Mueller}}, \bibinfo
  {author} {\bibfnamefont {S.}~\bibnamefont {McConkey}}, \ and\ \bibinfo
  {author} {\bibfnamefont {J.}~\bibnamefont {Tennyson}},\ }\href {\doibase
  10.1093/mnras/stx1211} {\bibfield  {journal} {\bibinfo  {journal} {Mon. Not.
  Roy. Astron. Soc.}\ }\textbf {\bibinfo {volume} {470}},\ \bibinfo {pages}
  {882} (\bibinfo {year} {2017})}\BibitemShut {NoStop}%
\bibitem [{\citenamefont {Lagerqvist}\ and\ \citenamefont
  {Miescher}(1958)}]{58LaMixx.NO}%
  \BibitemOpen
  \bibfield  {author} {\bibinfo {author} {\bibfnamefont {A.}~\bibnamefont
  {Lagerqvist}}\ and\ \bibinfo {author} {\bibfnamefont {E.}~\bibnamefont
  {Miescher}},\ }\href@noop {} {\bibfield  {journal} {\bibinfo  {journal}
  {Helv. Phys. Acta}\ }\textbf {\bibinfo {volume} {31}},\ \bibinfo {pages}
  {221} (\bibinfo {year} {1958})}\BibitemShut {NoStop}%
\bibitem [{\citenamefont {Danielak}\ \emph {et~al.}(1997)\citenamefont
  {Danielak}, \citenamefont {Domin}, \citenamefont {Kepa}, \citenamefont
  {Rytel},\ and\ \citenamefont {Zachwieja}}]{97DaDoKe.NO}%
  \BibitemOpen
  \bibfield  {author} {\bibinfo {author} {\bibfnamefont {J.}~\bibnamefont
  {Danielak}}, \bibinfo {author} {\bibfnamefont {U.}~\bibnamefont {Domin}},
  \bibinfo {author} {\bibfnamefont {R.}~\bibnamefont {Kepa}}, \bibinfo {author}
  {\bibfnamefont {M.}~\bibnamefont {Rytel}}, \ and\ \bibinfo {author}
  {\bibfnamefont {M.}~\bibnamefont {Zachwieja}},\ }\href {\doibase
  10.1006/jmsp.1996.7181} {\bibfield  {journal} {\bibinfo  {journal} {J. Mol.
  Spectrosc.}\ }\textbf {\bibinfo {volume} {181}},\ \bibinfo {pages} {394}
  (\bibinfo {year} {1997})}\BibitemShut {NoStop}%
\bibitem [{\citenamefont {Yoshino}\ \emph {et~al.}(2006)\citenamefont
  {Yoshino}, \citenamefont {Thorne}, \citenamefont {Murray}, \citenamefont
  {Cheung}, \citenamefont {Wong},\ and\ \citenamefont {Imajo}}]{06YoThMu.NO}%
  \BibitemOpen
  \bibfield  {author} {\bibinfo {author} {\bibfnamefont {K.}~\bibnamefont
  {Yoshino}}, \bibinfo {author} {\bibfnamefont {A.~P.}\ \bibnamefont {Thorne}},
  \bibinfo {author} {\bibfnamefont {J.~E.}\ \bibnamefont {Murray}}, \bibinfo
  {author} {\bibfnamefont {A.~S.-C.}\ \bibnamefont {Cheung}}, \bibinfo {author}
  {\bibfnamefont {A.~L.}\ \bibnamefont {Wong}}, \ and\ \bibinfo {author}
  {\bibfnamefont {T.}~\bibnamefont {Imajo}},\ }\href {\doibase
  10.1063/1.2138029} {\bibfield  {journal} {\bibinfo  {journal} {J. Chem.
  Phys.}\ }\textbf {\bibinfo {volume} {124}},\ \bibinfo {pages} {054323}
  (\bibinfo {year} {2006})}\BibitemShut {NoStop}%
\bibitem [{\citenamefont {Gordon}\ \emph {et~al.}(2017)\citenamefont {Gordon},
  \citenamefont {Rothman}, \citenamefont {Hill}, \citenamefont {Kochanov},
  \citenamefont {Tan}, \citenamefont {Bernath}, \citenamefont {Birk},
  \citenamefont {Boudon}, \citenamefont {Campargue}, \citenamefont {Chance},
  \citenamefont {Drouin}, \citenamefont {Flaud}, \citenamefont {Gamache},
  \citenamefont {Hodges}, \citenamefont {Jacquemart}, \citenamefont
  {Perevalov}, \citenamefont {Perrin}, \citenamefont {Shine}, \citenamefont
  {Smith}, \citenamefont {Tennyson}, \citenamefont {Toon}, \citenamefont
  {Tran}, \citenamefont {Tyuterev}, \citenamefont {Barbe}, \citenamefont
  {Cs{\'a}sz{\'a}r}, \citenamefont {Devi}, \citenamefont {Furtenbacher},
  \citenamefont {Harrison}, \citenamefont {Hartmann}, \citenamefont {Jolly},
  \citenamefont {Johnson}, \citenamefont {Karman}, \citenamefont {Kleiner},
  \citenamefont {Kyuberis}, \citenamefont {Loos}, \citenamefont {Lyulin},
  \citenamefont {Massie}, \citenamefont {Mikhailenko}, \citenamefont
  {Moazzen-Ahmadi}, \citenamefont {M{\"u}ller}, \citenamefont {Naumenko},
  \citenamefont {Nikitin}, \citenamefont {Polyansky}, \citenamefont {Rey},
  \citenamefont {Rotger}, \citenamefont {Sharpe}, \citenamefont {Sung},
  \citenamefont {Starikova}, \citenamefont {Tashkun}, \citenamefont {{Vander
  Auwera}}, \citenamefont {Wagner}, \citenamefont {Wilzewski}, \citenamefont
  {Wcis{\l}o}, \citenamefont {Yu},\ and\ \citenamefont {Zak}}]{jt691}%
  \BibitemOpen
  \bibfield  {author} {\bibinfo {author} {\bibfnamefont {I.~E.}\ \bibnamefont
  {Gordon}}, \bibinfo {author} {\bibfnamefont {L.~S.}\ \bibnamefont {Rothman}},
  \bibinfo {author} {\bibfnamefont {C.}~\bibnamefont {Hill}}, \bibinfo {author}
  {\bibfnamefont {R.~V.}\ \bibnamefont {Kochanov}}, \bibinfo {author}
  {\bibfnamefont {Y.}~\bibnamefont {Tan}}, \bibinfo {author} {\bibfnamefont
  {P.~F.}\ \bibnamefont {Bernath}}, \bibinfo {author} {\bibfnamefont
  {M.}~\bibnamefont {Birk}}, \bibinfo {author} {\bibfnamefont {V.}~\bibnamefont
  {Boudon}}, \bibinfo {author} {\bibfnamefont {A.}~\bibnamefont {Campargue}},
  \bibinfo {author} {\bibfnamefont {K.~V.}\ \bibnamefont {Chance}}, \bibinfo
  {author} {\bibfnamefont {B.~J.}\ \bibnamefont {Drouin}}, \bibinfo {author}
  {\bibfnamefont {J.-M.}\ \bibnamefont {Flaud}}, \bibinfo {author}
  {\bibfnamefont {R.~R.}\ \bibnamefont {Gamache}}, \bibinfo {author}
  {\bibfnamefont {J.~T.}\ \bibnamefont {Hodges}}, \bibinfo {author}
  {\bibfnamefont {D.}~\bibnamefont {Jacquemart}}, \bibinfo {author}
  {\bibfnamefont {V.~I.}\ \bibnamefont {Perevalov}}, \bibinfo {author}
  {\bibfnamefont {A.}~\bibnamefont {Perrin}}, \bibinfo {author} {\bibfnamefont
  {K.~P.}\ \bibnamefont {Shine}}, \bibinfo {author} {\bibfnamefont {M.-A.~H.}\
  \bibnamefont {Smith}}, \bibinfo {author} {\bibfnamefont {J.}~\bibnamefont
  {Tennyson}}, \bibinfo {author} {\bibfnamefont {G.~C.}\ \bibnamefont {Toon}},
  \bibinfo {author} {\bibfnamefont {H.}~\bibnamefont {Tran}}, \bibinfo {author}
  {\bibfnamefont {V.~G.}\ \bibnamefont {Tyuterev}}, \bibinfo {author}
  {\bibfnamefont {A.}~\bibnamefont {Barbe}}, \bibinfo {author} {\bibfnamefont
  {A.~G.}\ \bibnamefont {Cs{\'a}sz{\'a}r}}, \bibinfo {author} {\bibfnamefont
  {V.~M.}\ \bibnamefont {Devi}}, \bibinfo {author} {\bibfnamefont
  {T.}~\bibnamefont {Furtenbacher}}, \bibinfo {author} {\bibfnamefont {J.~J.}\
  \bibnamefont {Harrison}}, \bibinfo {author} {\bibfnamefont {J.-M.}\
  \bibnamefont {Hartmann}}, \bibinfo {author} {\bibfnamefont {A.}~\bibnamefont
  {Jolly}}, \bibinfo {author} {\bibfnamefont {T.~J.}\ \bibnamefont {Johnson}},
  \bibinfo {author} {\bibfnamefont {T.}~\bibnamefont {Karman}}, \bibinfo
  {author} {\bibfnamefont {I.}~\bibnamefont {Kleiner}}, \bibinfo {author}
  {\bibfnamefont {A.~A.}\ \bibnamefont {Kyuberis}}, \bibinfo {author}
  {\bibfnamefont {J.}~\bibnamefont {Loos}}, \bibinfo {author} {\bibfnamefont
  {O.~M.}\ \bibnamefont {Lyulin}}, \bibinfo {author} {\bibfnamefont {S.~T.}\
  \bibnamefont {Massie}}, \bibinfo {author} {\bibfnamefont {S.~N.}\
  \bibnamefont {Mikhailenko}}, \bibinfo {author} {\bibfnamefont
  {N.}~\bibnamefont {Moazzen-Ahmadi}}, \bibinfo {author} {\bibfnamefont
  {H.~S.~P.}\ \bibnamefont {M{\"u}ller}}, \bibinfo {author} {\bibfnamefont
  {O.~V.}\ \bibnamefont {Naumenko}}, \bibinfo {author} {\bibfnamefont {A.~V.}\
  \bibnamefont {Nikitin}}, \bibinfo {author} {\bibfnamefont {O.~L.}\
  \bibnamefont {Polyansky}}, \bibinfo {author} {\bibfnamefont {M.}~\bibnamefont
  {Rey}}, \bibinfo {author} {\bibfnamefont {M.}~\bibnamefont {Rotger}},
  \bibinfo {author} {\bibfnamefont {S.~W.}\ \bibnamefont {Sharpe}}, \bibinfo
  {author} {\bibfnamefont {K.}~\bibnamefont {Sung}}, \bibinfo {author}
  {\bibfnamefont {E.}~\bibnamefont {Starikova}}, \bibinfo {author}
  {\bibfnamefont {S.~A.}\ \bibnamefont {Tashkun}}, \bibinfo {author}
  {\bibfnamefont {J.}~\bibnamefont {{Vander Auwera}}}, \bibinfo {author}
  {\bibfnamefont {G.}~\bibnamefont {Wagner}}, \bibinfo {author} {\bibfnamefont
  {J.}~\bibnamefont {Wilzewski}}, \bibinfo {author} {\bibfnamefont
  {P.}~\bibnamefont {Wcis{\l}o}}, \bibinfo {author} {\bibfnamefont
  {S.}~\bibnamefont {Yu}}, \ and\ \bibinfo {author} {\bibfnamefont {E.~J.}\
  \bibnamefont {Zak}},\ }\href {\doibase 10.1016/j.jqsrt.2017.06.038}
  {\bibfield  {journal} {\bibinfo  {journal} {J. Quant. Spectrosc. Radiat.
  Transf.}\ }\textbf {\bibinfo {volume} {203}},\ \bibinfo {pages} {3} (\bibinfo
  {year} {2017})}\BibitemShut {NoStop}%
\bibitem [{\citenamefont {Jacquinet-Husson}\ \emph {et~al.}(2016)\citenamefont
  {Jacquinet-Husson}, \citenamefont {Armante}, \citenamefont {Scott},
  \citenamefont {Ch{\'e}din}, \citenamefont {Cr{\'e}peau}, \citenamefont
  {Boutammine}, \citenamefont {Bouhdaoui}, \citenamefont {Crevoisier},
  \citenamefont {Capelle}, \citenamefont {Boonne}, \citenamefont
  {Poulet-Crovisier}, \citenamefont {Barbe}, \citenamefont {Benner},
  \citenamefont {Boudon}, \citenamefont {Brown}, \citenamefont {Buldyreva},
  \citenamefont {Campargue}, \citenamefont {Coudert}, \citenamefont {Devi},
  \citenamefont {Down}, \citenamefont {Drouin}, \citenamefont {Fayt},
  \citenamefont {Fittschen}, \citenamefont {Flaud}, \citenamefont {Gamache},
  \citenamefont {Harrison}, \citenamefont {Hill}, \citenamefont {Hodnebrog},
  \citenamefont {Hu}, \citenamefont {Jacquemart}, \citenamefont {Jolly},
  \citenamefont {Jim{\'e}nez}, \citenamefont {Lavrentieva}, \citenamefont
  {Liu}, \citenamefont {Lodi}, \citenamefont {Lyulin}, \citenamefont {Massie},
  \citenamefont {Mikhailenko}, \citenamefont {M{\"u}ller}, \citenamefont
  {Naumenko}, \citenamefont {Nikitin}, \citenamefont {Nielsen}, \citenamefont
  {Orphal}, \citenamefont {Perevalov}, \citenamefont {Perrin}, \citenamefont
  {Polovtseva}, \citenamefont {Predoi-Cross}, \citenamefont {Rotger},
  \citenamefont {Ruth}, \citenamefont {Yu}, \citenamefont {Sung}, \citenamefont
  {Tashkun}, \citenamefont {Tennyson}, \citenamefont {Tyuterev}, \citenamefont
  {{Vander Auwera}}, \citenamefont {Voronin},\ and\ \citenamefont
  {Makie}}]{jt636}%
  \BibitemOpen
  \bibfield  {author} {\bibinfo {author} {\bibfnamefont {N.}~\bibnamefont
  {Jacquinet-Husson}}, \bibinfo {author} {\bibfnamefont {R.}~\bibnamefont
  {Armante}}, \bibinfo {author} {\bibfnamefont {N.~A.}\ \bibnamefont {Scott}},
  \bibinfo {author} {\bibfnamefont {A.}~\bibnamefont {Ch{\'e}din}}, \bibinfo
  {author} {\bibfnamefont {L.}~\bibnamefont {Cr{\'e}peau}}, \bibinfo {author}
  {\bibfnamefont {C.}~\bibnamefont {Boutammine}}, \bibinfo {author}
  {\bibfnamefont {A.}~\bibnamefont {Bouhdaoui}}, \bibinfo {author}
  {\bibfnamefont {C.}~\bibnamefont {Crevoisier}}, \bibinfo {author}
  {\bibfnamefont {V.}~\bibnamefont {Capelle}}, \bibinfo {author} {\bibfnamefont
  {C.}~\bibnamefont {Boonne}}, \bibinfo {author} {\bibfnamefont
  {N.}~\bibnamefont {Poulet-Crovisier}}, \bibinfo {author} {\bibfnamefont
  {A.}~\bibnamefont {Barbe}}, \bibinfo {author} {\bibfnamefont {D.~C.}\
  \bibnamefont {Benner}}, \bibinfo {author} {\bibfnamefont {V.}~\bibnamefont
  {Boudon}}, \bibinfo {author} {\bibfnamefont {L.~R.}\ \bibnamefont {Brown}},
  \bibinfo {author} {\bibfnamefont {J.}~\bibnamefont {Buldyreva}}, \bibinfo
  {author} {\bibfnamefont {A.}~\bibnamefont {Campargue}}, \bibinfo {author}
  {\bibfnamefont {L.~H.}\ \bibnamefont {Coudert}}, \bibinfo {author}
  {\bibfnamefont {V.~M.}\ \bibnamefont {Devi}}, \bibinfo {author}
  {\bibfnamefont {M.~J.}\ \bibnamefont {Down}}, \bibinfo {author}
  {\bibfnamefont {B.~J.}\ \bibnamefont {Drouin}}, \bibinfo {author}
  {\bibfnamefont {A.}~\bibnamefont {Fayt}}, \bibinfo {author} {\bibfnamefont
  {C.}~\bibnamefont {Fittschen}}, \bibinfo {author} {\bibfnamefont {J.-M.}\
  \bibnamefont {Flaud}}, \bibinfo {author} {\bibfnamefont {R.~R.}\ \bibnamefont
  {Gamache}}, \bibinfo {author} {\bibfnamefont {J.~J.}\ \bibnamefont
  {Harrison}}, \bibinfo {author} {\bibfnamefont {C.}~\bibnamefont {Hill}},
  \bibinfo {author} {\bibfnamefont {{\O}.}~\bibnamefont {Hodnebrog}}, \bibinfo
  {author} {\bibfnamefont {S.~M.}\ \bibnamefont {Hu}}, \bibinfo {author}
  {\bibfnamefont {D.}~\bibnamefont {Jacquemart}}, \bibinfo {author}
  {\bibfnamefont {A.}~\bibnamefont {Jolly}}, \bibinfo {author} {\bibfnamefont
  {E.}~\bibnamefont {Jim{\'e}nez}}, \bibinfo {author} {\bibfnamefont {N.~N.}\
  \bibnamefont {Lavrentieva}}, \bibinfo {author} {\bibfnamefont {A.~W.}\
  \bibnamefont {Liu}}, \bibinfo {author} {\bibfnamefont {L.}~\bibnamefont
  {Lodi}}, \bibinfo {author} {\bibfnamefont {O.~M.}\ \bibnamefont {Lyulin}},
  \bibinfo {author} {\bibfnamefont {S.~T.}\ \bibnamefont {Massie}}, \bibinfo
  {author} {\bibfnamefont {S.}~\bibnamefont {Mikhailenko}}, \bibinfo {author}
  {\bibfnamefont {H.~S.~P.}\ \bibnamefont {M{\"u}ller}}, \bibinfo {author}
  {\bibfnamefont {O.~V.}\ \bibnamefont {Naumenko}}, \bibinfo {author}
  {\bibfnamefont {A.}~\bibnamefont {Nikitin}}, \bibinfo {author} {\bibfnamefont
  {C.~J.}\ \bibnamefont {Nielsen}}, \bibinfo {author} {\bibfnamefont
  {J.}~\bibnamefont {Orphal}}, \bibinfo {author} {\bibfnamefont {V.~I.}\
  \bibnamefont {Perevalov}}, \bibinfo {author} {\bibfnamefont {A.}~\bibnamefont
  {Perrin}}, \bibinfo {author} {\bibfnamefont {E.}~\bibnamefont {Polovtseva}},
  \bibinfo {author} {\bibfnamefont {A.}~\bibnamefont {Predoi-Cross}}, \bibinfo
  {author} {\bibfnamefont {M.}~\bibnamefont {Rotger}}, \bibinfo {author}
  {\bibfnamefont {A.~A.}\ \bibnamefont {Ruth}}, \bibinfo {author}
  {\bibfnamefont {S.~S.}\ \bibnamefont {Yu}}, \bibinfo {author} {\bibfnamefont
  {K.}~\bibnamefont {Sung}}, \bibinfo {author} {\bibfnamefont {S.~A.}\
  \bibnamefont {Tashkun}}, \bibinfo {author} {\bibfnamefont {J.}~\bibnamefont
  {Tennyson}}, \bibinfo {author} {\bibfnamefont {V.~G.}\ \bibnamefont
  {Tyuterev}}, \bibinfo {author} {\bibfnamefont {J.}~\bibnamefont {{Vander
  Auwera}}}, \bibinfo {author} {\bibfnamefont {B.~A.}\ \bibnamefont {Voronin}},
  \ and\ \bibinfo {author} {\bibfnamefont {A.}~\bibnamefont {Makie}},\ }\href
  {\doibase 10.1016/j.jms.2016.06.007} {\bibfield  {journal} {\bibinfo
  {journal} {J. Mol. Spectrosc.}\ }\textbf {\bibinfo {volume} {327}},\ \bibinfo
  {pages} {31} (\bibinfo {year} {2016})}\BibitemShut {NoStop}%
\bibitem [{\citenamefont {Luque}\ and\ \citenamefont
  {Crosley}(1995)}]{95LuCrxx.NO}%
  \BibitemOpen
  \bibfield  {author} {\bibinfo {author} {\bibfnamefont {J.}~\bibnamefont
  {Luque}}\ and\ \bibinfo {author} {\bibfnamefont {D.~R.}\ \bibnamefont
  {Crosley}},\ }\href {\doibase 10.1016/0022-4073(95)90006-3} {\bibfield
  {journal} {\bibinfo  {journal} {J. Quant. Spectrosc. Radiat. Transf.}\
  }\textbf {\bibinfo {volume} {53}},\ \bibinfo {pages} {189} (\bibinfo {year}
  {1995})}\BibitemShut {NoStop}%
\bibitem [{\citenamefont {Luque}\ and\ \citenamefont
  {Crosley}(1999{\natexlab{a}})}]{99LuCrxx.NO}%
  \BibitemOpen
  \bibfield  {author} {\bibinfo {author} {\bibfnamefont {J.}~\bibnamefont
  {Luque}}\ and\ \bibinfo {author} {\bibfnamefont {D.~R.}\ \bibnamefont
  {Crosley}},\ }\href {\doibase 10.1063/1.480064} {\bibfield  {journal}
  {\bibinfo  {journal} {J. Chem. Phys.}\ }\textbf {\bibinfo {volume} {111}},\
  \bibinfo {pages} {7405} (\bibinfo {year} {1999}{\natexlab{a}})}\BibitemShut
  {NoStop}%
\bibitem [{\citenamefont {Luque}\ and\ \citenamefont
  {Crosley}(2000)}]{00LuCrxx.NO}%
  \BibitemOpen
  \bibfield  {author} {\bibinfo {author} {\bibfnamefont {J.}~\bibnamefont
  {Luque}}\ and\ \bibinfo {author} {\bibfnamefont {D.~R.}\ \bibnamefont
  {Crosley}},\ }\href {\doibase 10.1063/1.481560} {\bibfield  {journal}
  {\bibinfo  {journal} {J. Chem. Phys.}\ }\textbf {\bibinfo {volume} {112}},\
  \bibinfo {pages} {9411} (\bibinfo {year} {2000})}\BibitemShut {NoStop}%
\bibitem [{\citenamefont {Luque}\ and\ \citenamefont
  {Crosley}(1999{\natexlab{b}})}]{99LIFBASE}%
  \BibitemOpen
  \bibfield  {author} {\bibinfo {author} {\bibfnamefont {J.}~\bibnamefont
  {Luque}}\ and\ \bibinfo {author} {\bibfnamefont {D.~R.}\ \bibnamefont
  {Crosley}},\ }\href
  {https://archive.sri.com/engage/products-solutions/lifbase} {\bibfield
  {journal} {\bibinfo  {journal} {SRI international report MP}\ }\textbf
  {\bibinfo {volume} {99}} (\bibinfo {year} {1999}{\natexlab{b}})}\BibitemShut
  {NoStop}%
\bibitem [{\citenamefont {Yoshino}\ \emph {et~al.}(1998)\citenamefont
  {Yoshino}, \citenamefont {Esmond}, \citenamefont {Parkinson}, \citenamefont
  {Thorne}, \citenamefont {Murray}, \citenamefont {Learner}, \citenamefont
  {Cox}, \citenamefont {Cheung}, \citenamefont {Leung}, \citenamefont {Ito},
  \citenamefont {Matsui},\ and\ \citenamefont {Imajo}}]{98YoEsPa.NO}%
  \BibitemOpen
  \bibfield  {author} {\bibinfo {author} {\bibfnamefont {K.}~\bibnamefont
  {Yoshino}}, \bibinfo {author} {\bibfnamefont {J.~R.}\ \bibnamefont {Esmond}},
  \bibinfo {author} {\bibfnamefont {W.~H.}\ \bibnamefont {Parkinson}}, \bibinfo
  {author} {\bibfnamefont {A.~P.}\ \bibnamefont {Thorne}}, \bibinfo {author}
  {\bibfnamefont {J.~E.}\ \bibnamefont {Murray}}, \bibinfo {author}
  {\bibfnamefont {R.~C.~M.}\ \bibnamefont {Learner}}, \bibinfo {author}
  {\bibfnamefont {G.}~\bibnamefont {Cox}}, \bibinfo {author} {\bibfnamefont
  {A.~S.-C.}\ \bibnamefont {Cheung}}, \bibinfo {author} {\bibfnamefont
  {K.~W.-S.}\ \bibnamefont {Leung}}, \bibinfo {author} {\bibfnamefont
  {K.}~\bibnamefont {Ito}}, \bibinfo {author} {\bibfnamefont {T.}~\bibnamefont
  {Matsui}}, \ and\ \bibinfo {author} {\bibfnamefont {T.}~\bibnamefont
  {Imajo}},\ }\href {\doibase 10.1063/1.476750} {\bibfield  {journal} {\bibinfo
   {journal} {J. Chem. Phys.}\ }\textbf {\bibinfo {volume} {109}},\ \bibinfo
  {pages} {1751} (\bibinfo {year} {1998})}\BibitemShut {NoStop}%
\bibitem [{\citenamefont {Albritton}, \citenamefont {Schmeltekopf},\ and\
  \citenamefont {Zare}(1979)}]{79AlScZa.NOplus}%
  \BibitemOpen
  \bibfield  {author} {\bibinfo {author} {\bibfnamefont {D.~L.}\ \bibnamefont
  {Albritton}}, \bibinfo {author} {\bibfnamefont {A.~L.}\ \bibnamefont
  {Schmeltekopf}}, \ and\ \bibinfo {author} {\bibfnamefont {R.~N.}\
  \bibnamefont {Zare}},\ }\href {\doibase 10.1063/1.438757} {\bibfield
  {journal} {\bibinfo  {journal} {The Journal of Chemical Physics}\ }\textbf
  {\bibinfo {volume} {71}},\ \bibinfo {pages} {3271} (\bibinfo {year}
  {1979})}\BibitemShut {NoStop}%
\bibitem [{\citenamefont {Gallusser}\ and\ \citenamefont
  {Dressler}(1982)}]{82GaDrxx.NO}%
  \BibitemOpen
  \bibfield  {author} {\bibinfo {author} {\bibfnamefont {R.}~\bibnamefont
  {Gallusser}}\ and\ \bibinfo {author} {\bibfnamefont {K.}~\bibnamefont
  {Dressler}},\ }\href {\doibase 10.1063/1.443565} {\bibfield  {journal}
  {\bibinfo  {journal} {J. Chem. Phys.}\ }\textbf {\bibinfo {volume} {76}},\
  \bibinfo {pages} {4311} (\bibinfo {year} {1982})}\BibitemShut {NoStop}%
\bibitem [{\citenamefont {Furtenbacher}, \citenamefont {{Cs\'asz\'ar}},\ and\
  \citenamefont {Tennyson}(2007{\natexlab{a}})}]{jt412}%
  \BibitemOpen
  \bibfield  {author} {\bibinfo {author} {\bibfnamefont {T.}~\bibnamefont
  {Furtenbacher}}, \bibinfo {author} {\bibfnamefont {A.~G.}\ \bibnamefont
  {{Cs\'asz\'ar}}}, \ and\ \bibinfo {author} {\bibfnamefont {J.}~\bibnamefont
  {Tennyson}},\ }\href {\doibase 10.1016/j.jms.2007.07.005} {\bibfield
  {journal} {\bibinfo  {journal} {J. Mol. Spectrosc.}\ }\textbf {\bibinfo
  {volume} {245}},\ \bibinfo {pages} {115} (\bibinfo {year}
  {2007}{\natexlab{a}})}\BibitemShut {NoStop}%
\bibitem [{\citenamefont {T\'obi\'as}\ \emph {et~al.}(2019)\citenamefont
  {T\'obi\'as}, \citenamefont {Furtenbacher}, \citenamefont {Tennyson},\ and\
  \citenamefont {Cs\'asz\'ar}}]{jt750}%
  \BibitemOpen
  \bibfield  {author} {\bibinfo {author} {\bibfnamefont {R.}~\bibnamefont
  {T\'obi\'as}}, \bibinfo {author} {\bibfnamefont {T.}~\bibnamefont
  {Furtenbacher}}, \bibinfo {author} {\bibfnamefont {J.}~\bibnamefont
  {Tennyson}}, \ and\ \bibinfo {author} {\bibfnamefont {A.~G.}\ \bibnamefont
  {Cs\'asz\'ar}},\ }\href {\doibase 10.1039/c8cp05169k} {\bibfield  {journal}
  {\bibinfo  {journal} {Phys. Chem. Chem. Phys.}\ }\textbf {\bibinfo {volume}
  {21}},\ \bibinfo {pages} {3473} (\bibinfo {year} {2019})}\BibitemShut
  {NoStop}%
\bibitem [{\citenamefont {Werner}\ \emph {et~al.}(2015)\citenamefont {Werner},
  \citenamefont {Knowles}, \citenamefont {Knizia}, \citenamefont {Manby},
  \citenamefont {Sch{\"{u}}tz}, \citenamefont {Celani}, \citenamefont
  {Gy{\H{o}}rffy}, \citenamefont {Kats}, \citenamefont {Korona}, \citenamefont
  {Lindh}, \citenamefont {Mitrushenkov}, \citenamefont {Rauhut}, \citenamefont
  {Shamasundar}, \citenamefont {Adler}, \citenamefont {Amos}, \citenamefont
  {Bernhardsson}, \citenamefont {Berning}, \citenamefont {Cooper},
  \citenamefont {Deegan}, \citenamefont {Dobbyn}, \citenamefont {Eckert},
  \citenamefont {Goll}, \citenamefont {Hampel}, \citenamefont {Hesselmann},
  \citenamefont {Hetzer}, \citenamefont {Hrenar}, \citenamefont {Jansen},
  \citenamefont {K{\"{o}}ppl}, \citenamefont {Liu}, \citenamefont {Lloyd},
  \citenamefont {Mata}, \citenamefont {May}, \citenamefont {McNicholas},
  \citenamefont {Meyer}, \citenamefont {Mura}, \citenamefont {Nicklass},
  \citenamefont {O'Neill}, \citenamefont {Palmieri}, \citenamefont {Peng},
  \citenamefont {Pfl{\"{u}}ger}, \citenamefont {Pitzer}, \citenamefont
  {Reiher}, \citenamefont {Shiozaki}, \citenamefont {Stoll}, \citenamefont
  {Stone}, \citenamefont {Tarroni}, \citenamefont {Thorsteinsson},\ and\
  \citenamefont {Wang}}]{MOLPRO2015}%
  \BibitemOpen
  \bibfield  {author} {\bibinfo {author} {\bibfnamefont {H.~J.}\ \bibnamefont
  {Werner}}, \bibinfo {author} {\bibfnamefont {P.~J.}\ \bibnamefont {Knowles}},
  \bibinfo {author} {\bibfnamefont {G.}~\bibnamefont {Knizia}}, \bibinfo
  {author} {\bibfnamefont {F.~R.}\ \bibnamefont {Manby}}, \bibinfo {author}
  {\bibfnamefont {M.}~\bibnamefont {Sch{\"{u}}tz}}, \bibinfo {author}
  {\bibfnamefont {P.}~\bibnamefont {Celani}}, \bibinfo {author} {\bibfnamefont
  {W.}~\bibnamefont {Gy{\H{o}}rffy}}, \bibinfo {author} {\bibfnamefont
  {D.}~\bibnamefont {Kats}}, \bibinfo {author} {\bibfnamefont {T.}~\bibnamefont
  {Korona}}, \bibinfo {author} {\bibfnamefont {R.}~\bibnamefont {Lindh}},
  \bibinfo {author} {\bibfnamefont {A.}~\bibnamefont {Mitrushenkov}}, \bibinfo
  {author} {\bibfnamefont {G.}~\bibnamefont {Rauhut}}, \bibinfo {author}
  {\bibfnamefont {K.~R.}\ \bibnamefont {Shamasundar}}, \bibinfo {author}
  {\bibfnamefont {T.~B.}\ \bibnamefont {Adler}}, \bibinfo {author}
  {\bibfnamefont {R.~D.}\ \bibnamefont {Amos}}, \bibinfo {author}
  {\bibfnamefont {A.}~\bibnamefont {Bernhardsson}}, \bibinfo {author}
  {\bibfnamefont {A.}~\bibnamefont {Berning}}, \bibinfo {author} {\bibfnamefont
  {D.~L.}\ \bibnamefont {Cooper}}, \bibinfo {author} {\bibfnamefont {M.~J.~O.}\
  \bibnamefont {Deegan}}, \bibinfo {author} {\bibfnamefont {A.~J.}\
  \bibnamefont {Dobbyn}}, \bibinfo {author} {\bibfnamefont {F.}~\bibnamefont
  {Eckert}}, \bibinfo {author} {\bibfnamefont {E.}~\bibnamefont {Goll}},
  \bibinfo {author} {\bibfnamefont {C.}~\bibnamefont {Hampel}}, \bibinfo
  {author} {\bibfnamefont {A.}~\bibnamefont {Hesselmann}}, \bibinfo {author}
  {\bibfnamefont {G.}~\bibnamefont {Hetzer}}, \bibinfo {author} {\bibfnamefont
  {T.}~\bibnamefont {Hrenar}}, \bibinfo {author} {\bibfnamefont
  {G.}~\bibnamefont {Jansen}}, \bibinfo {author} {\bibfnamefont
  {C.}~\bibnamefont {K{\"{o}}ppl}}, \bibinfo {author} {\bibfnamefont
  {Y.}~\bibnamefont {Liu}}, \bibinfo {author} {\bibfnamefont {A.~W.}\
  \bibnamefont {Lloyd}}, \bibinfo {author} {\bibfnamefont {R.~A.}\ \bibnamefont
  {Mata}}, \bibinfo {author} {\bibfnamefont {A.~J.}\ \bibnamefont {May}},
  \bibinfo {author} {\bibfnamefont {S.~J.}\ \bibnamefont {McNicholas}},
  \bibinfo {author} {\bibfnamefont {W.}~\bibnamefont {Meyer}}, \bibinfo
  {author} {\bibfnamefont {M.~E.}\ \bibnamefont {Mura}}, \bibinfo {author}
  {\bibfnamefont {A.}~\bibnamefont {Nicklass}}, \bibinfo {author}
  {\bibfnamefont {D.~P.}\ \bibnamefont {O'Neill}}, \bibinfo {author}
  {\bibfnamefont {P.}~\bibnamefont {Palmieri}}, \bibinfo {author}
  {\bibfnamefont {D.}~\bibnamefont {Peng}}, \bibinfo {author} {\bibfnamefont
  {K.}~\bibnamefont {Pfl{\"{u}}ger}}, \bibinfo {author} {\bibfnamefont
  {R.}~\bibnamefont {Pitzer}}, \bibinfo {author} {\bibfnamefont
  {M.}~\bibnamefont {Reiher}}, \bibinfo {author} {\bibfnamefont
  {T.}~\bibnamefont {Shiozaki}}, \bibinfo {author} {\bibfnamefont
  {H.}~\bibnamefont {Stoll}}, \bibinfo {author} {\bibfnamefont {A.~J.}\
  \bibnamefont {Stone}}, \bibinfo {author} {\bibfnamefont {R.}~\bibnamefont
  {Tarroni}}, \bibinfo {author} {\bibfnamefont {T.}~\bibnamefont
  {Thorsteinsson}}, \ and\ \bibinfo {author} {\bibfnamefont {M.}~\bibnamefont
  {Wang}},\ }\href@noop {} {\enquote {\bibinfo {title} {Molpro, version 2015.1,
  a package of ab initio programs},}\ }\bibinfo {howpublished}
  {http://www.molpro.net} (\bibinfo {year} {2015})\BibitemShut {NoStop}%
\bibitem [{\citenamefont {Jenkins}, \citenamefont {Barton},\ and\ \citenamefont
  {Mulliken}(1927)}]{27JeBaMu.NO}%
  \BibitemOpen
  \bibfield  {author} {\bibinfo {author} {\bibfnamefont {F.~A.}\ \bibnamefont
  {Jenkins}}, \bibinfo {author} {\bibfnamefont {H.~A.}\ \bibnamefont {Barton}},
  \ and\ \bibinfo {author} {\bibfnamefont {R.~S.}\ \bibnamefont {Mulliken}},\
  }\href {\doibase 10.1103/PhysRev.30.150} {\bibfield  {journal} {\bibinfo
  {journal} {Phys. Rev.}\ }\textbf {\bibinfo {volume} {30}},\ \bibinfo {pages}
  {150} (\bibinfo {year} {1927})}\BibitemShut {NoStop}%
\bibitem [{\citenamefont {Amiot}(1982)}]{82Amiot.NO}%
  \BibitemOpen
  \bibfield  {author} {\bibinfo {author} {\bibfnamefont {C.}~\bibnamefont
  {Amiot}},\ }\href {\doibase 10.1016/0022-2852(82)90301-0} {\bibfield
  {journal} {\bibinfo  {journal} {J. Mol. Spectrosc.}\ }\textbf {\bibinfo
  {volume} {94}},\ \bibinfo {pages} {150} (\bibinfo {year} {1982})}\BibitemShut
  {NoStop}%
\bibitem [{\citenamefont {Cartwright}\ \emph {et~al.}(2000)\citenamefont
  {Cartwright}, \citenamefont {Brunger}, \citenamefont {Campbell},
  \citenamefont {Mojarrabi},\ and\ \citenamefont {Teubner}}]{00CaBrCa.NO}%
  \BibitemOpen
  \bibfield  {author} {\bibinfo {author} {\bibfnamefont {D.~C.}\ \bibnamefont
  {Cartwright}}, \bibinfo {author} {\bibfnamefont {M.~J.}\ \bibnamefont
  {Brunger}}, \bibinfo {author} {\bibfnamefont {L.}~\bibnamefont {Campbell}},
  \bibinfo {author} {\bibfnamefont {B.}~\bibnamefont {Mojarrabi}}, \ and\
  \bibinfo {author} {\bibfnamefont {P.~J.~O.}\ \bibnamefont {Teubner}},\ }\href
  {\doibase 10.1029/1999ja000333} {\bibfield  {journal} {\bibinfo  {journal}
  {J. Geophys. Res.}\ }\textbf {\bibinfo {volume} {105}},\ \bibinfo {pages}
  {20857} (\bibinfo {year} {2000})}\BibitemShut {NoStop}%
\bibitem [{\citenamefont {Werner}\ \emph {et~al.}(2012)\citenamefont {Werner},
  \citenamefont {Knowles}, \citenamefont {Knizia}, \citenamefont {Manby},\ and\
  \citenamefont {Sch\"utz}}]{MOLPRO}%
  \BibitemOpen
  \bibfield  {author} {\bibinfo {author} {\bibfnamefont {H.-J.}\ \bibnamefont
  {Werner}}, \bibinfo {author} {\bibfnamefont {P.~J.}\ \bibnamefont {Knowles}},
  \bibinfo {author} {\bibfnamefont {G.}~\bibnamefont {Knizia}}, \bibinfo
  {author} {\bibfnamefont {F.~R.}\ \bibnamefont {Manby}}, \ and\ \bibinfo
  {author} {\bibfnamefont {M.}~\bibnamefont {Sch\"utz}},\ }\href {\doibase
  10.1002/wcms.82} {\bibfield  {journal} {\bibinfo  {journal} {WIREs Comput.
  Mol. Sci.}\ }\textbf {\bibinfo {volume} {2}},\ \bibinfo {pages} {242}
  (\bibinfo {year} {2012})}\BibitemShut {NoStop}%
\bibitem [{\citenamefont {Grein}\ and\ \citenamefont
  {Kapur}(1982)}]{82GrKaxx.NO}%
  \BibitemOpen
  \bibfield  {author} {\bibinfo {author} {\bibfnamefont {F.}~\bibnamefont
  {Grein}}\ and\ \bibinfo {author} {\bibfnamefont {A.}~\bibnamefont {Kapur}},\
  }\href {\doibase 10.1063/1.443622} {\bibfield  {journal} {\bibinfo  {journal}
  {J. Chem. Phys.}\ }\textbf {\bibinfo {volume} {77}},\ \bibinfo {pages} {415}
  (\bibinfo {year} {1982})}\BibitemShut {NoStop}%
\bibitem [{\citenamefont {Devivie}\ and\ \citenamefont
  {Peyerimhoff}(1988)}]{88DePexx.NO}%
  \BibitemOpen
  \bibfield  {author} {\bibinfo {author} {\bibfnamefont {R.}~\bibnamefont
  {Devivie}}\ and\ \bibinfo {author} {\bibfnamefont {S.~D.}\ \bibnamefont
  {Peyerimhoff}},\ }\href {\doibase 10.1063/1.454958} {\bibfield  {journal}
  {\bibinfo  {journal} {J. Chem. Phys.}\ }\textbf {\bibinfo {volume} {89}},\
  \bibinfo {pages} {3028} (\bibinfo {year} {1988})}\BibitemShut {NoStop}%
\bibitem [{\citenamefont {Shi}\ and\ \citenamefont {East}(2006)}]{06ShEaxx.NO}%
  \BibitemOpen
  \bibfield  {author} {\bibinfo {author} {\bibfnamefont {H.}~\bibnamefont
  {Shi}}\ and\ \bibinfo {author} {\bibfnamefont {A.~L.~L.}\ \bibnamefont
  {East}},\ }\href {\doibase 10.1063/1.2336214} {\bibfield  {journal} {\bibinfo
   {journal} {J. Chem. Phys.}\ }\textbf {\bibinfo {volume} {125}},\ \bibinfo
  {pages} {7} (\bibinfo {year} {2006})}\BibitemShut {NoStop}%
\bibitem [{\citenamefont {Cheng}, \citenamefont {Zhang},\ and\ \citenamefont
  {Cheng}(2017)}]{17ChZhCh.NO}%
  \BibitemOpen
  \bibfield  {author} {\bibinfo {author} {\bibfnamefont {J.}~\bibnamefont
  {Cheng}}, \bibinfo {author} {\bibfnamefont {H.}~\bibnamefont {Zhang}}, \ and\
  \bibinfo {author} {\bibfnamefont {X.}~\bibnamefont {Cheng}},\ }\href
  {\doibase 10.1016/j.comptc.2017.05.038} {\bibfield  {journal} {\bibinfo
  {journal} {Comput. Theor. Chem.}\ }\textbf {\bibinfo {volume} {1114}},\
  \bibinfo {pages} {165} (\bibinfo {year} {2017})}\BibitemShut {NoStop}%
\bibitem [{\citenamefont {Cheng}\ \emph {et~al.}(2017)\citenamefont {Cheng},
  \citenamefont {Zhang}, \citenamefont {Cheng},\ and\ \citenamefont
  {Song}}]{17ChZhChS.NO}%
  \BibitemOpen
  \bibfield  {author} {\bibinfo {author} {\bibfnamefont {J.}~\bibnamefont
  {Cheng}}, \bibinfo {author} {\bibfnamefont {H.}~\bibnamefont {Zhang}},
  \bibinfo {author} {\bibfnamefont {X.}~\bibnamefont {Cheng}}, \ and\ \bibinfo
  {author} {\bibfnamefont {X.}~\bibnamefont {Song}},\ }\href {\doibase
  10.1080/00268976.2017.1336261} {\bibfield  {journal} {\bibinfo  {journal}
  {Mol. Phys.}\ }\textbf {\bibinfo {volume} {115}},\ \bibinfo {pages} {2577}
  (\bibinfo {year} {2017})}\BibitemShut {NoStop}%
\bibitem [{\citenamefont {Cooper}(1982)}]{82Cooper.NO}%
  \BibitemOpen
  \bibfield  {author} {\bibinfo {author} {\bibfnamefont {D.~M.}\ \bibnamefont
  {Cooper}},\ }\href {\doibase 10.1016/0022-4073(82)90080-2} {\bibfield
  {journal} {\bibinfo  {journal} {J. Quant. Spectrosc. Radiat. Transf.}\
  }\textbf {\bibinfo {volume} {27}},\ \bibinfo {pages} {459} (\bibinfo {year}
  {1982})}\BibitemShut {NoStop}%
\bibitem [{\citenamefont {Langhoff}, \citenamefont {Bauschlicher},\ and\
  \citenamefont {Partridge}(1988)}]{88LaBaPa.NO}%
  \BibitemOpen
  \bibfield  {author} {\bibinfo {author} {\bibfnamefont {S.~R.}\ \bibnamefont
  {Langhoff}}, \bibinfo {author} {\bibfnamefont {C.~W.}\ \bibnamefont
  {Bauschlicher}}, \ and\ \bibinfo {author} {\bibfnamefont {H.}~\bibnamefont
  {Partridge}},\ }\href {\doibase 10.1063/1.455661} {\bibfield  {journal}
  {\bibinfo  {journal} {J. Chem. Phys.}\ }\textbf {\bibinfo {volume} {89}},\
  \bibinfo {pages} {4909} (\bibinfo {year} {1988})}\BibitemShut {NoStop}%
\bibitem [{\citenamefont {Langhoff}\ \emph {et~al.}(1991)\citenamefont
  {Langhoff}, \citenamefont {Partridge}, \citenamefont {Bauschlicher},\ and\
  \citenamefont {Komornicki}}]{91LaPaBa.NO}%
  \BibitemOpen
  \bibfield  {author} {\bibinfo {author} {\bibfnamefont {S.~R.}\ \bibnamefont
  {Langhoff}}, \bibinfo {author} {\bibfnamefont {H.}~\bibnamefont {Partridge}},
  \bibinfo {author} {\bibfnamefont {C.~W.}\ \bibnamefont {Bauschlicher}}, \
  and\ \bibinfo {author} {\bibfnamefont {A.}~\bibnamefont {Komornicki}},\
  }\href {\doibase 10.1063/1.460291} {\bibfield  {journal} {\bibinfo  {journal}
  {J. Chem. Phys.}\ }\textbf {\bibinfo {volume} {94}},\ \bibinfo {pages} {6638}
  (\bibinfo {year} {1991})}\BibitemShut {NoStop}%
\bibitem [{\citenamefont {Polak}\ and\ \citenamefont
  {Fiser}(2003)}]{03PoFixx.NO}%
  \BibitemOpen
  \bibfield  {author} {\bibinfo {author} {\bibfnamefont {R.}~\bibnamefont
  {Polak}}\ and\ \bibinfo {author} {\bibfnamefont {J.}~\bibnamefont {Fiser}},\
  }\href {\doibase 10.1016/s0009-2614(03)01179-5} {\bibfield  {journal}
  {\bibinfo  {journal} {Chem. Phys. Lett.}\ }\textbf {\bibinfo {volume}
  {377}},\ \bibinfo {pages} {564} (\bibinfo {year} {2003})}\BibitemShut
  {NoStop}%
\bibitem [{\citenamefont {Dunning}(1989)}]{89Dunning.ai}%
  \BibitemOpen
  \bibfield  {author} {\bibinfo {author} {\bibfnamefont {T.~H.}\ \bibnamefont
  {Dunning}},\ }\href {\doibase 10.1063/1.456153} {\bibfield  {journal}
  {\bibinfo  {journal} {J. Chem. Phys.}\ }\textbf {\bibinfo {volume} {90}},\
  \bibinfo {pages} {1007} (\bibinfo {year} {1989})}\BibitemShut {NoStop}%
\bibitem [{\citenamefont {Huber}\ and\ \citenamefont
  {Herzberg}(1979)}]{79HeHuxx.book}%
  \BibitemOpen
  \bibfield  {author} {\bibinfo {author} {\bibfnamefont {K.~P.}\ \bibnamefont
  {Huber}}\ and\ \bibinfo {author} {\bibfnamefont {G.}~\bibnamefont
  {Herzberg}},\ }\href@noop {} {\emph {\bibinfo {title} {Molecular Spectra and
  Molecular Structure IV. Constants of Diatomic Molecules}}}\ (\bibinfo
  {publisher} {Van Nostrand Reinhold Company},\ \bibinfo {address} {New York},\
  \bibinfo {year} {1979})\BibitemShut {NoStop}%
\bibitem [{\citenamefont {Murray}\ \emph {et~al.}(1994)\citenamefont {Murray},
  \citenamefont {Yoshino}, \citenamefont {Esmond}, \citenamefont {Parkinson},
  \citenamefont {Sun}, \citenamefont {Dalgarno}, \citenamefont {Thorne},\ and\
  \citenamefont {Cox}}]{94MuYoEs.NO}%
  \BibitemOpen
  \bibfield  {author} {\bibinfo {author} {\bibfnamefont {J.~E.}\ \bibnamefont
  {Murray}}, \bibinfo {author} {\bibfnamefont {K.}~\bibnamefont {Yoshino}},
  \bibinfo {author} {\bibfnamefont {J.~R.}\ \bibnamefont {Esmond}}, \bibinfo
  {author} {\bibfnamefont {W.~H.}\ \bibnamefont {Parkinson}}, \bibinfo {author}
  {\bibfnamefont {Y.}~\bibnamefont {Sun}}, \bibinfo {author} {\bibfnamefont
  {A.}~\bibnamefont {Dalgarno}}, \bibinfo {author} {\bibfnamefont {A.~P.}\
  \bibnamefont {Thorne}}, \ and\ \bibinfo {author} {\bibfnamefont
  {G.}~\bibnamefont {Cox}},\ }\href {\doibase 10.1063/1.468118} {\bibfield
  {journal} {\bibinfo  {journal} {J. Chem. Phys.}\ }\textbf {\bibinfo {volume}
  {101}},\ \bibinfo {pages} {62} (\bibinfo {year} {1994})}\BibitemShut
  {NoStop}%
\bibitem [{\citenamefont {Imajo}\ \emph {et~al.}(2000)\citenamefont {Imajo},
  \citenamefont {Yoshino}, \citenamefont {Esmond}, \citenamefont {Parkinson},
  \citenamefont {Thorne}, \citenamefont {Murray}, \citenamefont {Learner},
  \citenamefont {Cox}, \citenamefont {Cheung}, \citenamefont {Ito},\ and\
  \citenamefont {Matsui}}]{00ImYoEs.NO}%
  \BibitemOpen
  \bibfield  {author} {\bibinfo {author} {\bibfnamefont {T.}~\bibnamefont
  {Imajo}}, \bibinfo {author} {\bibfnamefont {K.}~\bibnamefont {Yoshino}},
  \bibinfo {author} {\bibfnamefont {J.~R.}\ \bibnamefont {Esmond}}, \bibinfo
  {author} {\bibfnamefont {W.~H.}\ \bibnamefont {Parkinson}}, \bibinfo {author}
  {\bibfnamefont {A.~P.}\ \bibnamefont {Thorne}}, \bibinfo {author}
  {\bibfnamefont {J.~E.}\ \bibnamefont {Murray}}, \bibinfo {author}
  {\bibfnamefont {R.~C.~M.}\ \bibnamefont {Learner}}, \bibinfo {author}
  {\bibfnamefont {G.}~\bibnamefont {Cox}}, \bibinfo {author} {\bibfnamefont
  {A.~S.-C.}\ \bibnamefont {Cheung}}, \bibinfo {author} {\bibfnamefont
  {K.}~\bibnamefont {Ito}}, \ and\ \bibinfo {author} {\bibfnamefont
  {T.}~\bibnamefont {Matsui}},\ }\href {\doibase 10.1063/1.480790} {\bibfield
  {journal} {\bibinfo  {journal} {J. Chem. Phys.}\ }\textbf {\bibinfo {volume}
  {112}},\ \bibinfo {pages} {2251} (\bibinfo {year} {2000})}\BibitemShut
  {NoStop}%
\bibitem [{\citenamefont {Cheung}\ \emph {et~al.}(2002)\citenamefont {Cheung},
  \citenamefont {Lo}, \citenamefont {Leung}, \citenamefont {Yoshino},
  \citenamefont {Thorne}, \citenamefont {Murray}, \citenamefont {Ito},
  \citenamefont {Matsui},\ and\ \citenamefont {Imajo}}]{02ChLoLe.NO}%
  \BibitemOpen
  \bibfield  {author} {\bibinfo {author} {\bibfnamefont {A.~S.-C.}\
  \bibnamefont {Cheung}}, \bibinfo {author} {\bibfnamefont {D.~H.-Y.}\
  \bibnamefont {Lo}}, \bibinfo {author} {\bibfnamefont {K.~W.-S.}\ \bibnamefont
  {Leung}}, \bibinfo {author} {\bibfnamefont {K.}~\bibnamefont {Yoshino}},
  \bibinfo {author} {\bibfnamefont {A.~P.}\ \bibnamefont {Thorne}}, \bibinfo
  {author} {\bibfnamefont {J.~E.}\ \bibnamefont {Murray}}, \bibinfo {author}
  {\bibfnamefont {K.}~\bibnamefont {Ito}}, \bibinfo {author} {\bibfnamefont
  {T.}~\bibnamefont {Matsui}}, \ and\ \bibinfo {author} {\bibfnamefont
  {T.}~\bibnamefont {Imajo}},\ }\href {\doibase 10.1063/1.1421064} {\bibfield
  {journal} {\bibinfo  {journal} {J. Chem. Phys.}\ }\textbf {\bibinfo {volume}
  {116}},\ \bibinfo {pages} {155} (\bibinfo {year} {2002})}\BibitemShut
  {NoStop}%
\bibitem [{\citenamefont {Rufus}\ \emph {et~al.}(2002)\citenamefont {Rufus},
  \citenamefont {Yoshino}, \citenamefont {Thorne}, \citenamefont {Murray},
  \citenamefont {Imajo}, \citenamefont {Ito},\ and\ \citenamefont
  {Matsui}}]{02RuYoTh.NO}%
  \BibitemOpen
  \bibfield  {author} {\bibinfo {author} {\bibfnamefont {J.}~\bibnamefont
  {Rufus}}, \bibinfo {author} {\bibfnamefont {K.}~\bibnamefont {Yoshino}},
  \bibinfo {author} {\bibfnamefont {A.~P.}\ \bibnamefont {Thorne}}, \bibinfo
  {author} {\bibfnamefont {J.~E.}\ \bibnamefont {Murray}}, \bibinfo {author}
  {\bibfnamefont {T.}~\bibnamefont {Imajo}}, \bibinfo {author} {\bibfnamefont
  {K.}~\bibnamefont {Ito}}, \ and\ \bibinfo {author} {\bibfnamefont
  {T.}~\bibnamefont {Matsui}},\ }\href {\doibase 10.1063/1.1520535} {\bibfield
  {journal} {\bibinfo  {journal} {J. Chem. Phys.}\ }\textbf {\bibinfo {volume}
  {117}},\ \bibinfo {pages} {10621} (\bibinfo {year} {2002})}\BibitemShut
  {NoStop}%
\bibitem [{\citenamefont {Faris}\ and\ \citenamefont
  {Cosby}(1992)}]{92FaCoxx.NO}%
  \BibitemOpen
  \bibfield  {author} {\bibinfo {author} {\bibfnamefont {G.~W.}\ \bibnamefont
  {Faris}}\ and\ \bibinfo {author} {\bibfnamefont {P.~C.}\ \bibnamefont
  {Cosby}},\ }\href {\doibase 10.1063/1.463533} {\bibfield  {journal} {\bibinfo
   {journal} {J. Chem. Phys.}\ }\textbf {\bibinfo {volume} {97}},\ \bibinfo
  {pages} {7073} (\bibinfo {year} {1992})}\BibitemShut {NoStop}%
\bibitem [{\citenamefont {Drabbels}\ and\ \citenamefont
  {Wodtke}(1996)}]{96DrWoxx.NO}%
  \BibitemOpen
  \bibfield  {author} {\bibinfo {author} {\bibfnamefont {M.}~\bibnamefont
  {Drabbels}}\ and\ \bibinfo {author} {\bibfnamefont {A.~M.}\ \bibnamefont
  {Wodtke}},\ }\href {\doibase 10.1016/0009-2614(96)00409-5} {\bibfield
  {journal} {\bibinfo  {journal} {Chem. Phys. Lett.}\ }\textbf {\bibinfo
  {volume} {256}},\ \bibinfo {pages} {8} (\bibinfo {year} {1996})}\BibitemShut
  {NoStop}%
\bibitem [{\citenamefont {Amiot}\ and\ \citenamefont
  {Verges}(1982)}]{82AmVexx.NO}%
  \BibitemOpen
  \bibfield  {author} {\bibinfo {author} {\bibfnamefont {C.}~\bibnamefont
  {Amiot}}\ and\ \bibinfo {author} {\bibfnamefont {J.}~\bibnamefont {Verges}},\
  }\href {\doibase 10.1088/0031-8949/25/2/009} {\bibfield  {journal} {\bibinfo
  {journal} {Phys. Scr.}\ }\textbf {\bibinfo {volume} {25}},\ \bibinfo {pages}
  {302} (\bibinfo {year} {1982})}\BibitemShut {NoStop}%
\bibitem [{\citenamefont {Furtenbacher}, \citenamefont {{Cs\'asz\'ar}},\ and\
  \citenamefont {Tennyson}(2007{\natexlab{b}})}]{MARVEL}%
  \BibitemOpen
  \bibfield  {author} {\bibinfo {author} {\bibfnamefont {T.}~\bibnamefont
  {Furtenbacher}}, \bibinfo {author} {\bibfnamefont {A.~G.}\ \bibnamefont
  {{Cs\'asz\'ar}}}, \ and\ \bibinfo {author} {\bibfnamefont {J.}~\bibnamefont
  {Tennyson}},\ }\href@noop {} {\bibfield  {journal} {\bibinfo  {journal} {J.
  Mol. Spectrosc.}\ }\textbf {\bibinfo {volume} {245}},\ \bibinfo {pages} {115}
  (\bibinfo {year} {2007}{\natexlab{b}})}\BibitemShut {NoStop}%
\bibitem [{\citenamefont {Thorne}\ \emph {et~al.}(2005)\citenamefont {Thorne},
  \citenamefont {Rufus}, \citenamefont {Yoshino}, \citenamefont {Cheung},\ and\
  \citenamefont {Imajo}}]{05ThRuYo.NO}%
  \BibitemOpen
  \bibfield  {author} {\bibinfo {author} {\bibfnamefont {A.~P.}\ \bibnamefont
  {Thorne}}, \bibinfo {author} {\bibfnamefont {J.}~\bibnamefont {Rufus}},
  \bibinfo {author} {\bibfnamefont {K.}~\bibnamefont {Yoshino}}, \bibinfo
  {author} {\bibfnamefont {A.~S.-C.}\ \bibnamefont {Cheung}}, \ and\ \bibinfo
  {author} {\bibfnamefont {T.}~\bibnamefont {Imajo}},\ }\href {\doibase
  10.1063/1.1883648} {\bibfield  {journal} {\bibinfo  {journal} {J. Chem.
  Phys.}\ }\textbf {\bibinfo {volume} {122}},\ \bibinfo {pages} {179901}
  (\bibinfo {year} {2005})}\BibitemShut {NoStop}%
\bibitem [{\citenamefont {Ventura}\ and\ \citenamefont
  {Fellows}(2020)}]{20VeFexx.NO}%
  \BibitemOpen
  \bibfield  {author} {\bibinfo {author} {\bibfnamefont {L.~R.}\ \bibnamefont
  {Ventura}}\ and\ \bibinfo {author} {\bibfnamefont {C.~E.}\ \bibnamefont
  {Fellows}},\ }\href {\doibase 10.1016/j.jqsrt.2020.106900} {\bibfield
  {journal} {\bibinfo  {journal} {J. Quant. Spectrosc. Radiat. Transf.}\
  }\textbf {\bibinfo {volume} {246}},\ \bibinfo {pages} {2} (\bibinfo {year}
  {2020})}\BibitemShut {NoStop}%
\bibitem [{\citenamefont {Sulakshina}\ and\ \citenamefont
  {Borkov}(2018)}]{18SuBoxx.NO}%
  \BibitemOpen
  \bibfield  {author} {\bibinfo {author} {\bibfnamefont {O.~N.}\ \bibnamefont
  {Sulakshina}}\ and\ \bibinfo {author} {\bibfnamefont {Y.~G.}\ \bibnamefont
  {Borkov}},\ }\href {\doibase 10.1016/j.jqsrt.2018.01.020} {\bibfield
  {journal} {\bibinfo  {journal} {J. Quant. Spectrosc. Radiat. Transf.}\
  }\textbf {\bibinfo {volume} {209}},\ \bibinfo {pages} {171} (\bibinfo {year}
  {2018})}\BibitemShut {NoStop}%
\bibitem [{\citenamefont {Yurchenko}\ \emph {et~al.}(2016)\citenamefont
  {Yurchenko}, \citenamefont {Lodi}, \citenamefont {Tennyson},\ and\
  \citenamefont {Stolyarov}}]{jt609}%
  \BibitemOpen
  \bibfield  {author} {\bibinfo {author} {\bibfnamefont {S.~N.}\ \bibnamefont
  {Yurchenko}}, \bibinfo {author} {\bibfnamefont {L.}~\bibnamefont {Lodi}},
  \bibinfo {author} {\bibfnamefont {J.}~\bibnamefont {Tennyson}}, \ and\
  \bibinfo {author} {\bibfnamefont {A.~V.}\ \bibnamefont {Stolyarov}},\ }\href
  {\doibase 10.1016/j.cpc.2015.12.021} {\bibfield  {journal} {\bibinfo
  {journal} {Comput. Phys. Commun.}\ }\textbf {\bibinfo {volume} {202}},\
  \bibinfo {pages} {262} (\bibinfo {year} {2016})}\BibitemShut {NoStop}%
\bibitem [{\citenamefont {Tennyson}\ \emph
  {et~al.}(2016{\natexlab{b}})\citenamefont {Tennyson}, \citenamefont {Lodi},
  \citenamefont {McKemmish},\ and\ \citenamefont {Yurchenko}}]{jt632}%
  \BibitemOpen
  \bibfield  {author} {\bibinfo {author} {\bibfnamefont {J.}~\bibnamefont
  {Tennyson}}, \bibinfo {author} {\bibfnamefont {L.}~\bibnamefont {Lodi}},
  \bibinfo {author} {\bibfnamefont {L.~K.}\ \bibnamefont {McKemmish}}, \ and\
  \bibinfo {author} {\bibfnamefont {S.~N.}\ \bibnamefont {Yurchenko}},\
  }\href@noop {} {\bibfield  {journal} {\bibinfo  {journal} {J. Phys. B-At.
  Mol. Opt. Phys.}\ }\textbf {\bibinfo {volume} {49}},\ \bibinfo {pages}
  {102001} (\bibinfo {year} {2016}{\natexlab{b}})}\BibitemShut {NoStop}%
\bibitem [{\citenamefont {Colbert}\ and\ \citenamefont
  {Miller}(1992)}]{92CoMixx.method}%
  \BibitemOpen
  \bibfield  {author} {\bibinfo {author} {\bibfnamefont {D.~T.}\ \bibnamefont
  {Colbert}}\ and\ \bibinfo {author} {\bibfnamefont {W.~H.}\ \bibnamefont
  {Miller}},\ }\href {\doibase 10.1063/1.462100} {\bibfield  {journal}
  {\bibinfo  {journal} {J. Chem. Phys.}\ }\textbf {\bibinfo {volume} {96}},\
  \bibinfo {pages} {1982} (\bibinfo {year} {1992})}\BibitemShut {NoStop}%
\bibitem [{\citenamefont {Lee}\ \emph {et~al.}(1999)\citenamefont {Lee},
  \citenamefont {Seto}, \citenamefont {Hirao}, \citenamefont {Bernath},\ and\
  \citenamefont {Le~Roy}}]{EMO}%
  \BibitemOpen
  \bibfield  {author} {\bibinfo {author} {\bibfnamefont {E.~G.}\ \bibnamefont
  {Lee}}, \bibinfo {author} {\bibfnamefont {J.~Y.}\ \bibnamefont {Seto}},
  \bibinfo {author} {\bibfnamefont {T.}~\bibnamefont {Hirao}}, \bibinfo
  {author} {\bibfnamefont {P.~F.}\ \bibnamefont {Bernath}}, \ and\ \bibinfo
  {author} {\bibfnamefont {R.~J.}\ \bibnamefont {Le~Roy}},\ }\href {\doibase
  10.1006/jmsp.1998.7789} {\bibfield  {journal} {\bibinfo  {journal} {J. Mol.
  Spectrosc.}\ }\textbf {\bibinfo {volume} {194}},\ \bibinfo {pages} {197}
  (\bibinfo {year} {1999})}\BibitemShut {NoStop}%
\bibitem [{\citenamefont {Karman}\ \emph {et~al.}(2018)\citenamefont {Karman},
  \citenamefont {Besemer}, \citenamefont {van~der Avoird},\ and\ \citenamefont
  {Groenenboom}}]{18KaBeAv}%
  \BibitemOpen
  \bibfield  {author} {\bibinfo {author} {\bibfnamefont {T.}~\bibnamefont
  {Karman}}, \bibinfo {author} {\bibfnamefont {M.}~\bibnamefont {Besemer}},
  \bibinfo {author} {\bibfnamefont {A.}~\bibnamefont {van~der Avoird}}, \ and\
  \bibinfo {author} {\bibfnamefont {G.~C.}\ \bibnamefont {Groenenboom}},\
  }\href {\doibase 10.1063/1.5013091} {\bibfield  {journal} {\bibinfo
  {journal} {J. Chem. Phys.}\ }\textbf {\bibinfo {volume} {148}},\ \bibinfo
  {pages} {094105} (\bibinfo {year} {2018})}\BibitemShut {NoStop}%
\bibitem [{\citenamefont {Yurchenko}\ \emph {et~al.}(2018)\citenamefont
  {Yurchenko}, \citenamefont {Szabo}, \citenamefont {Pyatenko},\ and\
  \citenamefont {Tennyson}}]{jt736}%
  \BibitemOpen
  \bibfield  {author} {\bibinfo {author} {\bibfnamefont {S.~N.}\ \bibnamefont
  {Yurchenko}}, \bibinfo {author} {\bibfnamefont {I.}~\bibnamefont {Szabo}},
  \bibinfo {author} {\bibfnamefont {E.}~\bibnamefont {Pyatenko}}, \ and\
  \bibinfo {author} {\bibfnamefont {J.}~\bibnamefont {Tennyson}},\ }\href
  {\doibase 10.1093/mnras/sty2050} {\bibfield  {journal} {\bibinfo  {journal}
  {Mon. Not. Roy. Astron. Soc.}\ }\textbf {\bibinfo {volume} {480}},\ \bibinfo
  {pages} {3397} (\bibinfo {year} {2018})}\BibitemShut {NoStop}%
\bibitem [{\citenamefont {Brown}\ and\ \citenamefont
  {Merer}(1979)}]{79BrMexx.methods}%
  \BibitemOpen
  \bibfield  {author} {\bibinfo {author} {\bibfnamefont {J.~M.}\ \bibnamefont
  {Brown}}\ and\ \bibinfo {author} {\bibfnamefont {A.~J.}\ \bibnamefont
  {Merer}},\ }\href {\doibase 10.1016/0022-2852(79)90172-3} {\bibfield
  {journal} {\bibinfo  {journal} {J. Mol. Spectrosc.}\ }\textbf {\bibinfo
  {volume} {74}},\ \bibinfo {pages} {488} (\bibinfo {year} {1979})}\BibitemShut
  {NoStop}%
\end{thebibliography}%

\end{document}